\documentclass[12pt]{iopart}

\usepackage[utf8]{inputenc}
\usepackage{iopams}
\usepackage{graphicx}
\usepackage{tabularx}
\usepackage{color}
\usepackage{hyperref}
\begin{document}

\title[Particle Production via Strings and Baryon Stopping]{Particle Production via Strings and Baryon Stopping within a Hadronic Transport Approach}

\author{J Mohs$^{1,2,3}$, S Ryu$^{2,1}$ and H Elfner$^{3,2,1}$}

\address{$^1$ Frankfurt Institute For Advanced Studies, Ruth-Moufang-Straße 1, 60438 Frankfurt am Main}
\address{$^2$ Institute for Theoretical Physics, Goethe University, Max-von-Laue-Strasse 1, 60438 Frankfurt am Main, Germany}
\address{$^3$ GSI Helmholtzzentrum f\"ur Schwerionenforschung, Planckstr. 1, 64291 Darmstadt, Germany}
\ead{mohs@fias.uni-frankfurt.de}
\vspace{10pt}
\begin{indented}
\item[]\today
\end{indented}

\begin{abstract}
The stopping of baryons in heavy ion collisions at beam momenta of $p_{\rm lab} = 20-160A$ GeV  is lacking a quantitative description within theoretical calculations.  Heavy ion reactions at these energies are experimentally explored at the Super Proton Synchrotron (SPS) and the Relativistic Heavy Ion Collider (RHIC) and will be studied at future facilities such as FAIR and NICA. Since the net baryon density is determined by the amount of stopping, this is the pre-requisiste for any investigation of other observables related to structures in the QCD phase diagram such as a first-order phase transition or a critical endpoint. 
In this work we employ a string model for treating hadron-hadron interactions within a hadronic transport approach (SMASH, Simulating Many Accelerated Strongly-interacting Hadrons). Free parameters of the string excitation and decay are tuned to match experimental measurements in elementary proton-proton collisions, where some mismatch in the $x_F$ distribution of protons is still present. Afterwards, the model is applied to heavy ion collisions, where the experimentally observed change of the shape of the proton rapidity spectrum from a single peak structure to a double peak structure with increasing beam energy is reproduced. Heavy ion collisions provide the opportunity to study the formation process of string fragments in terms of formation times and reduced interaction cross-sections for pre-formed hadrons. A good agreement with the measured rapidity spectra of protons and pions is achieved while insights on the fragmentation process are obtained. In the future, the presented approach can be used to create event-by-event initial conditions for hybrid calculations.
\end{abstract}


%
%
%

\section{Introduction}

Understanding the properties of strongly interacting matter has been a long standing problem that can be addressed by studying the QCD phase diagram.
In the case of high temperature and vanishing baryonic chemical potential, it was demonstrated that there can be a crossover phase transition instead of a first-order one, depending on the number of quark flavor and their masses \cite{Brown:1990ev}.
More recent lattice QCD computations \cite{Borsanyi:2013bia, Bazavov:2014pvz} show that there is a crossover transition between a hadronic gas and a quark-gluon plasma phase, if one goes to higher temperature ($T$) keeping the net baryon chemical potential ($\mu_B$) near zero.
On the other hand, the QCD phase transition at $T = 0$ and finite-$\mu_B$ has been studied based on effective models, such as the NJL model \cite{Asakawa:1989bq} and the composite-operator formalism \cite{Barducci:1989eu}.
It was demonstrated that the phase transition in this regime is first-order.
The existence of a critical end point (CEP) is justified by the fact that the QCD matter exhibits different types of phase transition in two limiting cases \cite{Stephanov:2004wx}.
Many theoretical and experimental studies in heavy ion physics have been aiming to find where the CEP is located in the $T$-$\mu_B$ plane.
On the experimental side, heavy ion collisions at various collision energies and several system sizes are carried out in order to probe a wide range in both temperature and baryon chemical potential.
Those include the beam energy scan performed at RHIC \cite{Adamczyk:2013dal, Adamczyk:2014fia, Adare:2015aqk} and the CERN-SPS \cite{Appelshauser:1998yb, Alt:2006dk, Blume:2007kw}.
In the future, this region will be studied further by CBM at FAIR and at NICA.

To connect the final state observables on particle yields and spectra with the properties of hot and dense QCD matter, sophisticated dynamical approaches are indespensable. 
The bulk observables in ultra-relativistic heavy ion collisions at RHIC and the LHC are successfully described by solving the hydrodynamic equations \cite{Gale:2013da, deSouza:2015ena}.
Hybrid approaches, which separate the non-equilibrium dynamics in the early stages from a hydrodynamic evolution of the thermalized medium, have proven to give a realistic description of heavy ion collisions also at lower beam energies  \cite{Petersen:2008dd,Karpenko:2015xea}.
The dynamical initialization of hybrid approaches is explored as well in \cite{Du:2018mpf,Shen:2017bsr,Murase:2019gro}.

In particular, the dynamics of baryon stopping has received some attention more recently, since it has been realized that the mean of the proton distribution should be understood before investigating higher order cumulants that are associated with a critical endpoint \cite{Adamczyk:2013dal}. The experimentally observed stopping of baryons \cite{Blume:2007kw} is still lacking a quantitative description within theoretical models. In principle, there are three different options:
\begin{enumerate}
\item Push the gluon saturation picture down in energy and extend it to three dimensions as explored in \cite{McLerran:2018avb}
\item Study the source terms of projectile and target into the fireball fluid in a 3-fluid hydrodynamics approach \cite{Ivanov:2005yw,Batyuk:2016qmb}
\item Investigate the details of a string hadron transport approach for the initial non-equilibrium evolution \cite{Steinheimer:2007iy,Bialas:2016epd,Bialas:2017lkm}
\end{enumerate}
Here, we follow the last point and apply the hadronic transport model SMASH to understand the stopping of baryons in the SPS energy range.
This approach can be employed for the description of the early stages of a heavy ion collision, since microscopic transport is applicable to non-equilibrium circumstances.
In this work, SMASH is employed to simulate the full evolution of a heavy ion collision.
In the relevant energy range for this work, it is extremely important to have three-dimensional initial conditions
for starting a hydrodynamic evolution, since the system cannot be assumed to be boost invariant and the
colliding nuclei are too slow to reasonably neglect their longitudinal extent because of length contraction.

The paper is structured as follows:
Details of the transport model with a focus on the implementation of cross sections and the particle production at intermediate energies are given in section \ref{sec_model_description}.
Section \ref{sec_proton_proton_collisions} continues with calculations for proton-proton collisions, where the influence of varying the free parameters is investigated and the best possible set of parameters is determined.
In section \ref{sec_heavy_ion_collisions} we advance to heavy ion collisions, where one has the opportunity to study the interactions of string fragments and their formation process.
Finally we provide calculations for the time of the collision where the colliding nuclei just passed through each other in section \ref{sec_initial_state} which can serve as event-by-event initial state profiles for hydrodynamic calculations.

\section{Model Description}
\label{sec_model_description}
In this work, we investigate baryon stopping within the transport model SMASH \cite{Weil:2016zrk}. The code is publicly available on Github, see \url{https://smash-transport.github.io/}.
The degrees of freedom within the calculation are hadronic. The properties of the hadrons are adopted from the Particle Data Group 2018 \cite{PDG}, where the  more established resonances up to a mass of $m\approx2\,\mathrm{GeV}$ are included. SMASH has been tested against an analytic solution of the Boltzmann equation \cite{Tindall:2016try} and strangeness as well as dilepton production has been confronted with experimental data at GSI-SIS energies \cite{Steinberg:2018jvv,Staudenmaier:2017vtq}.

\begin{figure}[h]
  \centering
  \includegraphics[width=0.65\textwidth]{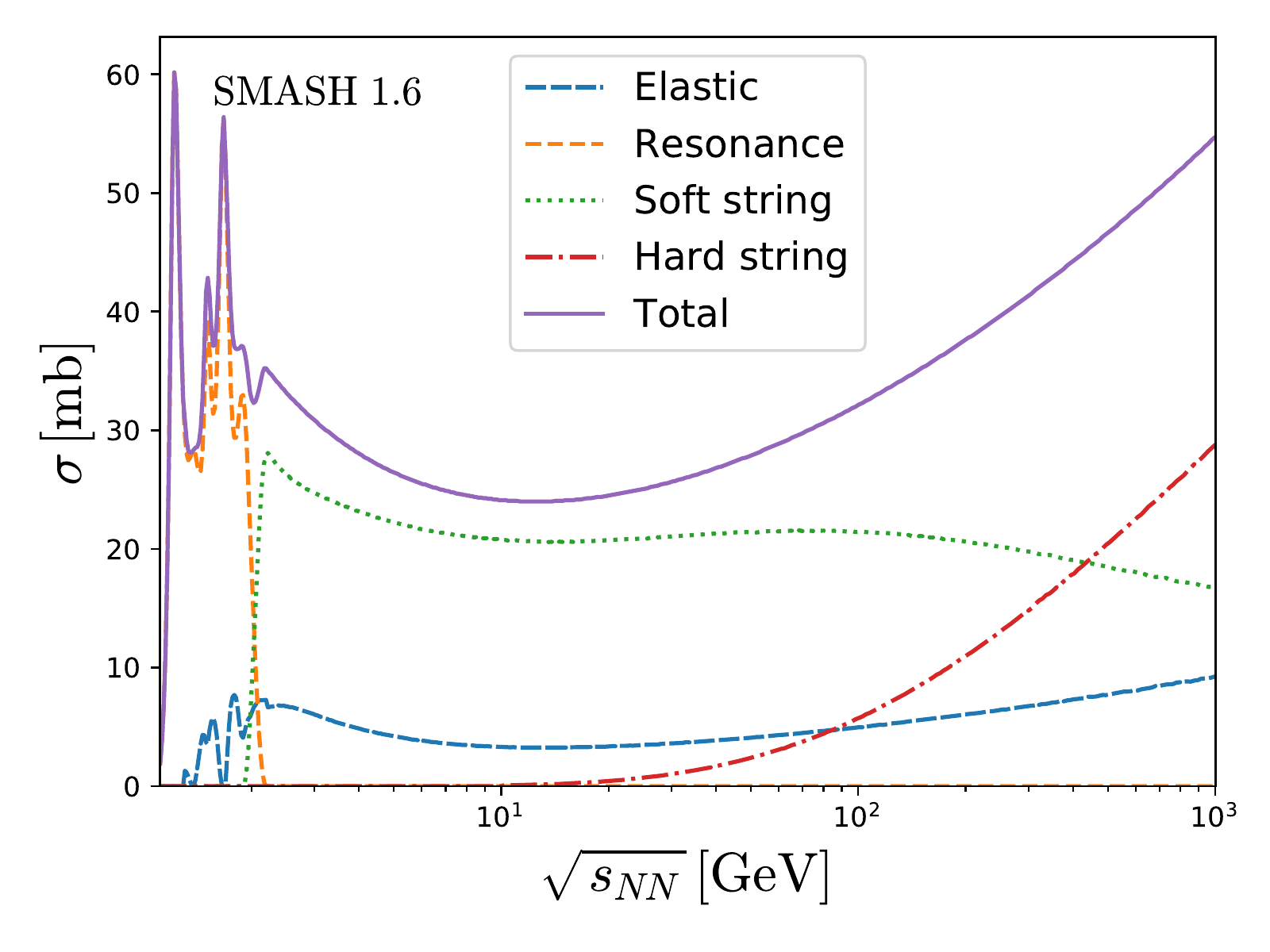}
  \caption{Cross section of a proton interacting with a negatively charged pion as a function of the center of mass energy of the colliding hadrons within SMASH. The total cross section is split into contributions from elastic collisions, resonance formations, soft string excitations and hard string excitations via \textsc{Pythia}.}
  \label{xsec_plot}
\end{figure}

The inelastic scatterings between hadrons at low energies are described via resonance formations and decays.
Since there are no resonances with masses larger than $m\approx 2 \,\mathrm{GeV}$ in the calculation, the cross section for resonance formations fades out when the center of mass energy of the interacting hadrons grows larger.
This can be seen as an example for the proton-pion cross section shown in figure \ref{xsec_plot}. 

Let us start with an overview of the general setup of our approach while more details are presented in the following sub-sections. 
In order to investigate baryon stopping at higher incident energies, in this work, a string model is employed, where colliding hadrons are excited to strings which then fragment producing new particles. In the transition region between resonances and string processes, the respective cross-sections are weighted with a linear function to achieve a smooth interpolation between both regimes. This is important to avoid artificial high mass resonances that are suppressed in this way.

The transition region is chosen such that it starts at a large enough energy to still include the resonant structures in the cross section and ends at a small enough energy so that the cross section from resonances does not fade out in the transition region.
For most combinations of particle species, the transition region starts at the sum of the masses of the colliding hadrons plus $0.9 \,\mathrm{GeV}$ and has a width of $1\,\mathrm{GeV}$.
For two very important special cases, the transition region is individually specified.
The first one is nucleon-pion collisions, where the transition from resonances to strings takes place between $1.9\,\mathrm{GeV}$ and $2.2\,\mathrm{GeV}$. In this case, the transition region is shorter, because the contribution from resonances is too small above $\sqrt{s}=2.2\,\mathrm{GeV}$.
The second special case is in collisions between two nucleons.
Here, the transition region spans from $4\,\mathrm{GeV}$ to $5\,\mathrm{GeV}$.
Compared to the default, the transition region is shifted to higher $\sqrt{s}$ because up to $4\,\mathrm{GeV}$, the total cross section from resonances reproduces the measurement and the exclusive cross sections from resonances are more realistic at low energies.

The calculation for the string excitation is split into hard and soft processes.
The hard string processes are relevant for very  high energetic binary interactions as can be seen in figure \ref{xsec_plot},  where perturbative QCD is applicable.
For the description of the pQCD scatterings, the string excitation and the string fragmentation, \textsc{Pythia} 8.235 \cite{Andersson:1983ia, Sjostrand:2014zea} is used. The hard string routine is described in more detail in section \ref{sec_hard_string_routine}.
The hard string process, where pQCD interactions are involved, is based on the $p_T$-ordered multiparton interaction (MPI) framework with initial and final state radiations \cite{Corke:2009tk}.
Given that pQCD is not applicable at low momentum scale, the lower $p_T$ threshold of those partonic interactions is chosen to be $1.5\,\textrm{GeV}$ and the pQCD cross section is computed accordingly.

In the transition region where the energy is too large to have cross sections via resonances but too low to apply pQCD, a phenomenological model for the excitation of strings is implemented.
In single diffractive, double diffractive and non-diffractive processes, strings are excited in hadronic interactions.
Using the calculated mass and momentum of the string as well as the flavor of the leading quarks, \textsc{Pythia} is employed only for the fragmentation of the string. Details of the string excitation at intermediate energies can be found in section \ref{sec_soft_string_routine}.

Figure \ref{xsec_plot} also shows the contribution of elastic scatterings to the total cross section.
Elastic collisions play an important role at all beam energies, since in heavy ion collisions, a large amount of nucleons will scatter elastically.
Especially for baryon stopping, the angular distribution of the final state particles in elastic scatterings plays an important role as shown in section \ref{sec_heavy_ion_collisions}.

\subsection{Cross Sections for String Processes}
\label{sec_cross_sections_for_string_processes}

The total cross section for each pair of hadrons is required within SMASH to determine, if two particles will scatter. Subsequently, the actual process is decided randomly according to the underlying partial cross-sections for the different channels. 

The total $\sigma_\mathrm{tot}$ and the elastic $\sigma_\mathrm{el}$ cross section, which is shown for the example of $p$-$\pi^-$ collisions in figure \ref{xsec_plot}, is parameterized to fit the experimental data. 
The inelastic cross section $\sigma_\mathrm{inel}$ is the difference between the two
\begin{equation}
\sigma_\mathrm{inel}=\sigma_\mathrm{tot}-\sigma_\mathrm{el}\,.
\end{equation}
The parametrizations for the total and elastic cross sections are taken from \cite{Tanabashi:2018oca} and \cite{Buss:2011mx,Weil:2013mya} respectively. 
For the process selection, cross sections for both the single $\sigma_\mathrm{SD}$ and double diffractive $\sigma_\mathrm{DD}$ processes are necessary. They are estimated in \cite{Schuler:1993wr} and implemented in \textsc{Pythia}. From $\sigma_\mathrm{SD}$ and $\sigma_\mathrm{DD}$ the non-diffractive cross section $\sigma_\mathrm{ND}$ is derived
\begin{equation}
\sigma_\mathrm{ND}=\sigma_\mathrm{inel}-\sigma_\mathrm{SD}-\sigma_\mathrm{DD}\,.
\end{equation}
The non-diffractive cross section includes the hard and soft non-diffractive processes. The cross section for hard non-diffractive processes is based on the pQCD cross section $\sigma_\mathrm{hard}$ from partonic interactions.
It is given by
\begin{equation}
  \sigma_\mathrm{hard}=\sum_{i,j,k}\int dx_1\int dx_2f_i(x_1) f_j(x_2) \,
    \sigma_{i,j}^k |_{\scriptsize \hat{p}_{\tiny T, \textrm{min}}} \,,
  \label{eq_hard_xsec}
\end{equation}
where $\sigma_{i,j}^k |_{\scriptsize \hat{p}_{\tiny T, \textrm{min}}}$ is the cross section for a subprocess $k$ between two partonic flavors $i$ and $j$ with minimum transverse momentum transfer $\hat{p}_{\scriptsize T, \textrm{min}}$, which is chosen to be $1.5\,\textrm{GeV}$.
The parton distribution function $f_i(x)$ provides the average number of flavor $i$ carrying the momentum fraction $x$ of the incoming hadron.
The NNPDF 2.3 parton distribution function with QED correction \cite{Ball:2013hta} is used in this work.
The sum takes each possible combination of partons from each ingoing hadron into account.

This pQCD cross section can therefore be larger than $\sigma_\mathrm{ND}$, incorporating the information of multi-parton scattering.
We take the multi-partion interaction (MPI) picture \cite{Sjostrand:1987su} and interpret the ratio $\sigma_\mathrm{hard} / \sigma_\mathrm{ND}$ to be the number of partonic interactions involved in a hadronic interaction, rather than the probability to have a hard non-diffractive interaction.
In addition, the number of parton interactions is assumed to follow a Poissonian distribution, where the average is given by $\sigma_\mathrm{hard}/\sigma_\mathrm{ND}$.
The probability of having no hard interaction is calculated according to the Poissonian distribution as
\begin{equation}
P(0)=\exp\left(-\frac{\sigma_\mathrm{hard}}{\sigma_\mathrm{ND}}\right)\,.
\end{equation}
In this case, the process is assumed to be soft non-diffractive, leading to a soft non-diffractive cross section $\sigma_\mathrm{ND,soft}$ of
\begin{equation}
\sigma_\mathrm{ND,soft}=\sigma_\mathrm{ND}\exp\left(-\frac{\sigma_\mathrm{hard}}{\sigma_\mathrm{ND}}\right)\,.
\end{equation}
Finally, the cross section $\sigma_\mathrm{ND,hard}$ for the hard string process follows as
\begin{equation}
\sigma_\mathrm{ND,hard} = \sigma_\mathrm{ND} -\sigma_\mathrm{ND,soft}\,.
\end{equation}
Since the pQCD cross sections can only be applied at sufficiently large energies, there is no contribution from hard non-diffractive processes below collision energies of $10\,\mathrm{GeV}$. If the energy is smaller, all non-diffractive processes will be soft.

Due to the fact that \textsc{Pythia 8} accepts only (anti-)nucleons and pions as the incoming hadrons, it is necessary to extrapolate these processes to handle arbitrary pairs of incoming hadrons. This is done by mapping hadronic species onto pions and nucleons and then rescaling cross sections based on the additive quark model.
If a baryon has positive electric charge, it is mapped onto a proton. Otherwise it is mapped onto a neutron. Similarly, if a meson has positive or vanishing electric charge, it is mapped onto $\pi^+$. Otherwise it is mapped onto $\pi^-$.
The additive quark model \cite{Goulianos:1982vk} is implemented in a similar manner as in the UrQMD model \cite{Bass:1998ca, Bleicher:1999xi}.
The total, elastic, diffractive and pQCD cross sections are multiplied by a constant factor, which depends on the valence quark/antiquark contents of the incoming hadrons.
\begin{equation}
  \sigma_{\scriptsize h_1 h_2} = \left( 1 - 0.4 \, \frac{n_{\scriptsize s / \bar{s}, 1}}{n_{\scriptsize q / \bar{q}, 1}} \right)
    \left( 1 - 0.4 \, \frac{n_{\scriptsize s / \bar{s}, 2}}{n_{\scriptsize q / \bar{q}, 2}} \right)
    \sigma_{\scriptsize h'_1 h'_2}\,,
  \label{eq_xsec_aqm}
\end{equation}
where $n_{q / \bar{q}}$ is the number of quark/antiquark constituents, while $n_{s / \bar{s}}$ is the number of strange quark/antiquark constituents.
$h$ and $h'$ stand for the incoming and mapped hadronic species respectively.
\subsection{Hard String Routine}
\label{sec_hard_string_routine}
Hard non-diffractive string excitations dominate the hadronic cross section at large center of mass energies.
As mentioned in section \ref{sec_cross_sections_for_string_processes}, \textsc{Pythia} 8 accepts only a limited number of species as incoming hadrons, and it is necessary to extrapolate hard non-diffractive scattering handled by \textsc{Pythia} 8 to all other hadronic species.
This is particularly crucial in high-energy heavy ion collisions, where plenty of hadrons other than (anti-)nucleons and pions are produced by primary nucleon-nucleon collisions.
To do this extrapolation, we rely on the assumption that the structure functions (or parton distribution functions) of all mesons and (anti-)baryons look similar to respectively those of pions and (anti-)nucleons once we swap the valence quark flavors.

Technically, this is achieved by mapping different hadron species to (anti-)nucleons and pions, where the quantum numbers of the original and mapped particle are as similar as possible.
This is done in the same way as the mapping for the cross sections, which is described in section \ref{sec_cross_sections_for_string_processes}.
Before the produced strings are fragmented within the \textsc{Pythia} calculation, light (anti-)quarks are exchanged with quarks of the original flavor. The momenta of all particles are rescaled in order to conserve the energy of the system, since the constituent masses are affected by the flavor exchange. Due to annihilation processes, it is not always possible to find a quark with the flavor of the mapped quark.
In this case, a gluon is split into a quark-antiquark pair with the flavor of the mapped quark or anti-quark.

\subsection{Soft String Routine}
\label{sec_soft_string_routine}

The soft string excitations are the most abundant processes in the intermediate energy range as can be seen in figure \ref{xsec_plot}. As in UrQMD \cite{Bass:1998ca,Bleicher:1999xi}, the excitation of a soft string can be performed according to one of three subprocesses, the single diffractive 
(see section \ref{sec_single_diffractive}), the double diffractive (see section \ref{sec_double_diffractive}) and the non-diffractive case (see section \ref{sec_non_diffractive}), which is the most common case. All soft string processes rely on \textsc{Pythia} for the fragmentation of the strings into final-state hadrons.

\subsubsection{Single Diffractive}
\label{sec_single_diffractive}
The single diffractive process describes the interaction between two hadrons, where exactly one of the two colliding hadrons $A$ and $B$ is excited to form a string $X$
\begin{equation}
  A+B\rightarrow A+X\quad\mathrm{or}\quad A+B\rightarrow X+B \,.
\end{equation}
The excited string $X$ has a larger mass than the incoming hadron.
The differential cross section, as a function of the string mass $M_X$ from diffractive excitation, is given by \cite{Ingelman:1984ns}
\begin{equation}
  \frac{d \sigma_{\scriptsize \textrm{SD}}}{dM_X^2} \propto\frac{1}{M_X^2}\,.
\end{equation}
Once the string mass is sampled, the three-momenum $p_{\scriptsize \textrm{CM}}$ of the string in the center-of-mass frame can be evaluated by solving
\begin{equation}
  \sqrt{s} = \left(p_{\scriptsize \textrm{CM}}^2 + M_X^2\right)^{1/2} + \left(p_{\scriptsize \textrm{CM}}^2 + m_H^2\right)^{1/2}\,,
  \label{eq_sqrts_SD}
\end{equation}
where $m_H$ is the mass of the incoming hadron, which remains intact.
Following the UrQMD approach \cite{Bass:1998ca}, the transverse momentum transfer $\textbf{p}_{\perp}$ between the incoming hadrons is assumed to follow a Gaussian distribution
\begin{equation}
\frac{d \sigma_{\scriptsize \textrm{SD}}}{d^2 \textbf{p}_{\perp}} \propto \exp\left(-\frac{p_{\perp}^2}{\sigma_T^2}\right)\,,
\label{eq_gaussian_momentum_transfer}
\end{equation}
where $\sigma_T$ is a free parameter that is constrained by observables in proton-proton collisions in section \ref{sec_transverse_momentum}.
To completely determine kinematics of the string-hadron system, we sample the transverse momentum transfer $\textbf{p}_{\perp}$ with a maximum of $p_{\scriptsize \perp, \textrm{max}} = p_{\scriptsize \textrm{CM}}$.
The string has a longitudinal momentum $p_{\parallel} = \left( p_{\scriptsize \textrm{CM}}^2 - p_{\perp}^2 \right)^{1/2}$, which is parallel to the collision axis.
Knowing the mass and the momenta of the reaction products, one can calculate the velocity of the string in order to boost into its rest frame, where the fragmentation machinery from \textsc{Pythia} is employed to obtain the particles in the final state of the interaction.

\subsubsection{Double Diffractive}
\label{sec_double_diffractive}
The double diffractive subprocess is a collision in which the two incoming hadrons $A$ and $B$ are both excited to strings
\begin{equation}
A+B\rightarrow X+X \, .
\end{equation}
The dynamics of the interaction is determined in the center of mass frame of the incoming hadrons, where the collision axis is set to be the longitudinal direction.
Kinematics of the double-diffractive excitation is modeled via pomeron exchange between gluons from two incoming hadrons.
Those gluons exchange transverse and lightcone momenta, such that they remain on-shell after the momentum exchange \cite{Bass:1998ca,Bleicher:1999xi}.
The light cone momentum fraction $x$ of each gluon is sampled from the parton distribution function for gluons, which is assumed to be of the form
\begin{equation}
\mathrm{PDF}\propto\frac{1}{x}\left(1-x\right)^{\beta+1}\,,
\label{gluon_pdf}
\end{equation}
where $\beta$ is set to be $0.5$.
The light cone momenta $p^\pm$ of the hadrons are given by
\begin{equation}
p^\pm=\frac{E\pm p_{\parallel}}{\sqrt{2}}\,,
\end{equation}
where $E$ is the energy of the hadron and $p_{\parallel}$ is the projection of the momentum on the collision axis of the colliding hadrons. The light cone momenta of the exchanged gluons are calculated as $p_g^\pm=x^\pm p^\pm$.
The distribution for the transverse momentum transfer $\textbf{p}_{\perp}$ between gluons is taken to be Gaussian, whose width is the same as in the single-diffractive case (see equation (\ref{eq_gaussian_momentum_transfer})).
The lightcone momenta $q_A^-$ and $q_B^+$, which are gained by the gluons from hadrons $A$ and $B$, are given by
\begin{eqnarray}
  2 \, \left( x_A^+ p_A^+ \right) \, q_A^- - p_{\perp}^2 & = & 0\,, \\
  2 \, \left( x_B^- p_B^- \right) \, q_B^+ - p_{\perp}^2 & = & 0\,.
\end{eqnarray}
Note that the collision axis is defined as the direction in which the hadron $A$ is moving.
The lightcone momenutum $Q^{\pm}$ transferred from the hadron $B$ to $A$ is given by
\begin{eqnarray}
  Q^+ & = & - q_B^+ = -\frac{p_{\perp}^2}{2 \, x_B^- p_B^-}\,, \\
  Q^- & = & q_A^- = \frac{p_{\perp}^2}{2 \, x_A^+ p_A^+}\,,
\end{eqnarray}
and it leads to evaluation of the energy $\Delta E$ and longitudinal momentum $\Delta p_{\parallel}$ transferred from $B$ to $A$ as
\begin{equation}
\Delta E = \frac{Q^+ + Q^-}{\sqrt{2}}\,,
\qquad
\Delta p_{\parallel} = \frac{Q^+ - Q^-}{\sqrt{2}}\,.
\label{eq_delta_e_delta_p}
\end{equation}
The mass of both excited strings can be calculated individually using the energy-momentum relation.
Each string is then fragmented in the rest frame of that string using the implementation of the fragmentation within \textsc{Pythia}.

\subsubsection{Non-Diffractive}
\label{sec_non_diffractive}
The non-diffractive string excitation is the most probable soft process, and therefore has the largest impact on the dynamics of the produced particles in the SPS energy region.
During the interaction, each hadron emits one valence quark, which is adopted by the other hadron.
The exchanged valence quark carries a fraction of the longitudinal momentum of the hadron it is emitted from.
The light cone momentum fraction carried by the exchanged quark is sampled according to the parton distribution function for quarks, which is assumed to have the following functional form:
\begin{equation}
\mathrm{PDF}\propto x^{\alpha-1}(1-x)^{\beta-1}\,,
\label{eq_quark_pdf}
\end{equation}
where $\alpha$ and $\beta$ are in general free parameters. In section \ref{sec_parton_distribution_function}, the dependence of the particle production on the PDF is studied in detail and the parameters are adjusted such that  the measured dynamics is reproduced as well as possible while supporting the physical picture of a valence quark exchange.

The momentum transfer in the transverse direction is sampled according to the same Gaussian as in the single diffractive and double diffractive case, using equation (\ref{eq_gaussian_momentum_transfer}).
With the light cone momentum fraction each exchanged quark carries and the transferred transverse momentum, the light cone momentum transfer is written as \cite{Bass:1998ca, Bleicher:1999xi}
    \begin{equation}
Q^+=-x^+_Ap^+_A+\frac{p_T^2}{2x_B^-p_B^-}
\end{equation}
\begin{equation}
Q^-=x_B^-p_B^--\frac{p_T^2}{2x_A^+p_A^+}\,.,
\end{equation}
where $x_A$ and $x_B$ are the light cone momentum fractions for the exchanged quarks, $p_A^\pm$ and $p_B^\pm$ are the light cone momenta of the colliding hadrons before the collision and $p_T$ is the transferred transverse momentum.
The exchanged energy and longitudinal momentum can be calculated using equation (\ref{eq_delta_e_delta_p}). The masses of the strings are obtained using the relativistic energy-momentum relation and each string is fragmented individually in the rest frame of the string using \textsc{Pythia}.

\subsection{String Fragmentation}
Once the mass of the excited string and the flavor of the quarks spanning the string is determined, the string is fragmented into hadrons by employing \textsc{Pythia}.
Within \textsc{Pythia}, the species of the fragmented hadron follows from the flavour of the quark-antiquark or diquark-antidiquark pair that is produced. While the light quarks have the same probability to be produced, there are empirical suppression factors for producing heavier quarks and diquarks. 

The transverse momentum of each string fragment is sampled from a Gaussian distribution with a width of $\sigma_{T,\mathrm{string}}$ which is a free parameter that is tuned to experimental data in section \ref{sec_transverse_momentum}. 
The longitudinal momentum of each string fragment is determined using the fragmentation function.
\textsc{Pythia} is based on the symmetric Lund fragmentation function \cite{Andersson:1983ia}, which has the following shape
\begin{equation}
	f(z) \propto \frac{1}{z} (1-z)^a \exp\left(-b\frac{m_T^2}{z}\right)\,.
	\label{eq_fragmentation_function}
\end{equation}
$m_T$ is the transverse mass of the string fragment while $a$ and $b$ are free model parameters.
For the fragmentation of leading baryons produced in soft non-diffractive processes, the parameters $a$ and $b$ are chosen differently from \textsc{Pythia}. The consequences of this treatment are discussed in more detail in section \ref{sec_fragmentation_function}.
Note that both the different string excitation in the soft case and the mentioned modification to the string fragmentation necessitate retuning the parameters for the fragmentation and the available tunes for \textsc{Pythia} are not necessarily compatible.
The process of finding a new tune is described in section \ref{sec_proton_proton_collisions}.

\subsection{Particle Formation}
\label{sec_particle_formation}
\begin{figure}[h]
  \centering
  \includegraphics[width=0.6\textwidth]{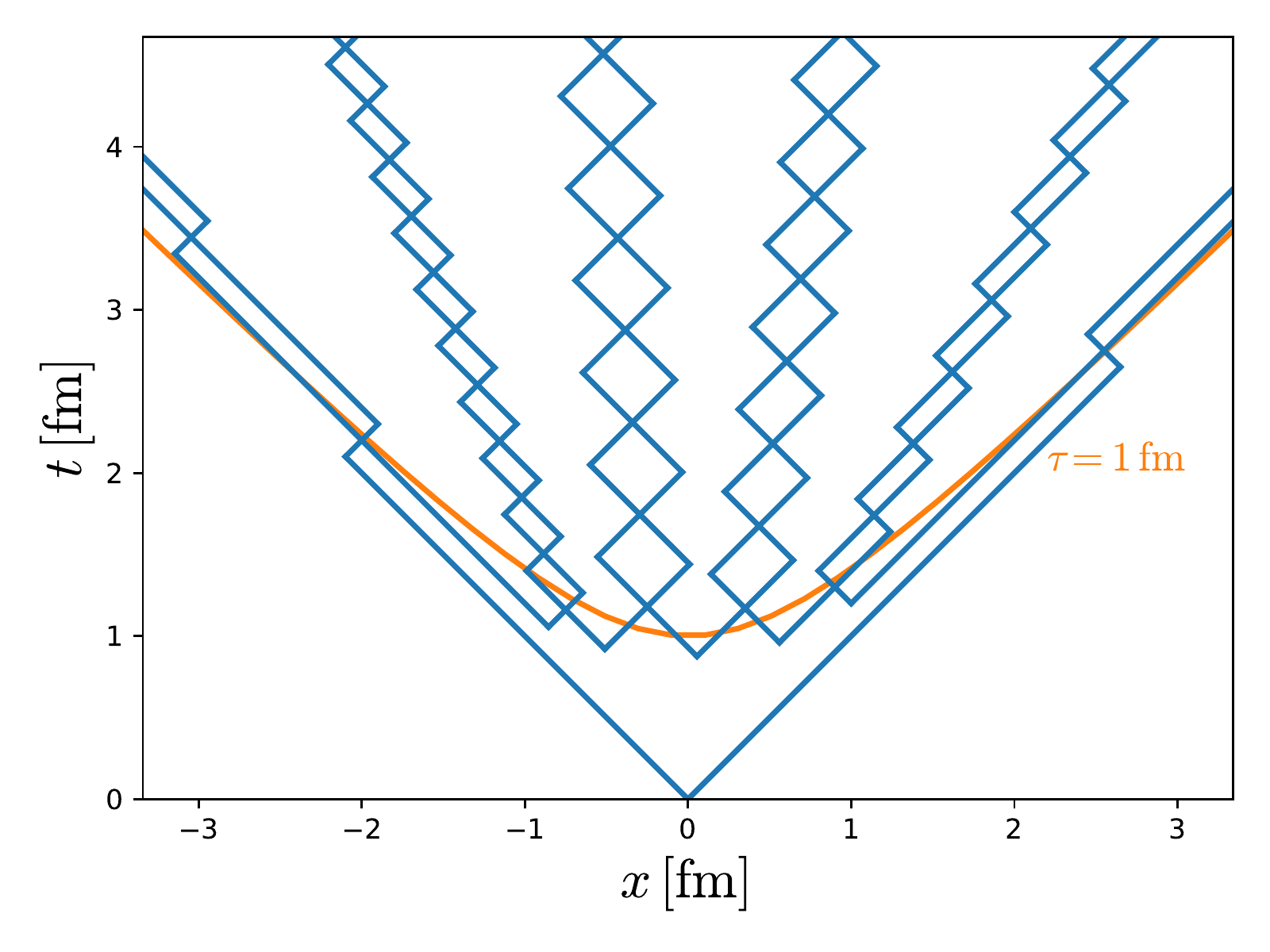}
  \caption{Sketch of a string fragmenting within the yoyo model.}
  \label{fig_string_drawing}
\end{figure}

A string fragments into hadrons by producing quark-antiquark pairs.
In a dynamical picture, the pair production does not happen simultaneously but at different points in time.
Figure \ref{fig_string_drawing} illustrates how a string fragments in coordinate space within the yoyo model.
The straight lines indicate the trajectories of (anti)quarks or (anti)diquarks.
While the pair production occurs at different points in coordinate time, the time when they recombine to hadrons fluctuates around a constant proper time.
In principle, the formation time and position of all string fragments can be calculated using the yoyo model and the momenta of the fragments obtained from \textsc{Pythia}.
For simplicity, the fluctuations in the formation time are neglected so that all string fragments form at a constant proper time in SMASH. The effect of changing the formation time is investigated in section \ref{sec_heavy_ion_collisions}.

In practice, all particles in SMASH are immediately produced once the colliding hadrons reach their point of closest approach.
Until the formation time has passed, the cross section of the string fragments are multiplied by a cross section scaling factor $f_\sigma$.
For most string fragments, this factor is initially 0.
However, since the leading string fragments contain quarks that do not originate from a pair production but from the initially colliding hadrons, the initial cross section scaling factor is not zero for leading string fragments.
The initial cross section scaling factor for each string fragment is set to be the number of quarks from the initially colliding hadrons contained in the fragment divided by the total number of quarks of that fragment.
For example, a leading baryon that contains a diquark from the initially colliding hadrons is assigned a scaling factor of $f_\sigma=2/3$ and a meson at the other end of the string that contains another quark from the initially colliding hadrons is assigned a scaling factor of $f_\sigma=1/2$.

Instead of having a constant cross section scaling factor until the time of formation, where the particle suddenly is allowed to interact, it is possible to mimic a continuous formation process by increasing the cross section scaling factor with time.
Timely increasing cross sections have been studied in a similiar fashion within the GiBUU model \cite{Gallmeister:2007an}.
To realize a continuous formation, the cross section scaling factor becomes a function of time $f_\sigma=f_\sigma(t)$.
This function needs to have the initially assigned scaling factor $f_0$ as described above at the time $t_\mathrm{prod}$ when the particle is produced and $f_\sigma(t_\mathrm{form})=1$ at the formation time $t_\mathrm{form}$.
Between the two points, the cross section scaling factor grows with a given power $\alpha$ in order to have a simple but flexible functional shape.
Using the three conditions, the function $f_\sigma(t)$ is written as follows
\begin{equation}
    f_\sigma(t)=(1-f_0)\left(\frac{t-t_\mathrm{prod}}{t_\mathrm{form}-t_\mathrm{prod}}\right)^\alpha  +f_0\,.
\end{equation}
This function is only used for $t_\mathrm{prod}<t<t_\mathrm{form}$, since it has no meaning before the particle is produced and the scaling factor is $f_\sigma(t)=1$ for $t>t_\mathrm{form}$, when the particle is fully formed.
\begin{figure}[h]
  \centering
  \includegraphics[width=0.5\textwidth]{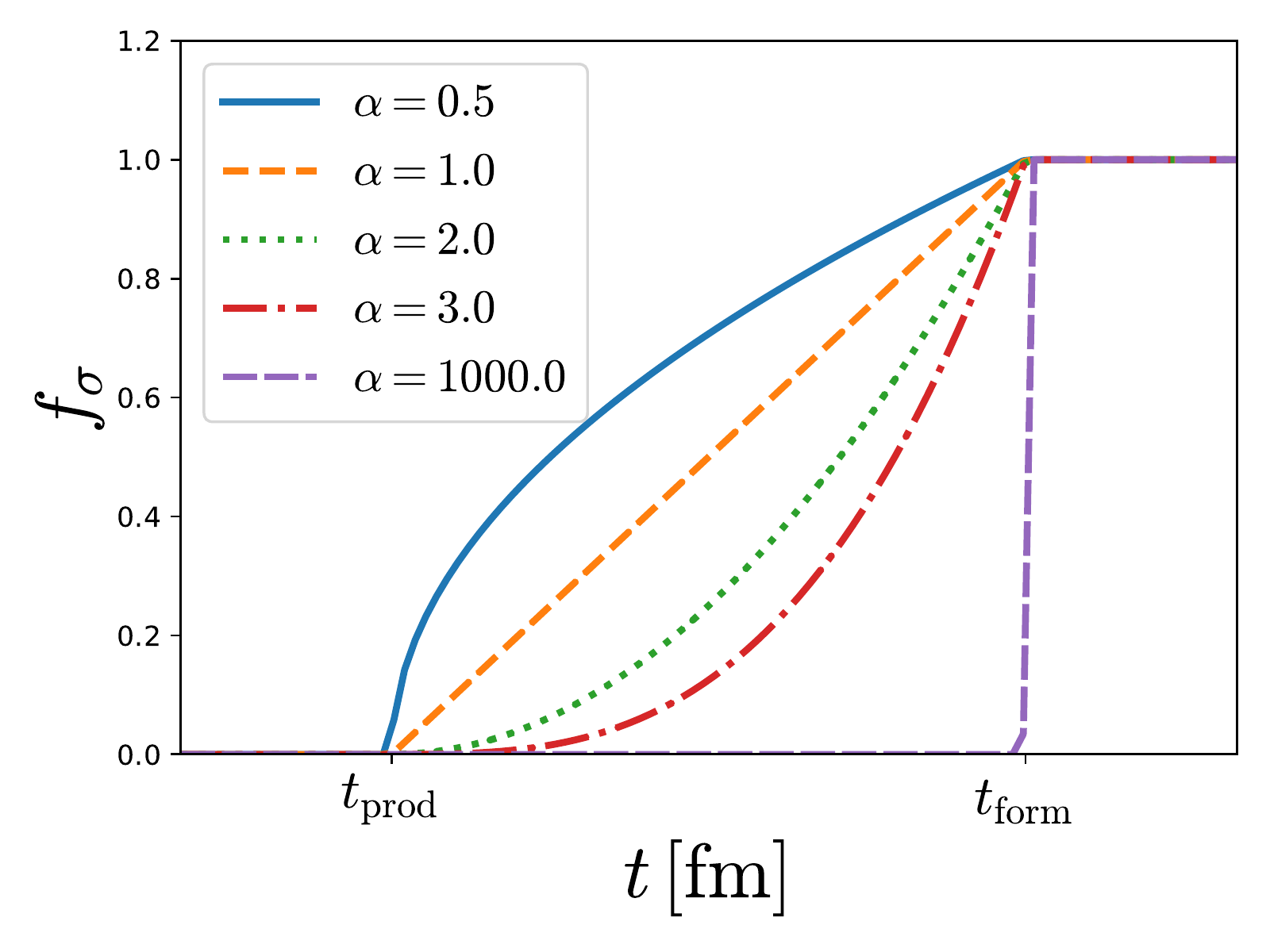}
  \caption{Cross section scaling factor $f_\sigma$ as a function of time for different powers $\alpha$ with which the cross section grows in time. In this example, the initial cross section scaling factor is set to be $f_\sigma(t_\mathrm{prod})=0$.}
  \label{fig_formation_power_scematic}
\end{figure}
The cross section scaling factor as a function of time for different values of $\alpha$ is shown in figure \ref{fig_formation_power_scematic}. The initial cross section scaling factor is set to $f_0=0$ in this figure.
In the limit of $\alpha$ going to infinity, one recovers a step function, while for small positive values of $\alpha$, the particles form immediately. In section \ref{sec_heavy_ion_collisions}, the effect of the details of the particle formation on particle spectra in heavy ion collisions is investigated.

\subsection{Elastic Collisions}
\label{sec_elastic_collisions}
Elastic collisions play an important role in describing heavy ion collisions, since the contribution of elastic collisions to the total hadronic cross sections is at all energies significant as can be seen for example for proton-$\pi^-$ collisions in figure \ref{xsec_plot}.
In SMASH, all hadrons have a finite cross section to interact elastically.
The most important elastic cross sections are parametrizations of the experimental data, if available.
If the elastic cross section is not measured for a pair of particle species, the additive quark model is applied to obtain a cross section for that pair of particles \cite{Bass:1998ca} as shown in equation (\ref{eq_xsec_aqm}).

While the angular distributions are close to being isotropic at low collision energies, they are more forward-backward peaked at larger collision energies \cite{Baglin:1975tg}.
Especially going to higher collision energies, including anisotropic angular distributions for elastic collisions is, therefore, of major importance for describing the dynamics of heavy ion collisions.
This is shown in a comparison between calculations with isotropic and anisotropic angular distributions provided in section \ref{sec_heavy_ion_collisions}.

For the elastic collisions of nucleons at relatively low collision energies, parametrizations for the angular distributions have been proposed in \cite{Cugnon:1996kh}.
Since there is few experimental data for elastic collisions of hadronic pairs other than nucleon-nucleon ones, we implement the Cugnon parametrizations \cite{Cugnon:1996kh} for nucleon-nucleon elastic collisions and extrapolate them to all other hadronic pairs.
Given that the Cugnon parametrization is a function of $p_{\scriptsize \textrm{lab}}$ of fixed-target experiments, there is an ambiguity for incoming particles with different masses.
To circumvent this, we first compute the center-of-mass momentum $p_{\scriptsize \textrm{cm}}$ of the collision.
Then a new lab-frame momentum $p_{\scriptsize \textrm{lab}}^*$ is evaluated from a new mandelstam variable $s^*$, which yields the original center-of-mass momentum when the nucleon mass is assumed
\begin{eqnarray}
  s^* & = & 4 \sqrt{p_{\scriptsize \textrm{cm}}^2 + m_N^2} \\
  p_{\scriptsize \textrm{lab}}^* & = & \frac{1}{2 \, m_N} \sqrt{s^* \left( s^* - 4\, m_N^2 \right)}\,.
\end{eqnarray}
Lastly, the differential cross section of elastic proton-proton collisions is extrapolated to all hadronic pairs such that
\begin{equation}
  \left. \frac{1}{\sigma_{\scriptsize \textrm{el}}} \frac{d\sigma_{\scriptsize \textrm{el}}}{d\theta} \right|_{\scriptsize h_1 h_2} =
    \left. \frac{1}{\sigma_{\scriptsize \textrm{el}}} \frac{d\sigma_{\scriptsize \textrm{el}}}{d\theta} \right|_{\scriptsize pp} (p_{\scriptsize \textrm{lab}}^*)\,,
\end{equation}
where $\theta$ is the scattering angle.
In addition, these angular distributions are extrapolated to arbitrary collision energies in order to obtain forward-backward peaked angular distributions for elastic collisions at large collision energies.

\section{Proton-Proton Collisions}
\label{sec_proton_proton_collisions}
In a heavy ion collision, the stopping of baryons is mostly determined by the first interactions of the participants. Therefore, experimental data for elementary proton-proton collisions is used to adjust the parameters of the string approach. Even though most of the parameters influence multiple observables, this section introduces the most important parameters and demonstrates their effect on the particle production of (anti)protons, pions and kaons to give some insight on how the value of each parameter is chosen.

Within this section, each parameter is varied separately, while all others are kept constant.
A lot of additional parameter combinations have been tried to account for potential correlations and here we only present a subset to indicate the reasoning for our final choices.
If not further specified, the value of each parameter can be found in table \ref{tab_defaults}. 
We mainly concentrate on the highest SPS energy, since the experimental data set is the largest at that energy and in the end of the section show results for all other beam energies as well.

\begin{table}
	\centering
    \renewcommand\tabularxcolumn[1]{m{#1}} 
	\begin{tabularx}{\textwidth}{|l|X|l|}
		\hline
		Name & Meaning & Default value \\ \hline \hline
		$\beta_\mathrm{quark}$ & Parameter $\beta$ in PDF for quarks as defined in equation (\ref{eq_quark_pdf}) & $7.0$ \\ \hline
		$a_\mathrm{leading}$ & Parameter $a$ in fragmentation function for leading baryons as defined in equation (\ref{eq_fragmentation_function}) & $0.2$ \\ \hline
		$b_\mathrm{leading}$ & Parameter $b$ in fragmentation function for leading baryons as defined in equation (\ref{eq_fragmentation_function}) & $2.0\,\mathrm{GeV^{-2}}$\\ \hline
		$a_\mathrm{string}$ & Parameter $a$ in fragmentation function for remaining hadrons as defined in equation (\ref{eq_fragmentation_function})  & $2.0$ \\ \hline
		$b_\mathrm{string}$ & Parameter $b$ in fragmentation function for remaining hadrons as defined in equation (\ref{eq_fragmentation_function}) & $0.55\,\mathrm{GeV}^{-2}$ \\ \hline
		$\sigma_T$ & Width of the Gaussian used to sample transverse momentum transfer between colliding hadrons in soft string routine as defined in equation (\ref{eq_gaussian_momentum_transfer}) & $0.42 \,\mathrm{GeV}$\\ \hline
		$\sigma_{T,\mathrm{string}}$ & Width of the Gaussian used to sample the transverse momentum of string fragments  & $0.5\,\mathrm{GeV}$\\ \hline
		$\lambda_s$ & Strangeness suppression factor as defined in equation (\ref{eq_strange_supp}) & $0.16$ \\ \hline
		$\lambda_\mathrm{diquark}$ & Diquark suppression factor as defined in equation (\ref{eq_diquark_supp}) & $0.036$ \\ \hline
		$\mathrm{Popcorn\ rate}$ & Probability for popcorn processes as described in section \ref{sec_popcorn_rate}& $0.15$ \\ \hline
	\end{tabularx}
	\caption{Default set of parameters of SMASH-1.6 tuned to reproduce the experimental data for proton-proton collisions at SPS energies with brief description.}
	\label{tab_defaults}
\end{table}

\subsection{Parton Distribution Function}
\label{sec_parton_distribution_function}
\begin{figure}[h]
	\centering
	\includegraphics[width=0.5\textwidth]{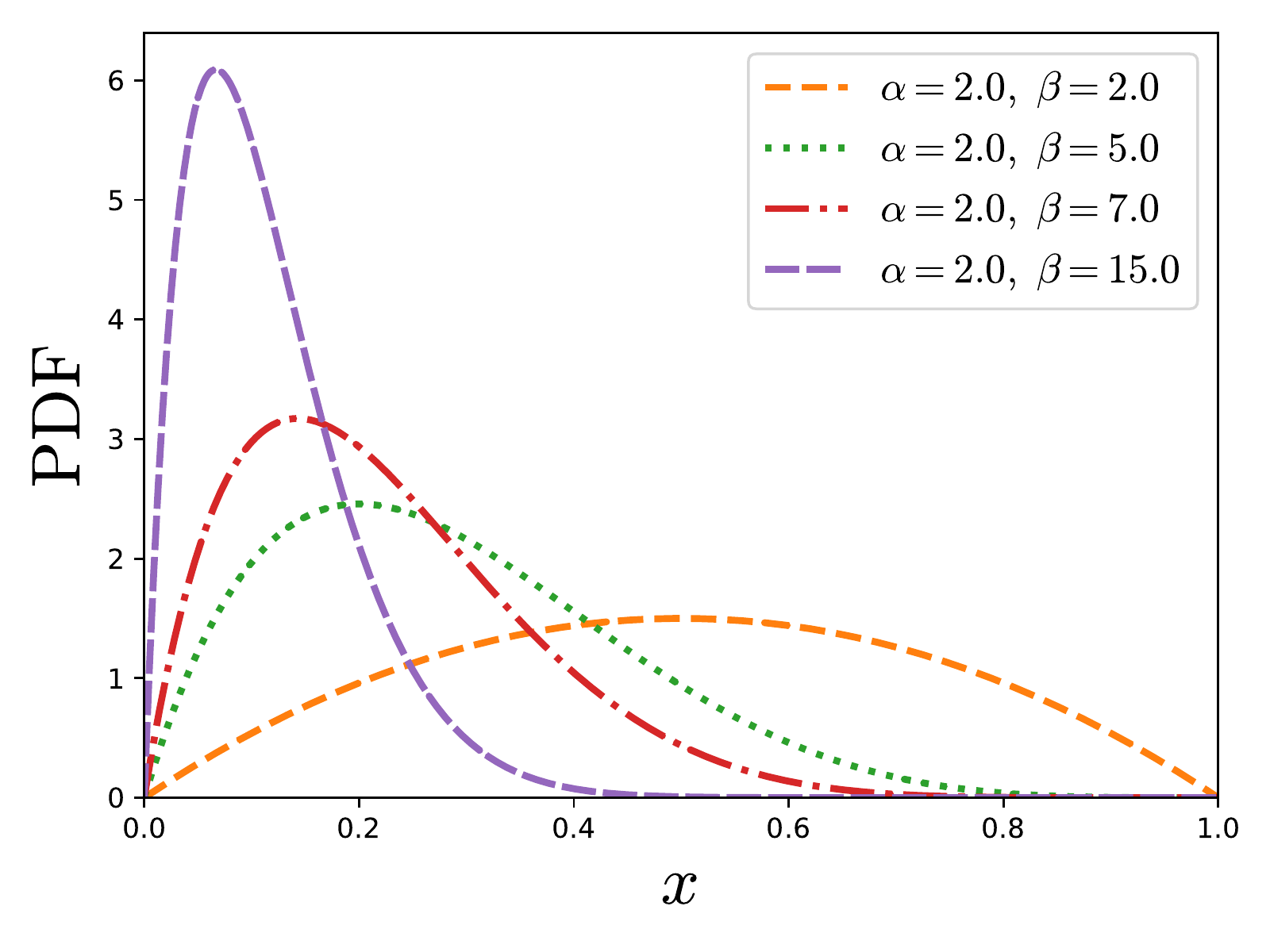}%
	\includegraphics[width=0.5\textwidth]{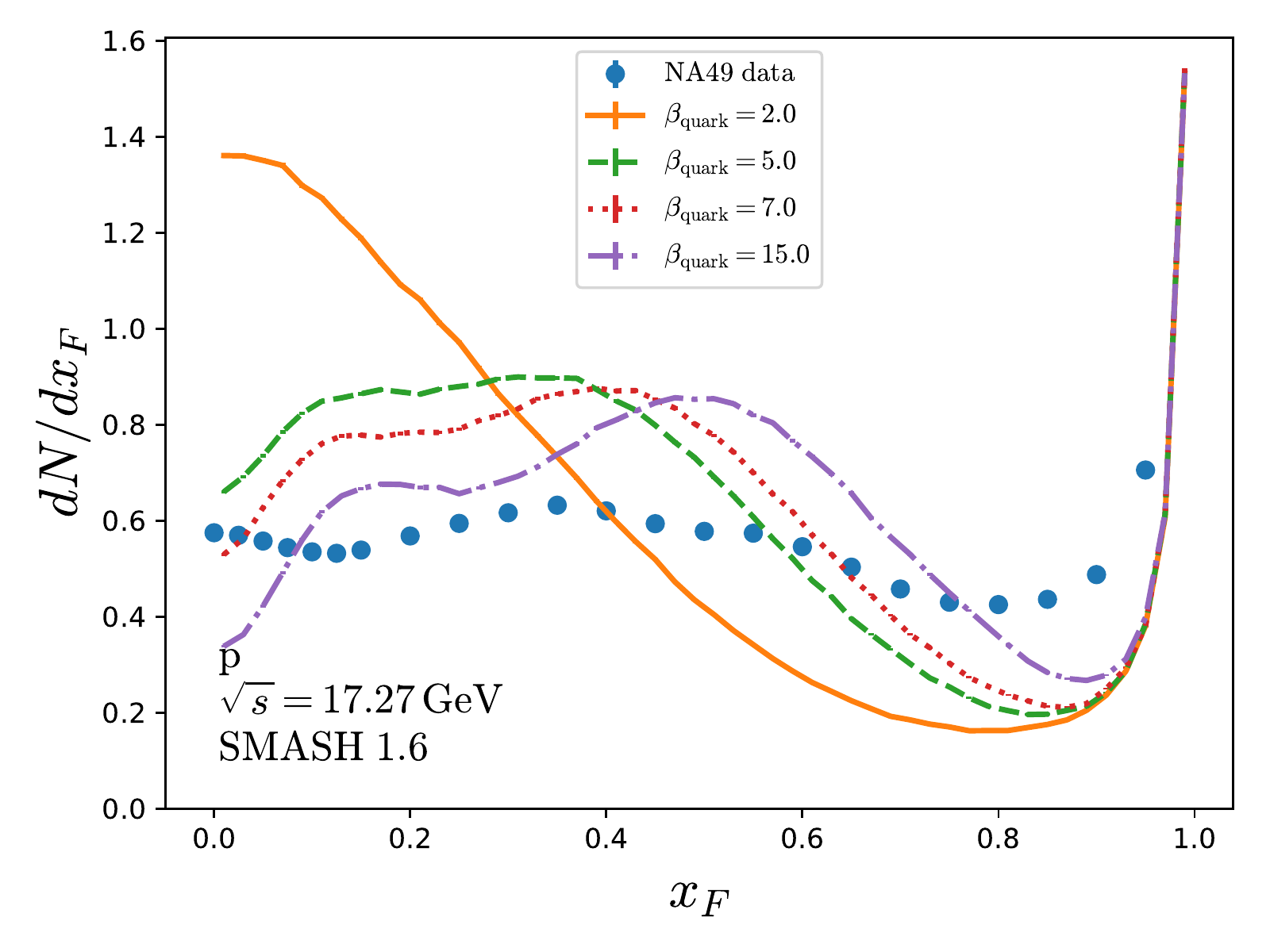}
	\caption{Left: The parton distribution function that is used for the different calculations. Right: Proton $x_F$ distribution in proton-proton collisions at $\sqrt{s}=17.27\,\mathrm{GeV}$ for different values of the parameter $\beta$ of the parton distribution function compared to experimental data \cite{NA49protons}.}
	\label{fig_quark_beta}
\end{figure}

In the SPS energy range, the soft non-diffractive string processes are the dominant interactions in proton-proton collisions. As described in section \ref{sec_non_diffractive}, the amount of exchanged longitudinal momentum is determined by the momentum fraction $x$ the exchanged valence quark carries. This does not only affect the dynamics of the string before the fragmentation, but also the mass of the string and therefore the energy available for producing string fragments. The PDFs used for the calculations are shown in figure \ref{fig_quark_beta} (left). Figure \ref{fig_quark_beta} (right) shows the dependence of the longitudinal momentum of protons in proton-proton collisions on the value of the parameter $\beta$ of the parton distribution function as defined in equation (\ref{eq_quark_pdf}). The longitudinal momentum distribution of protons is clearly the most relevant quantity to understand baryon stopping in heavy-ion collisions. 

The larger momentum transfer in the longitudinal direction is reflected in the $x_F=p_z/p_{z,\mathrm{beam}}$ distribution of protons as they are shifted to higher $x_F$ for larger values of $\beta_\mathrm{quark}$.
With a softer PDF, there is more energy available for the production of new particles.
This reflects in the proton yield at low $x_F$, which increases by a factor of 2 when using a very soft PDF.
Even though the proton $x_F$ distribution is not reproduced quantitatively, the best possible agreement is found for $\beta_\mathrm{quark}=7$.

Note that the description of the longitudinal momentum of protons in proton-proton collisions proves to be very challenging within theoretical models \cite{Uzhinsky:2014kxa}.
Modifications to improve the agreement between a different model and the data have been suggested \cite{Uzhinsky:2014jba}.
One of the suggestions from \cite{Uzhinsky:2014jba} is to modify how a proton is split into a quark and a diquark. 
An option to specify the probability to split a proton into $uu+d$ rather than $ud+u$ is implemented in SMASH but does not improve the overall agreement with the measurement.
Setting $\beta_\mathrm{quark}=7$ and $\alpha_\mathrm{quark}=2$, the mean value of the PDF is $2/9$, which is close to the expectation of $1/3$ which is assuming that there are three valence quarks sharing the full momentum of the proton.

\subsection{Fragmentation Function}
\label{sec_fragmentation_function}
The longitudinal momentum of each string fragment is determined by the shape of the fragmentation function. Starting at the forward and backward ends of the string, the fraction of the remaining lightcone momentum is sampled from the fragmentation function. \textsc{Pythia} employs the symmetric Lund fragmentation function defined in equation (\ref{eq_fragmentation_function}).

\begin{figure}[h]
  \centering
  \includegraphics[width=0.5\textwidth]{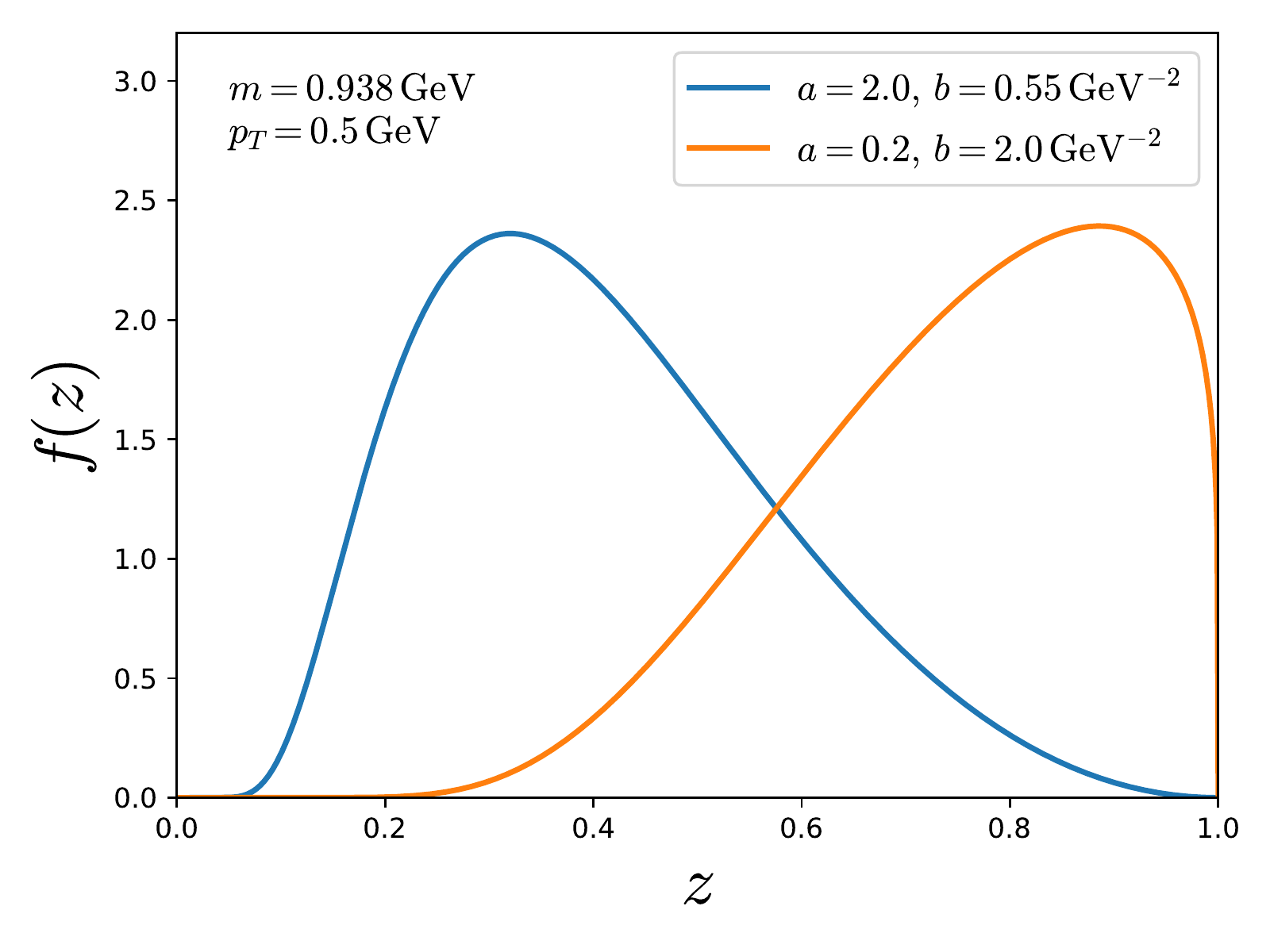}%
  \includegraphics[width=0.5\textwidth]{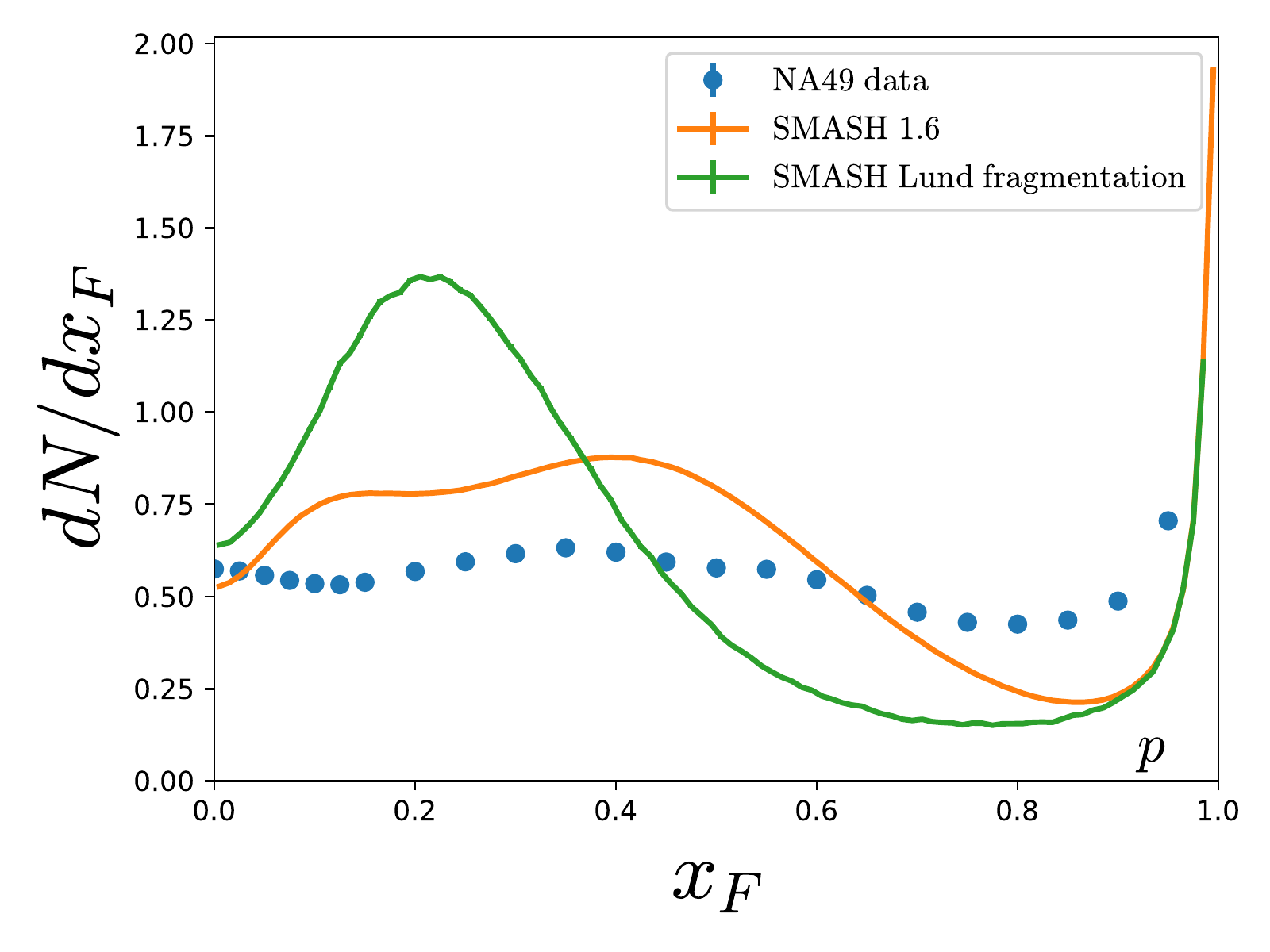}
  \caption{Left: Fragmentation function $f(z)$ according to equation (\ref{eq_fragmentation_function}) for protons. The harder fragmentation function is used for leading baryons, while the soft fragmentation function is employed to fragment all other particles. The transverse momentum is set to $p_T=0.5\,\mathrm{GeV}$. Right: $x_F$ distribution of protons in proton-proton collisions at $\sqrt{s}=17.27\,\rm GeV$ compared to experimental data \cite{NA49protons}. The curve labeled Lund fragmentation is calculated using the soft fragmentation function from the left panel consistently. The other curve shows results employing the harder fragmentation function from the left panel for leading baryons from soft non-diffractive string processes.}
  \label{fig_p_xF_old_new}
\end{figure}

Figure \ref{fig_p_xF_old_new} (left) shows the Lund fragmentation function for two different values of the parameter $b$. On the right panel, the distribution of $x_F$ for protons is plotted for two different settings.
The curve labeled Lund fragmentation refers to a calculation where the softer fragmentation from the left part of the figure is used consistently throughout the fragmentation.
Comparing to the experimental data shows that protons obtain too little longitudinal momentum that way.
Therefore, the protons require a different fragmentation mechanism.
To increase the longitudinal momentum of protons without producing harder light mesons than before, the harder fragmentation function shown in the left panel is used for leading baryons in soft non-diffractive string processes.
The other curve on the right panel of figure \ref{fig_p_xF_old_new} shows the result of the calculation after that adjustment.
A drastic improvement of the agreement with the experimental data is observed.

Even though the fragmentation function for non-leading hadrons is considered as an intrinsic property of a string which does not depend on what happens outside, the leading diquark holds information on the initial state kinematics.
That legitimates having a separate fragmentation function to determine lightcone momenta of leading baryons.

\begin{figure}
  \centering
  \includegraphics[width=0.5\textwidth]{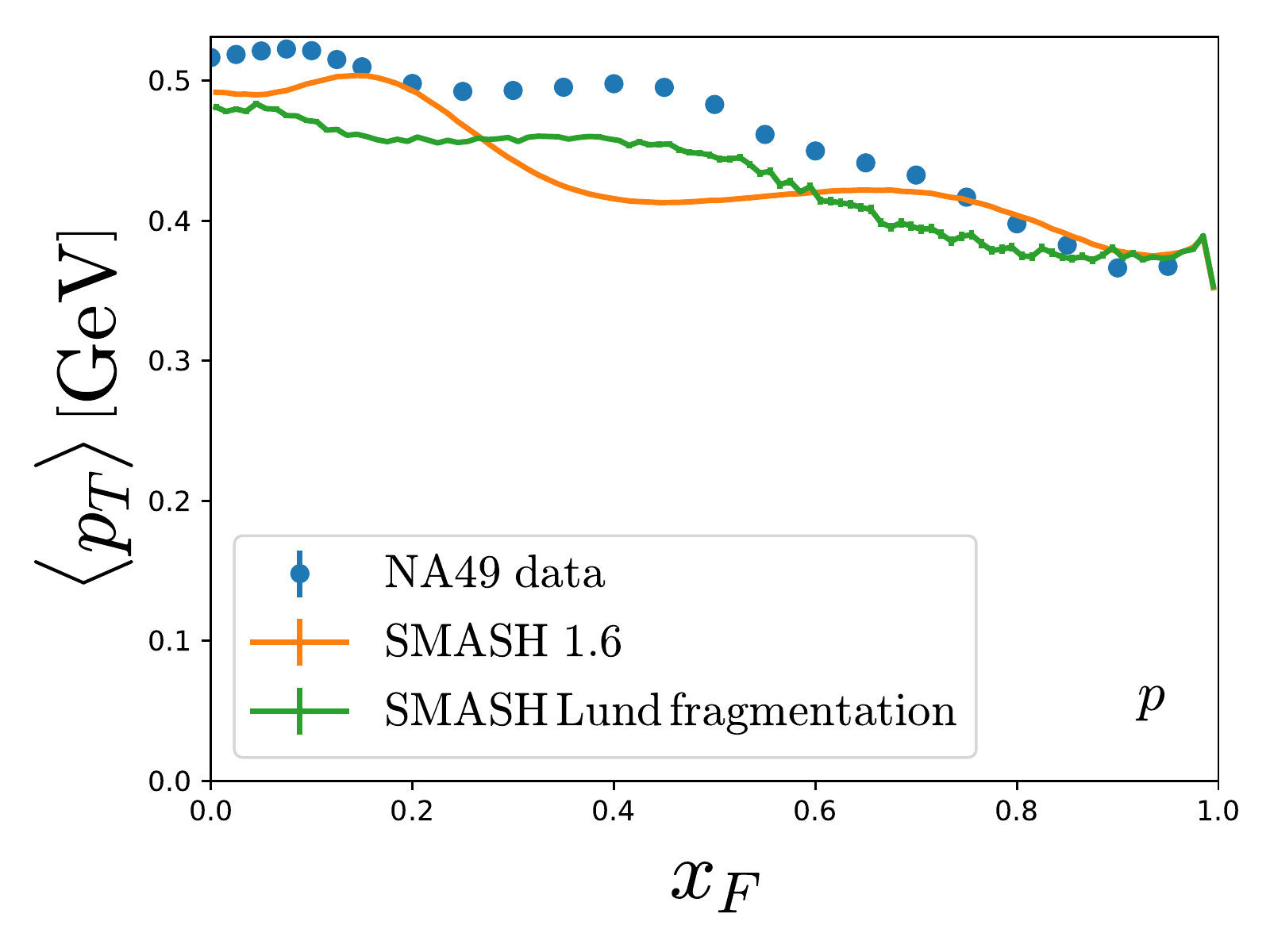}%
  \includegraphics[width=0.5\textwidth]{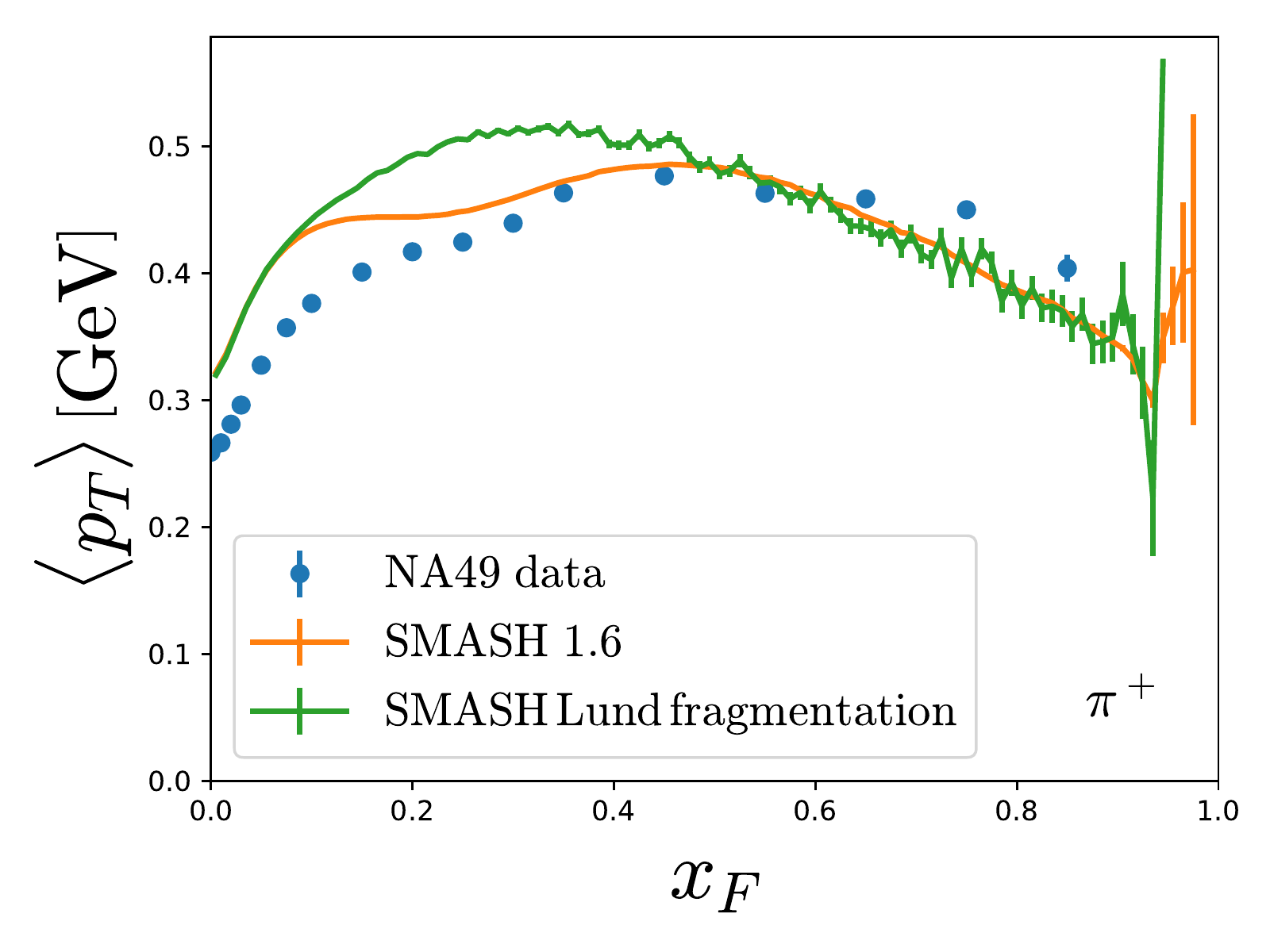}
  \caption{Mean transverse momentum of protons (left) and positively charged pions (right) as a function of $x_F$ in proton-proton collisions at $\sqrt{s}=17.27\,\mathrm{GeV}$ compared to experimental data \cite{NA49protons,NA49pions}. The curve labeled Lund fragmentation is obtained using the standard \textsc{Pythia} fragmentation mechanism employing the same fragmentation function for each string fragment, while for the other curve, a separate harder fragmentation function is used for leading baryons from soft non-diffractive string processes.}
  \label{fig_mpt_old_new}
\end{figure}

The influence of using a separate fragmentation function for leading baryons on the transverse momentum is shown in figure \ref{fig_mpt_old_new}, where the mean transverse momentum as a function of $x_F$ is shown for protons and pions. The shape of the transverse momentum of protons as a function of the longitudinal momentum fraction is better reproduced with the unmodified Lund fragmentation function. The curve employing a separate fragmentation function reflects the expected change that has been observed in figure \ref{fig_p_xF_old_new}. To understand the stopping in heavy-ion collisions, the match of the transverse momentum at midrapidity corresponding to low values of $x_F$ is most important for our present work. Please refer to  section \ref{sec_transverse_momentum} for a more detailed discussion of transverse momenta. 

\begin{figure}[h]
	\centering
	\includegraphics[width=0.5\textwidth]{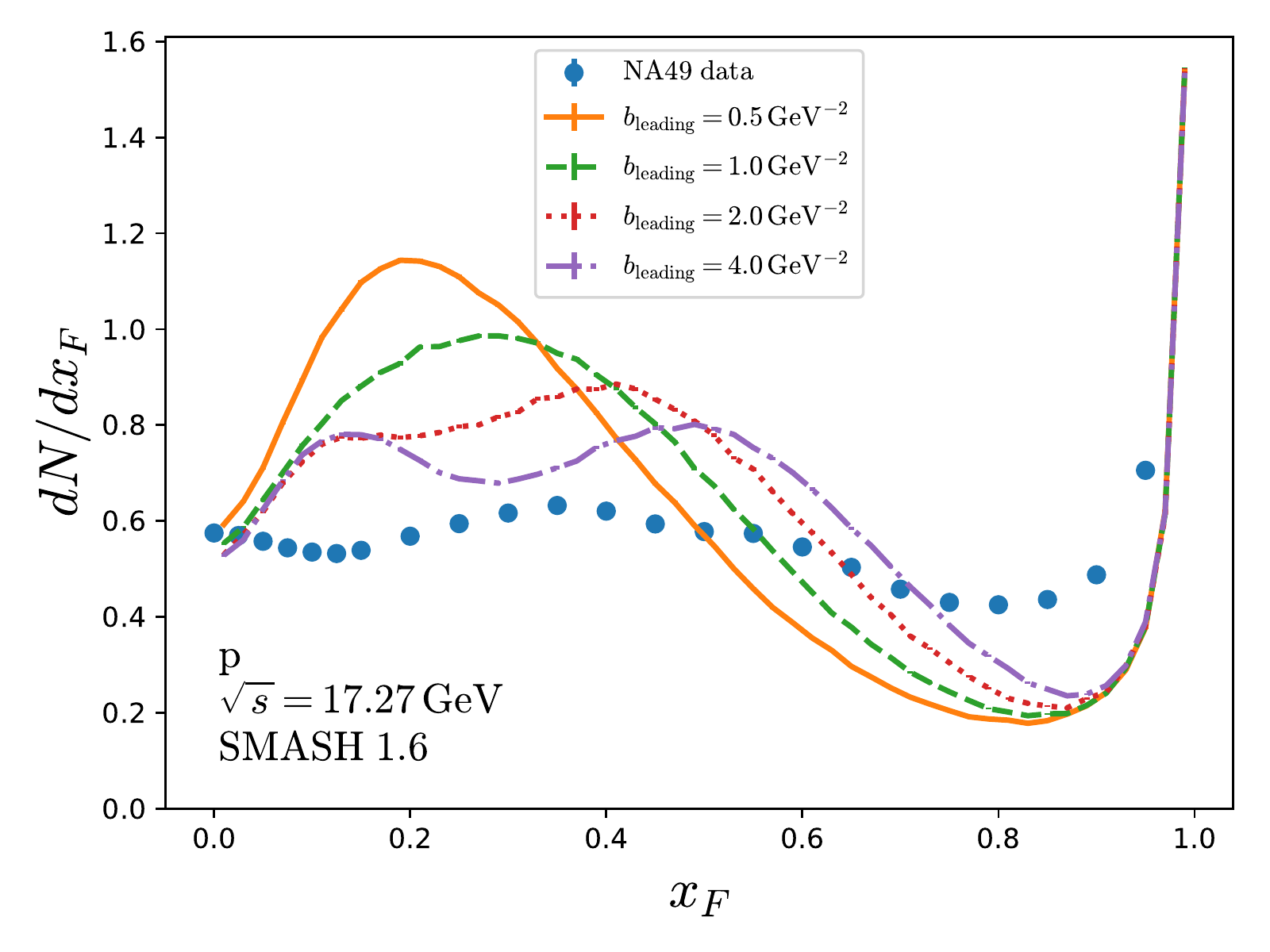}%
	\includegraphics[width=0.5\textwidth]{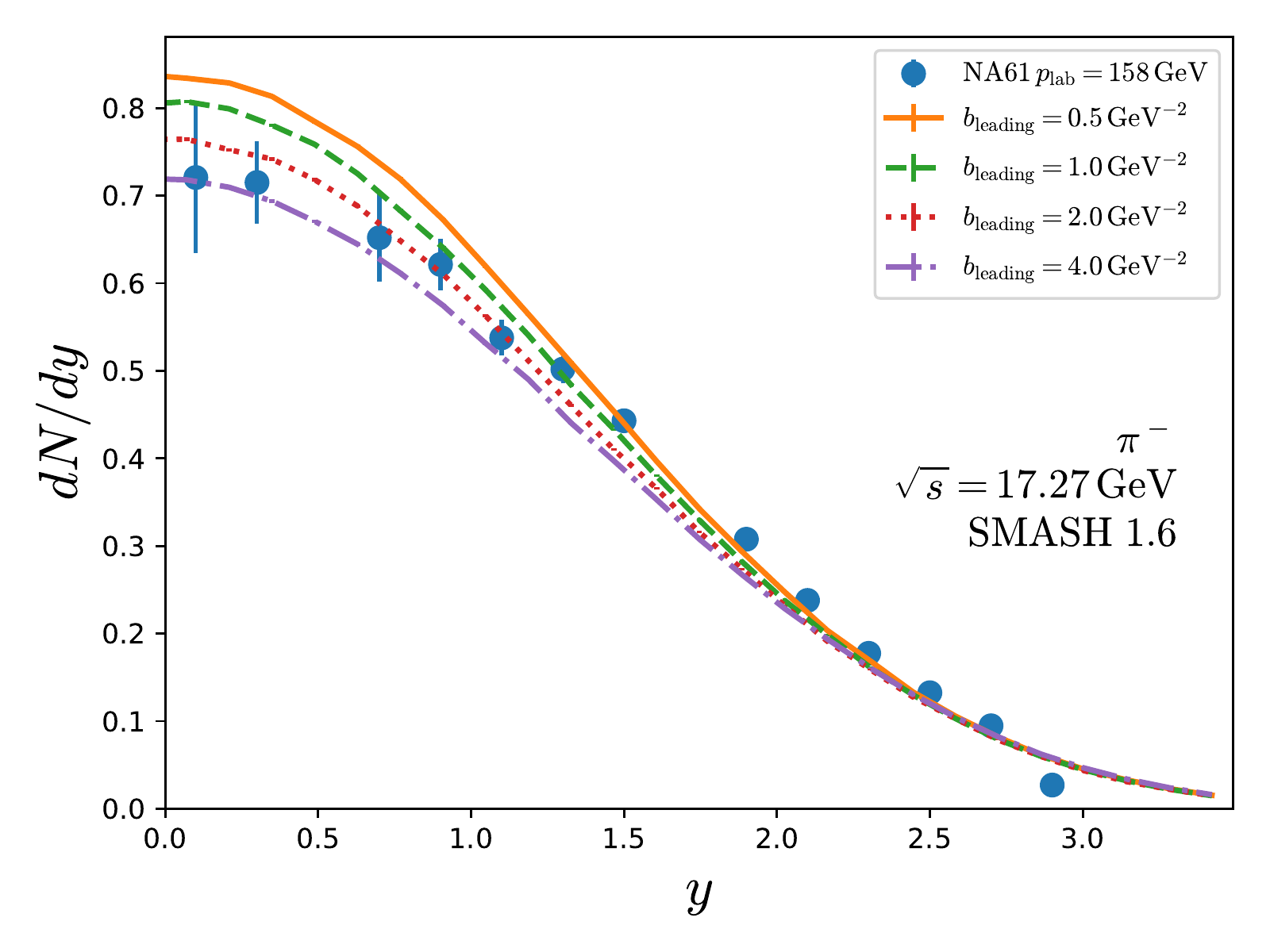}
	\caption{Left: $x_F$ distribution of protons in proton-proton collisions at $\sqrt{s}=17.27\,\mathrm{GeV}$ for different values of $b_\mathrm{leading}$ compared to experimental data \cite{NA49protons}. Right: Rapidity spectra of negatively charged pions in proton-proton collisions at $\sqrt{s}=17.27\,\mathrm{GeV}$ for different values of $b_\mathrm{leading}$ compared to experimental data \cite{Aduszkiewicz:2017sei}.}
	\label{fig_stringz_b_leading_p_pi}
\end{figure}

Let us now demonstrate in detail how the parameters for the fragmentation function for leading baryons from soft non-diffractive processes ($a_{\rm leading}$, $b_{\rm leading}$) and for all other particles ($a_{\rm string}$, and $b_{\rm string}$) have been determined. Figure \ref{fig_stringz_b_leading_p_pi} shows the longitudinal momentum distribution for protons and pions in pp collisions at the highest SPS energy for different values of $b_\mathrm{leading}$. In general higher values of $b_{\rm leading}$ are prefered by the proton $x_F$ distribution, but there needs to be enough energy for particle production as well. Therefore, $b_\mathrm{leading}=2.0\,\mathrm{GeV}^{-2}$ provides the best compromise to generate hard enough protons, while still producing a reasonable amount of pions. In addition, higher values of $b_{\rm leading}$ lead to a double-peak structure in the $x_F$ distribution that is not supported by the experimental data. 

\begin{figure}[h]
	\centering
	\includegraphics[width=0.5\textwidth]{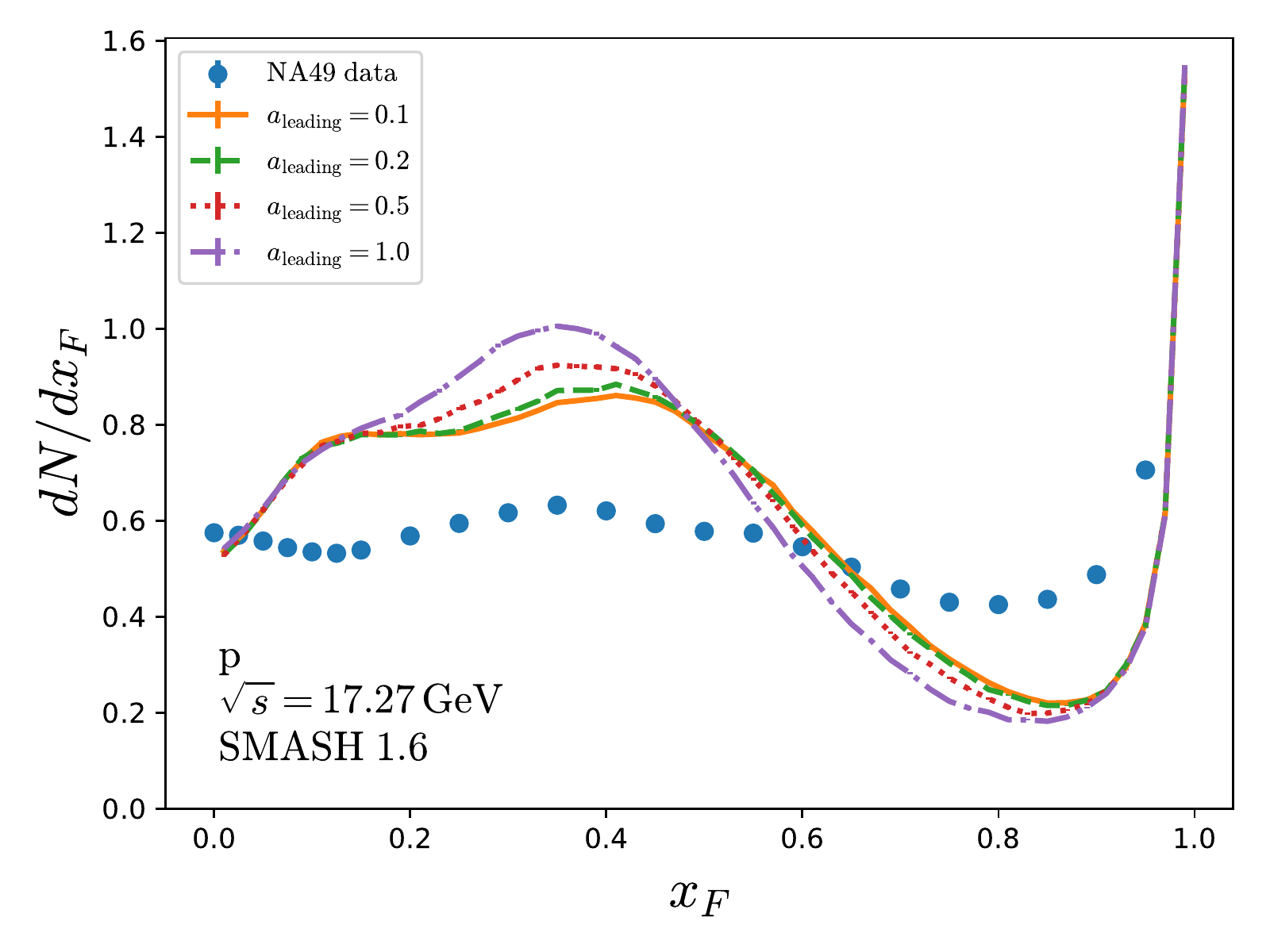}%
	\includegraphics[width=0.5\textwidth]{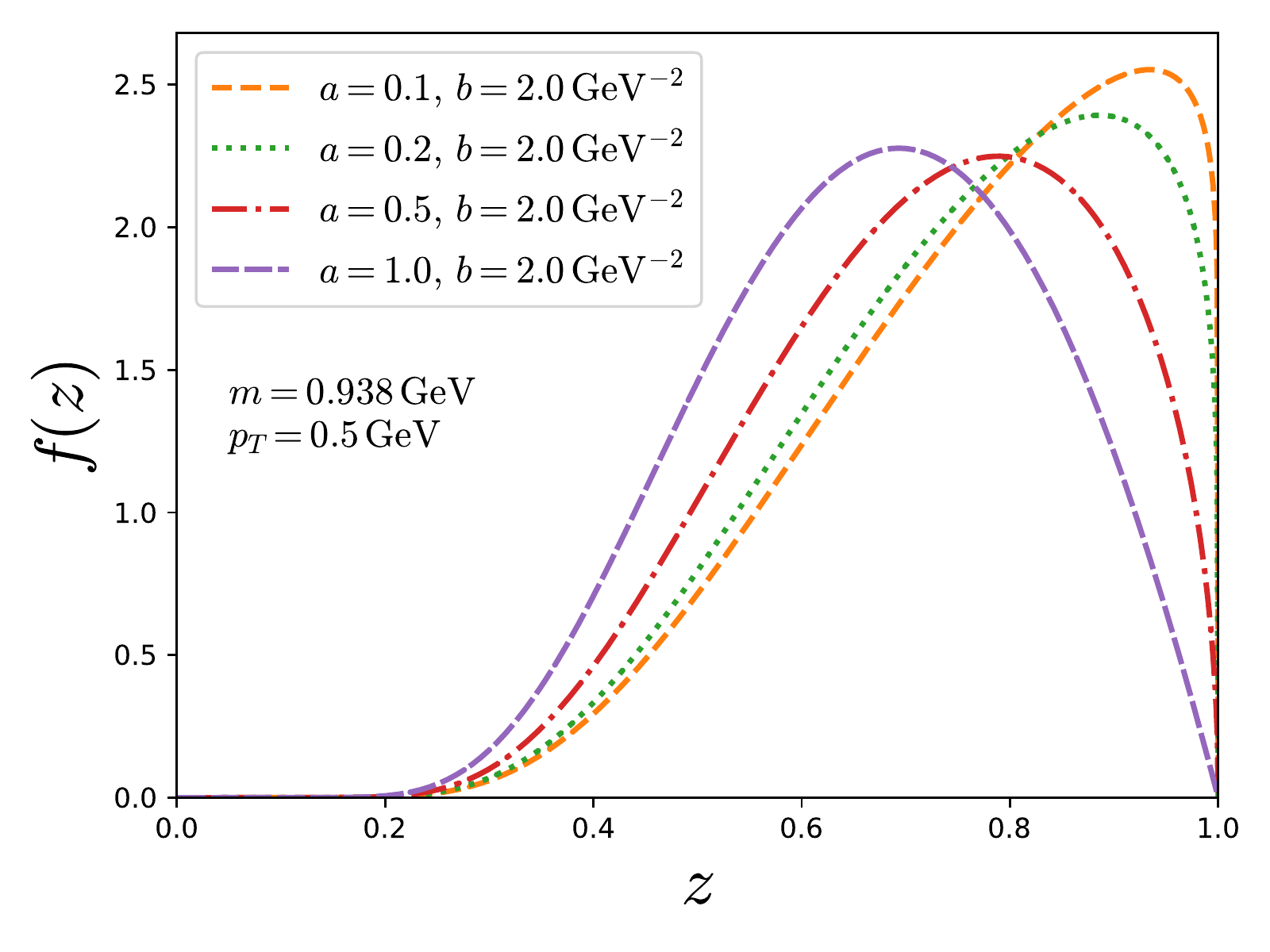}
	\caption{$x_F$ distribution of protons  in proton-proton collisions at $\sqrt{s}=17\,\mathrm{GeV}$ for different values of $a_\mathrm{leading}$ compared to experimental data \cite{NA49protons} (left) and the fragmentation functions for protons with a transverse momentum of $p_T=0.5\,\mathrm{GeV}$ used for the respective calculations (right).}
	\label{fig_stringz_a_leading}
\end{figure}

The effect on the longitudinal momentum distribution of protons and pions of changing the value of $a_\mathrm{leading}$ in the fragmentation function for the leading baryons from soft non-diffractive string processes is shown in figure \ref{fig_stringz_a_leading}. 
The main difference in the $x_F$ distribution of protons is observed in the height of the bump at $x_F\approx 0.3$. For larger $a_\mathrm{leading}$, the peak is more pronounced, since small values of $a_\mathrm{leading}$ correspond to harder fragmentation functions. The similarity  between the curves with the hardest fragmentation function is caused by the fact, that the string fragmentation fails numerically, if there is not enough energy left to produce new particles. Therefore, very high momentum fractions are rejected more often and therefore, the difference in the fragmenattion function is not visible anymore in the observable. As a compromise between data comparison and computational effort to determine the kinematics, the value of $a_\mathrm{leading}$ is set to $0.2$.

\begin{figure}[h]
	\centering
	\includegraphics[width=0.5\textwidth]{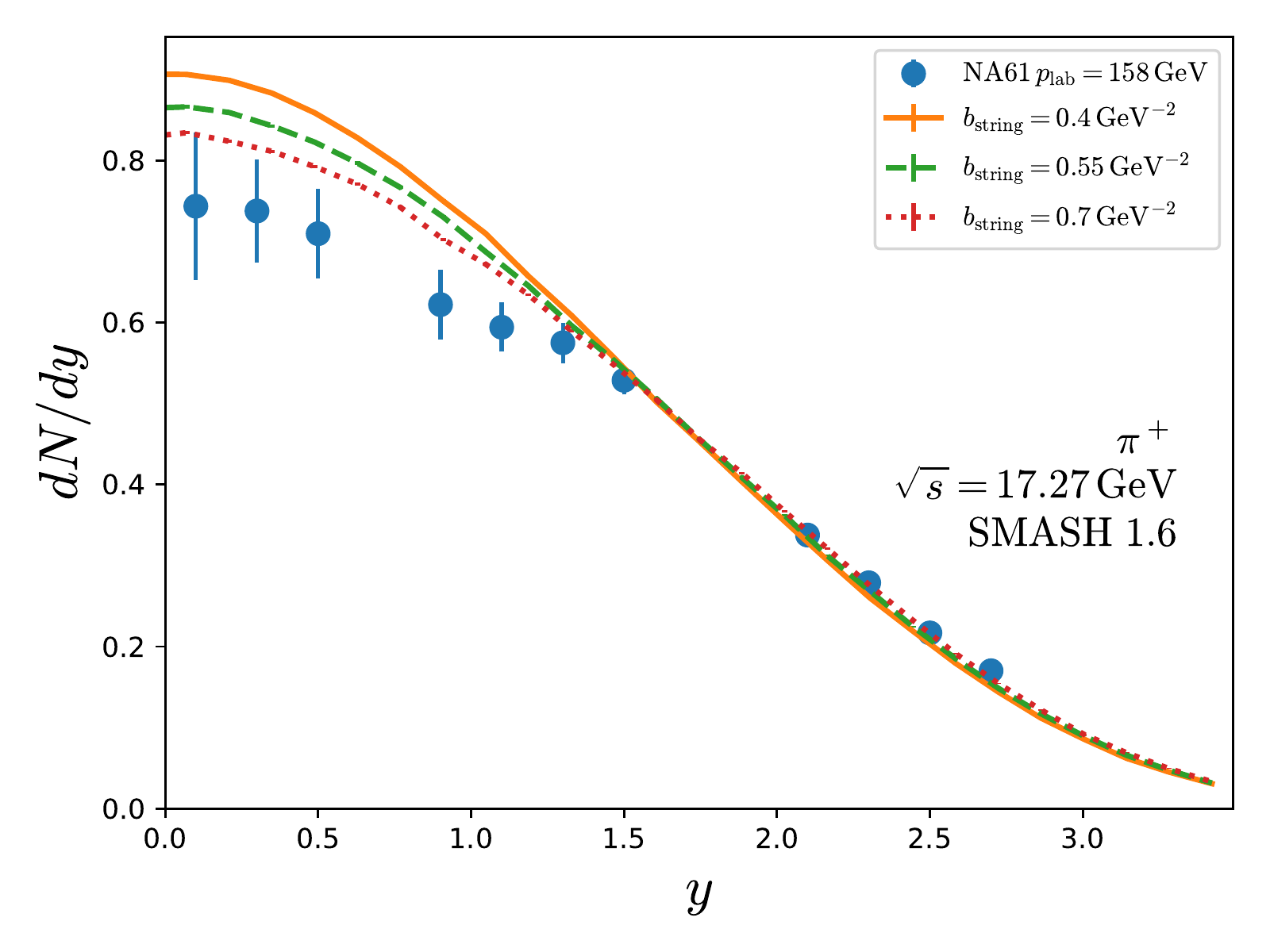}%
	\includegraphics[width=0.5\textwidth]{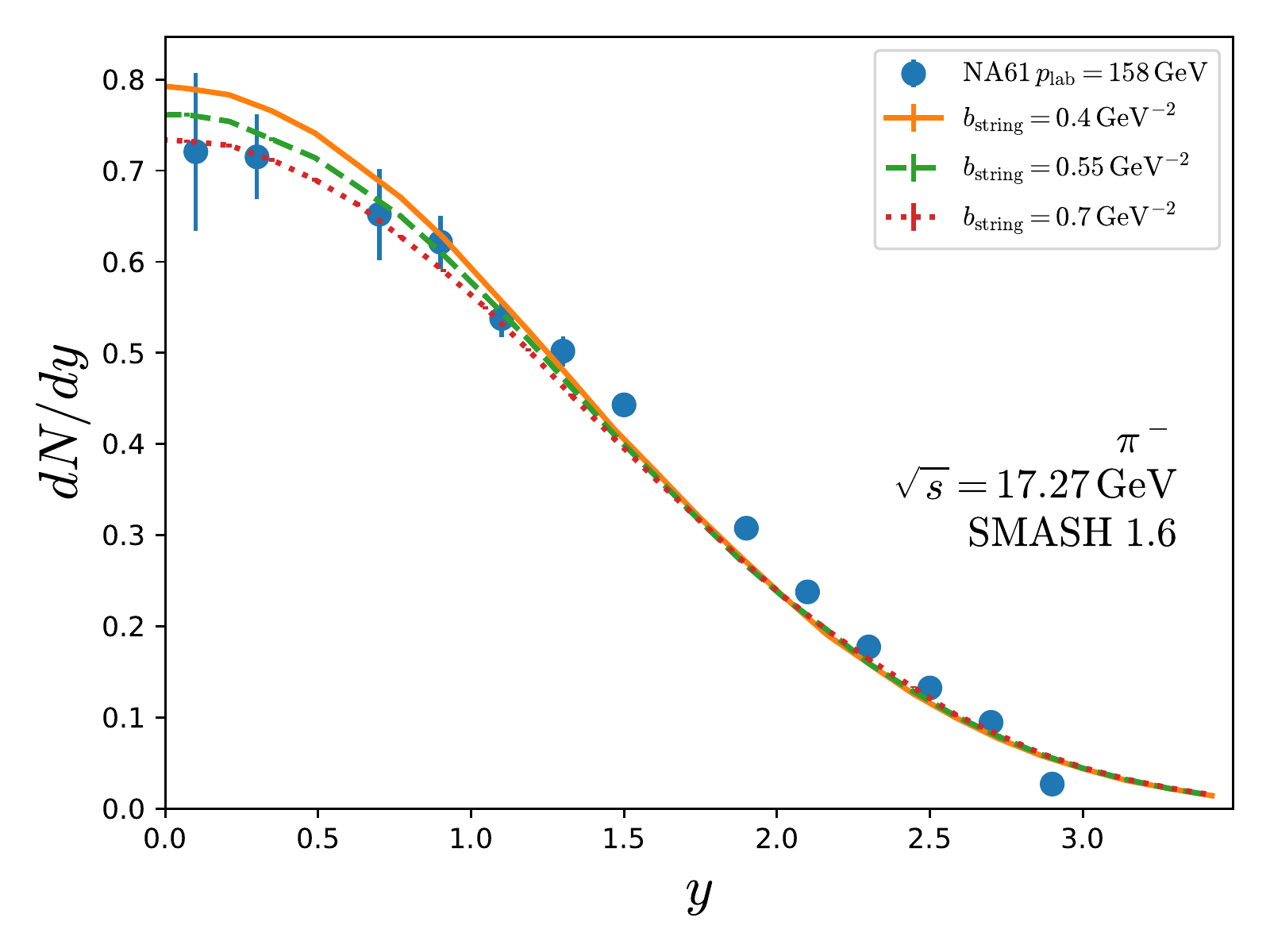}
	\caption{Rapidity spectra of positively (left) and negatively (right) charged pions in proton-proton collisions at $\sqrt{s}=17.27\,\mathrm{GeV}$ for different values of $b_\mathrm{string}$ compared to experimental data \cite{Aduszkiewicz:2017sei}.}
	\label{fig_stringz_b}
\end{figure}

The fragmentation function that is used for all other particles has a strong effect on the production of light mesons.
The value of the parameter $b$, which will be referred to as $b_\mathrm{string}$ in the following, is varied in figure \ref{fig_stringz_b}.
The rapidity spectra of positively and negatively charged pions are sensitive to small changes in the parameter $b_\mathrm{string}$. A softer fragmentation function will lead to more low-energetic pions, while a harder fragmentation function produces pions with larger momenta.

While the mid-rapidity yield of positively charged pions is overestimated for all three values of $b_\mathrm{string}$, the production of negatively charged pions is well described.
At lower beam energies, the pion multiplicity is slightly lower compared to the data as can be seen in figure \ref{fig_pp_summary_y}.
Therefore, the overall best agreement is obtained with $b_\mathrm{string}=0.55\,\mathrm{GeV^{-2}}$.  

\begin{figure}[h]
	\centering
	\includegraphics[width=0.5\textwidth]{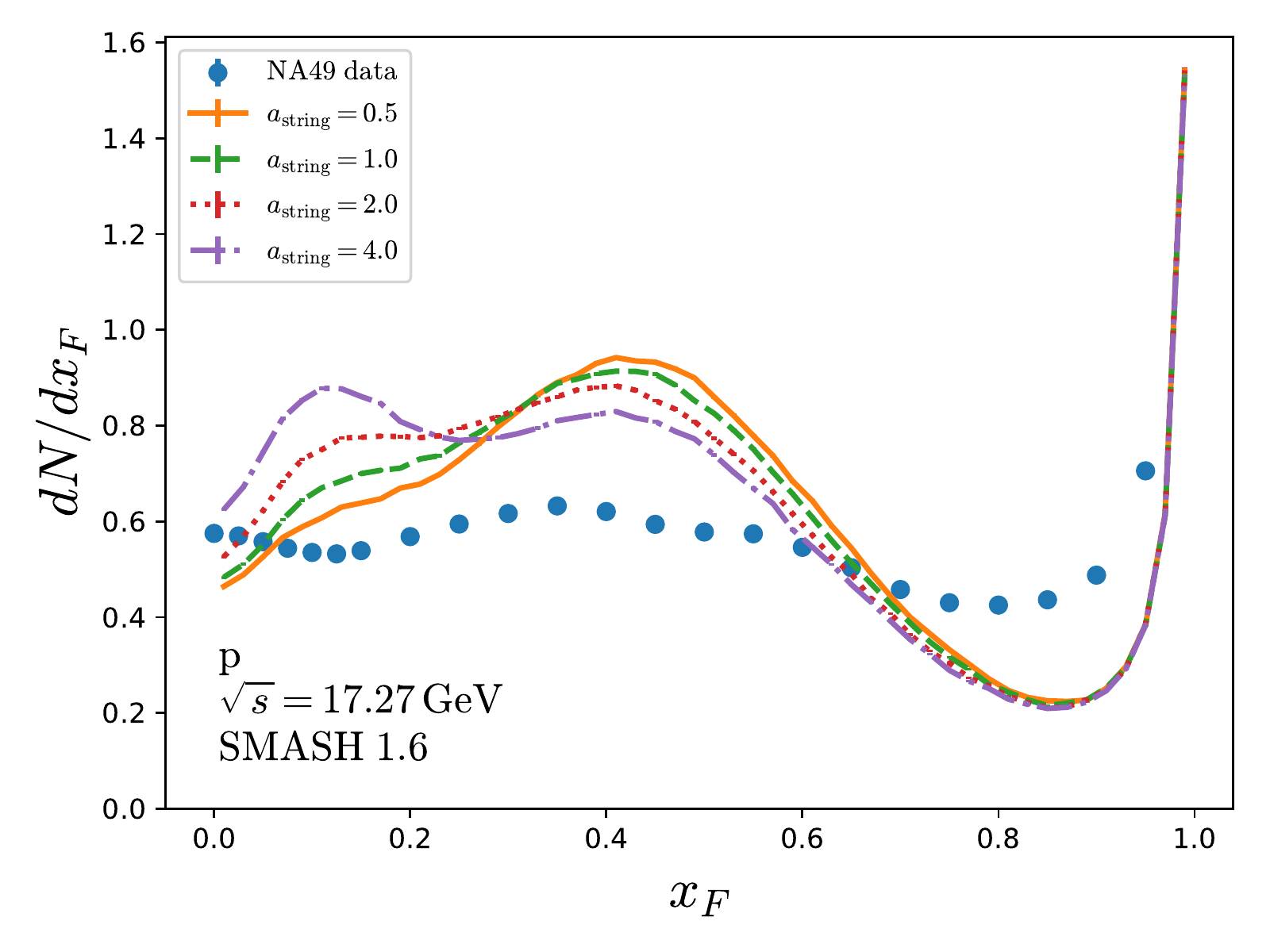}%
	\includegraphics[width=0.5\textwidth]{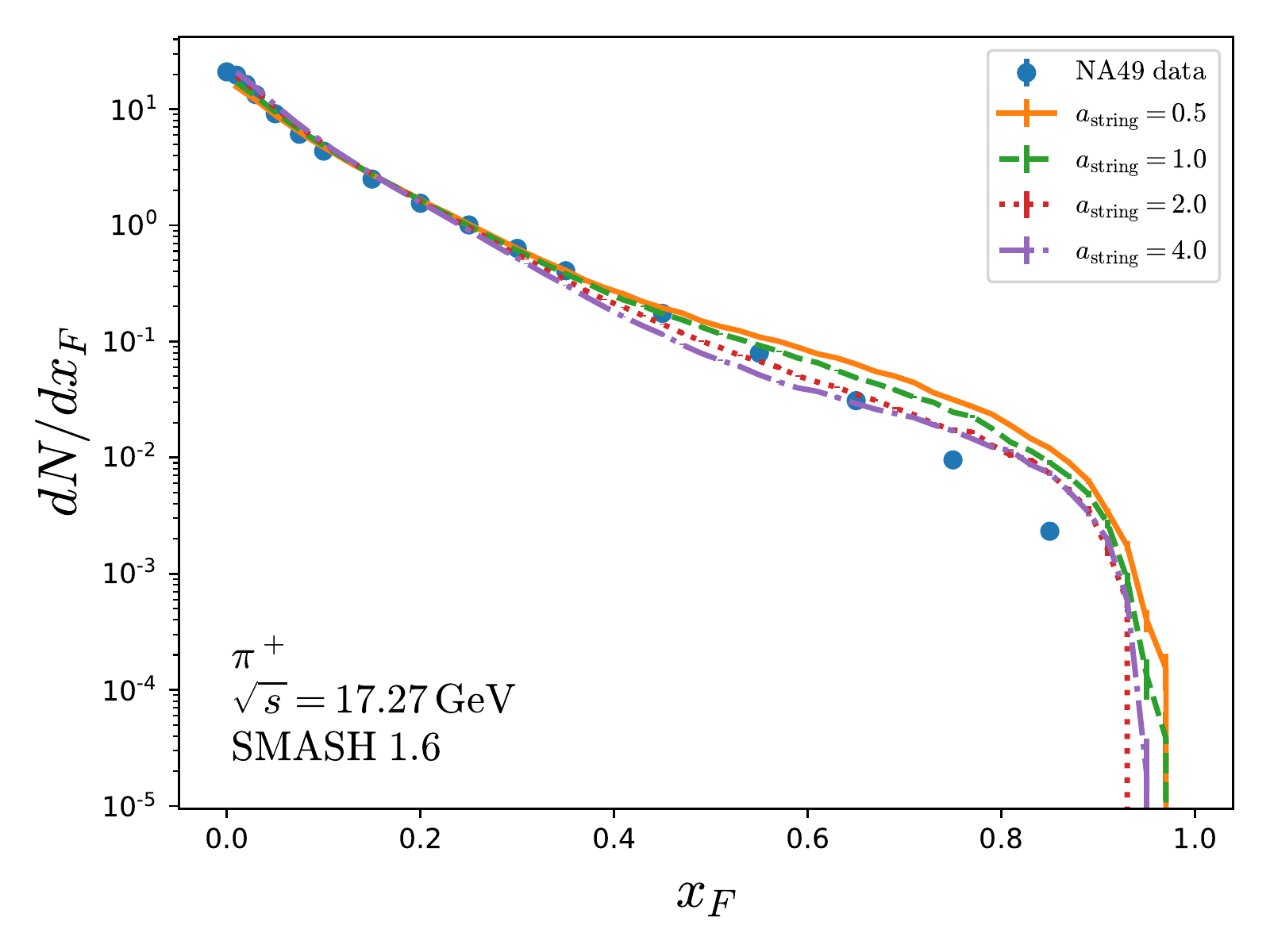}
	\caption{$x_F$ distributions of protons (left) and positively charged pions (right) in proton-proton collisions at $\sqrt{s}=17.27\,\mathrm{GeV}$ compared to experimental data \cite{NA49protons,NA49pions}.}
	\label{fig_stringz_a}
\end{figure}

The final parameter $a$ of the fragmentation function used for particles that are not leading baryons is called $a_\mathrm{string}$ in the following.
It is varied in figure \ref{fig_stringz_a} which shows the $x_F$ distribution of protons and positively charged pions.
Higher $a_\mathrm{string}$ corresponds to a harder fragmentation function which is reflected in the soft proton sector and the pion $x_F$ distribution.
Overall, the best agreement, considering both distributions, is found with $a_\mathrm{string} = 2.0$. 

\begin{figure}[h]
	\centering
	\includegraphics[width=0.5\textwidth]{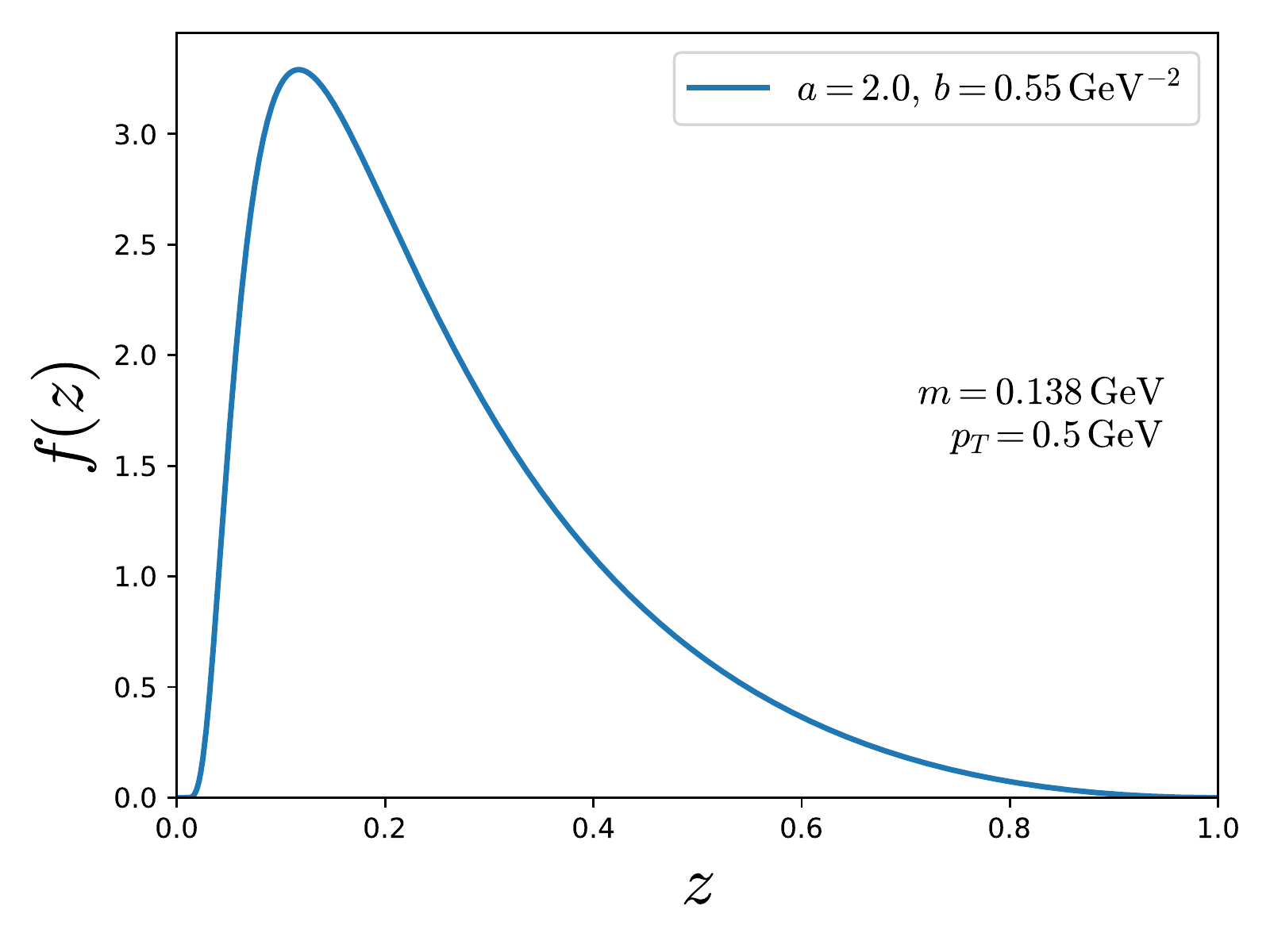}
	\caption{Fragmentation function $f(z)$ according to equation (\ref{eq_fragmentation_function}) for pions. The transverse momentum is set to $p_T=0.5\,\mathrm{GeV}$, which is approximately the average transverse momentum produced during the fragmentation.}
	\label{fig_fragmentation_function}
\end{figure}

For completeness, figure \ref{fig_fragmentation_function} shows the fragmentation function used for all string fragments apart from leading baryons in soft non-diffractive processes.
For the other string fragments, softer fragmentation functions are applied, where the difference between the particle species originates exclusively from the $m_T$ dependence in equation (\ref{eq_fragmentation_function}).

\subsection{Transverse Momentum Production}
\label{sec_transverse_momentum}

\begin{figure}[h]
	\centering
	\includegraphics[width=0.5\textwidth]{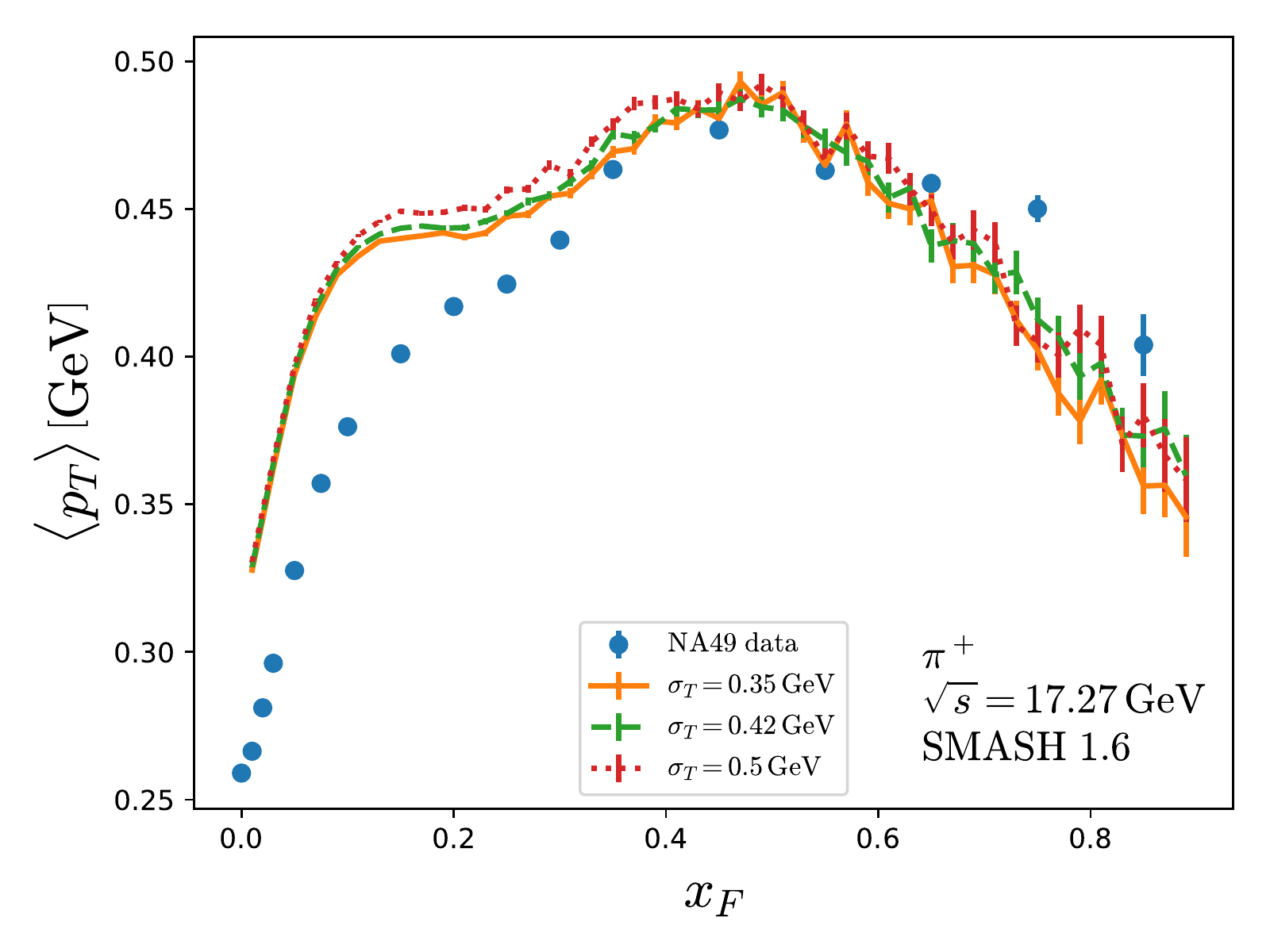}%
	\includegraphics[width=0.5\textwidth]{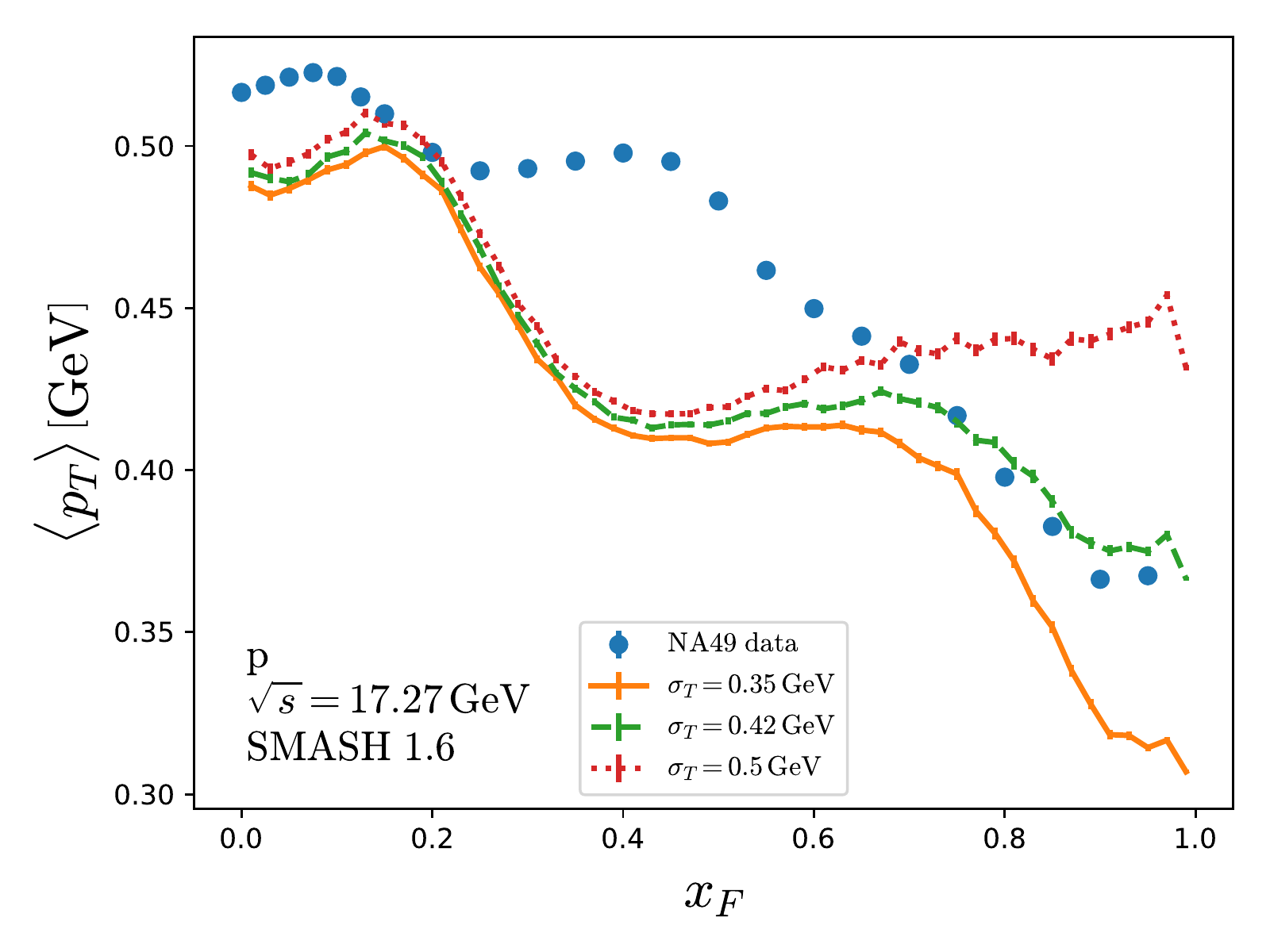}
	\caption{Mean transverse momentum of positively charged pions (left) and protons (right) as a function of $x_F$ in proton-proton collisions at $\sqrt{s}=17.27\,\mathrm{GeV}$ compared to experimental data \cite{NA49pions,NA49protons}, note that the zero is suppressed to zoom in to the region of interest. }
	\label{fig_sigma_perp}
\end{figure}

Transverse momentum is produced in two steps in proton-proton collisions: first in the excitation process adjusted by $\sigma_T$ and afterwards during the fragmentation tuned by changing $\sigma_{T,\mathrm{string}}$. The initial transverse momentum transfer between the interacting hadrons is sampled according to a Gaussian with a width of $\sigma_T$ as described in section \ref{sec_soft_string_routine}.
Figure \ref{fig_sigma_perp} shows the dependence of the mean transverse momentum of protons and pions as a function of $x_F$ on the value of $\sigma_T$. 

The mean transverse momentum of pions shows only a weak dependence on the value of $\sigma_T$. At low $x_F$, the protons are fragmented from a string. Therefore they show the same behavior as pions. Most large $x_F$ protons in proton-proton collisions are however not fragmented from a string, but only took part in a singe diffractive process. Their transverse momentum is directly sampled from the Gaussian with a width of $\sigma_T$, which explains the strong dependence on $\sigma_T$. $\sigma_T=0.42\,\mathrm{GeV}$ is therefore fixed to match the mean transverse momentum at large $x_F$.

\begin{figure}[h]
	\centering
	\includegraphics[width=0.5\textwidth]{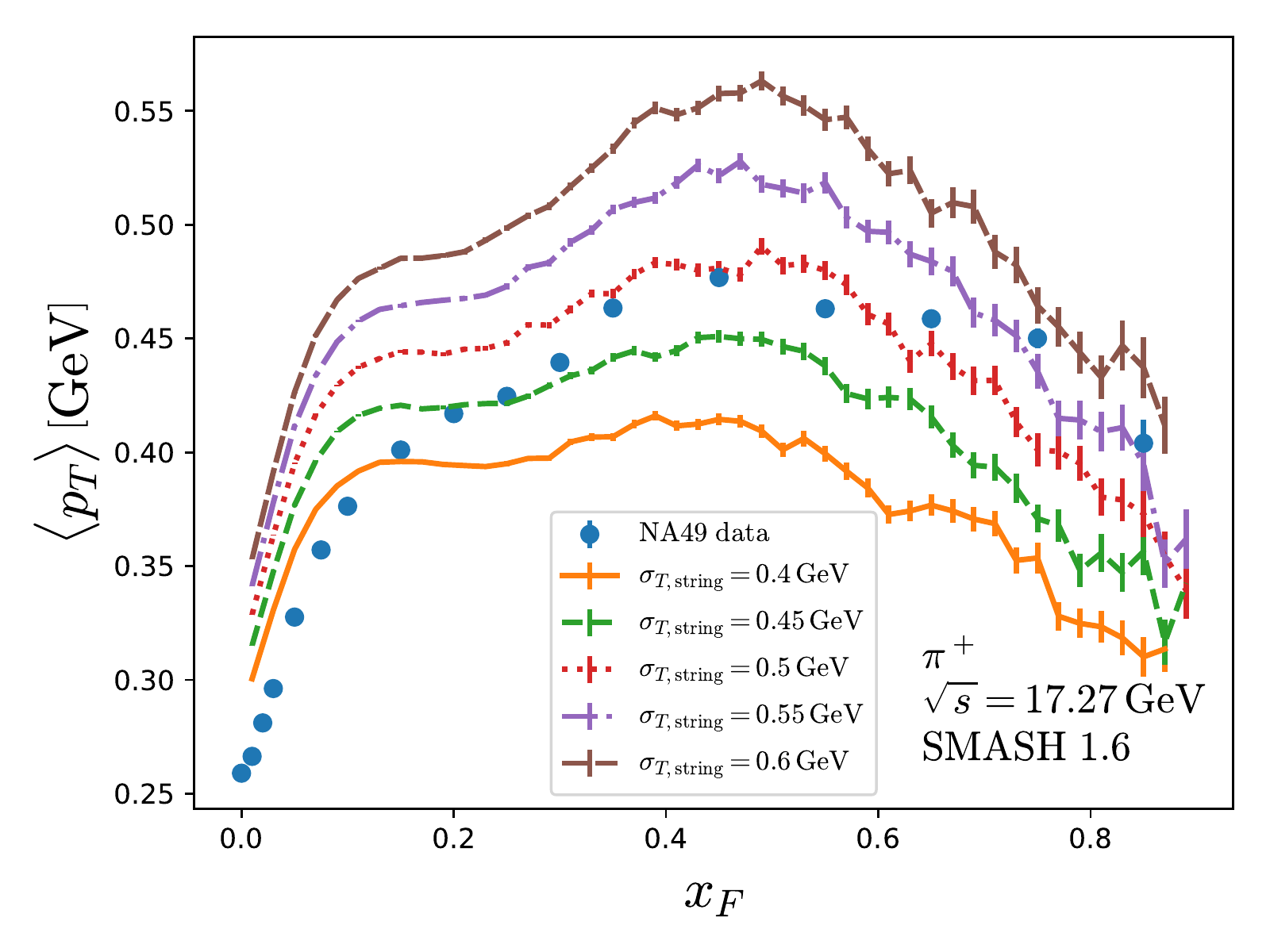}%
	\includegraphics[width=0.5\textwidth]{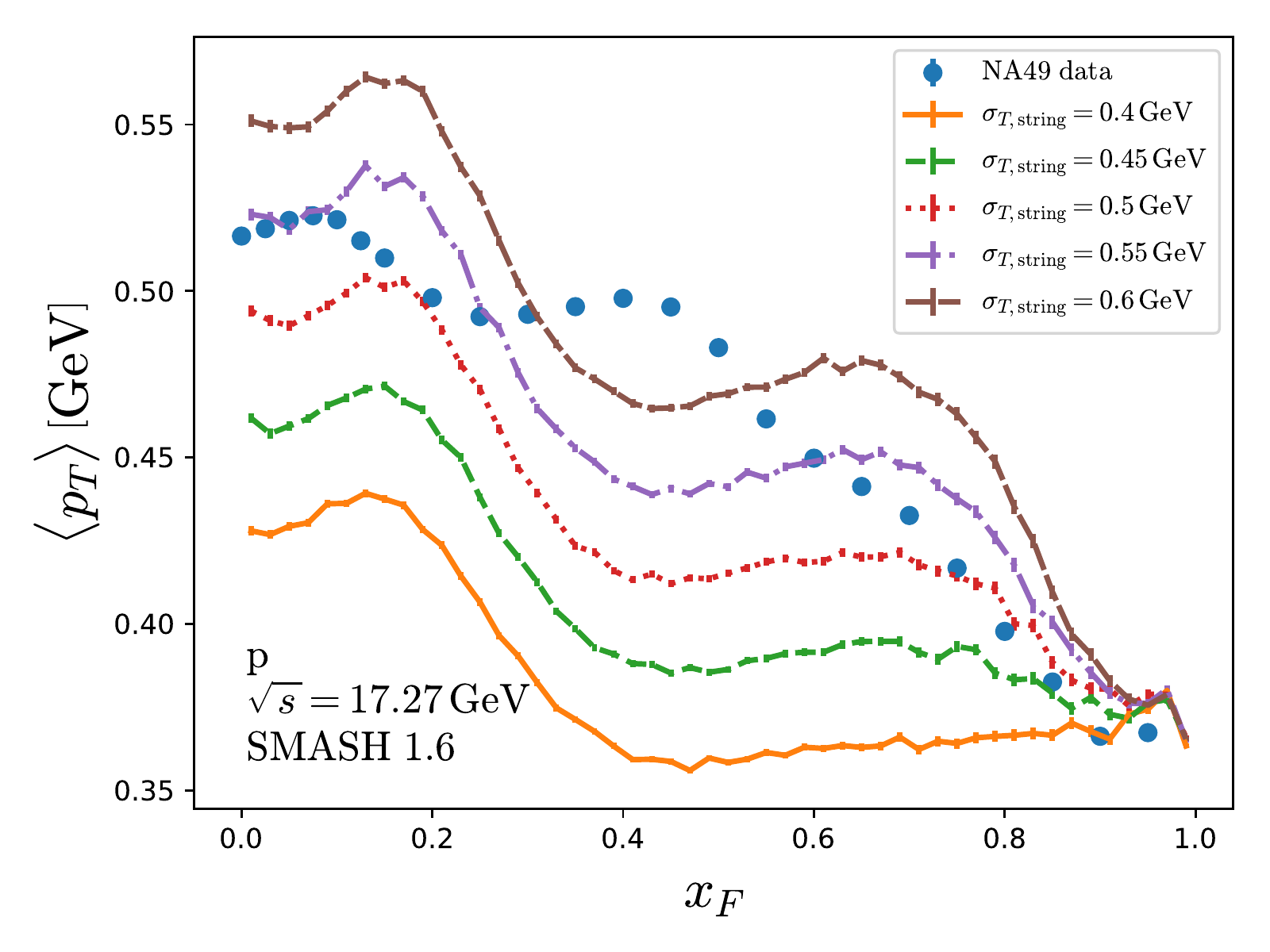}
	\caption{Mean transverse momentum of positively charged pions (left) and protons (right) as a function of $x_F$ in proton-proton collisions at $\sqrt{s}=17.27\,\mathrm{GeV}$ compared to experimental data \cite{NA49pions,NA49protons}.}
	\label{fig_string_sigma_t}
\end{figure}

The production of transverse momentum during the fragmentation of a string is regulated by $\sigma_{T,\mathrm{string}}$.
In \textsc{Pythia}, a Gaussian with a width of $\sigma_{T,\mathrm{string}}$ is used to sample the transverse momentum of each individual string fragment.
The influence of changing $\sigma_{T,\mathrm{string}}$ on the mean transverse momentum of pions and protons is shown in figure \ref{fig_string_sigma_t}. The transverse momentum distribution for the fragmentation is much more important for particle production than the one in the string excitation process. 

In the case of pions, the transverse momentum is scaled up for all bins of $x_F$. For protons, the opposite behavior to what is seen when varying $\sigma_T$ is observed.
At small $x_F$, where the protons originate from a string fragmentation, the proton $\langle p_T\rangle$ is strongly dependent on $\sigma_{T,\mathrm{string}}$, while at $x_F\approx 1$ all curves lie on top of each other. The value of $\sigma_{T,\mathrm{string}}$ therefore needs to be tuned to multiple particle species simultaneously.
The best agreement is found for $\sigma_{T,\mathrm{string}}=0.5\,\mathrm{GeV}$, where for protons too little transverse momentum is produced, while the pions at low $x_F$ obtain too much $p_T$.

\begin{figure}[h]
  \centering
  \includegraphics[width=0.5\textwidth]{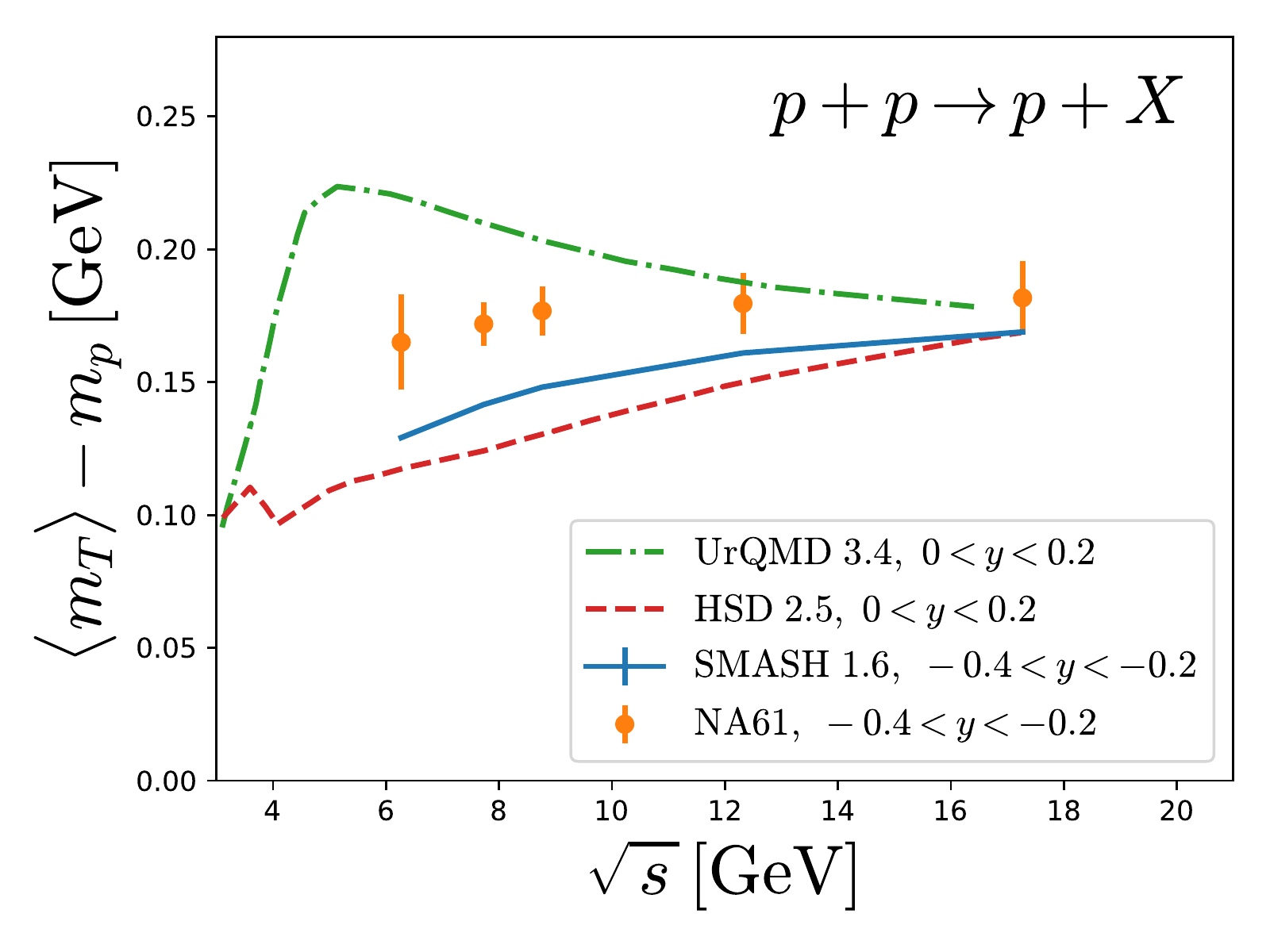}
  \caption{Difference between mean transverse mass of protons at mid-rapidity and the proton mass in proton-proton collisions as a function of the center of mass energy $\sqrt{s}$ of the collision compared to experimental data and other hadronic transport calculations \cite{Pulawski:2015tka,Vovchenko:2014vda,Bleicher:1999xi,Bass:1998ca,Bratkovskaya:2011wp}}
  \label{fig_mean_mt}
\end{figure}

Since the NA61 collaboration recently measured the transverse mass of protons at mid-rapidity as a function of collision energy, calculations from SMASH are compared with the data and other transport approaches in figure \ref{fig_mean_mt}. Please note that the rapidity ranges do not match exactly, since the HSD and UrQMD calculations were performed before the experiment was carried out, therefore the comparison is not fully quantitative. 
The UrQMD calculation overshoots the data at low $\sqrt{s}$ due to the transition from resonances to strings which is located at higher energies for the binary collisions than in HSD and SMASH. The shape of the HSD curve and the SMASH calculation qualitatively follow the trend of experimental data, while both underpredict the mean transverse momentum slightly. 

\subsection{Strangeness Production}
\begin{figure}[h]
	\centering
	\includegraphics[width=0.5\textwidth]{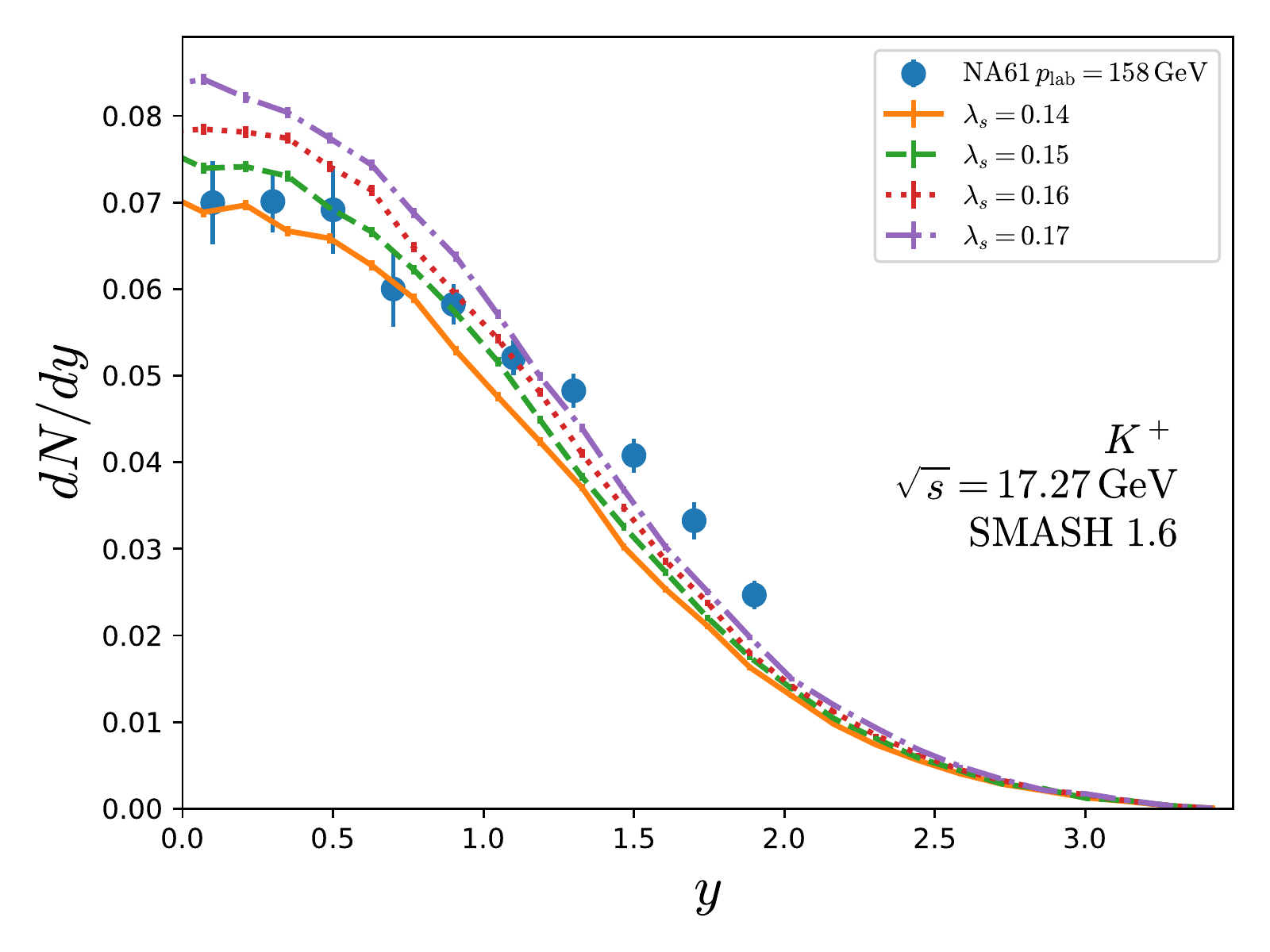}
	\caption{Rapidity spectra of positively charged kaons in proton-proton collisions at $\sqrt{s}=17.27\,\mathrm{GeV}$ for different values of $\lambda_s$ compared to experimental data \cite{Aduszkiewicz:2017sei}.}
	\label{fig_strange_supp}
\end{figure}

The production of strange quarks heavily relies on the probability of producing an $s\bar{s}$ pair compared to the probability of producing a light $q\bar{q}$ pair during the string fragmentation. To suppress the production of strange quark pairs according to their higher mass, the strangeness suppression factor $\lambda_s$ is introduced:
\begin{equation}
	\lambda_s= \frac{P(s\bar{s})}{P(u\bar{u})}=\frac{P(s\bar{s})}{P(d\bar{d})}\,,
	\label{eq_strange_supp}
\end{equation}
where $P(u\bar{u})$, $P(d\bar{d})$ and $P(s\bar{s})$ denote the probabilities to produce a up-antiup, down-antidown and a strange-antistrange quark pair respectively.
The impact of varying this parameter on the kaon rapidity spectra is shown in figure \ref{fig_strange_supp}.
Without affecting the dynamics of the system much, the strangeness suppression factor regulates the multiplicity of strange hadrons.
Since the rapidity distribution in our calculation is slightly steeper than the measured one, the strangeness suppression factor is set to $\lambda_s=0.16$ in order to obtain a good agreement for the total kaon multiplicity. For tuning $\lambda_s$, only the positively charged kaons are considered, since the energy dependence of the negatively charged kaons is not as well understood as can be seen in the bottom right panel of figure \ref{fig_pp_summary_y}.
 
 \subsection{Diquark Production}
 Similar to the description for the production of strange quarks, a diquark suppression factor $\lambda_\mathrm{diquark}$ is introduced to quantify the likelihood of producing diquarks:
 \begin{equation}
 	\lambda_\mathrm{diquark} = \frac{P(qq\bar{q}\bar{q})}{P(q\bar{q})}
	\label{eq_diquark_supp}
 \end{equation}
 
\begin{figure}[h]
	\centering
	\includegraphics[width=0.5\textwidth]{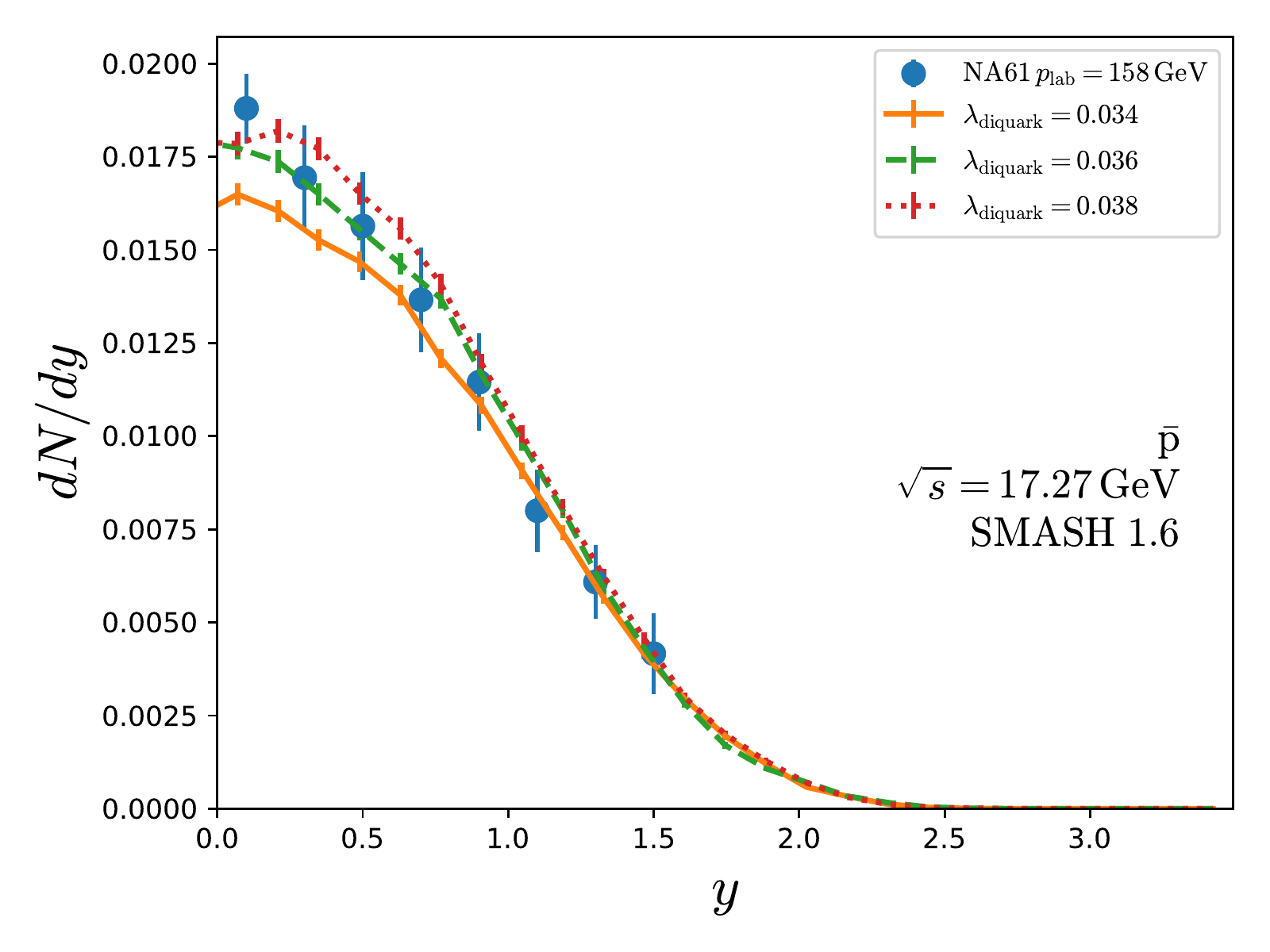}
	\caption{Rapidity spectra of antiprotons in proton-proton collisions at $\sqrt{s}=17.27\,\mathrm{GeV}$ for different values of $\lambda_\mathrm{diquark}$ compared to experimental data \cite{Aduszkiewicz:2017sei}.}
	\label{fig_diquark_supp}
\end{figure}

A diquark and an antidiquark always combine to a baryon and an antibaryon, since a meson cannot contain two (anti)quarks but only one quark and one antiquark.
Since diquarks are present in a much larger fraction than the newly produced pairs as valence quarks in the excited baryons, the antiproton production constrains the diquark suppression factor much more directly. The comparison of the rapidity spectrum of antiprotons for different values of $\lambda_\mathrm{diquark}$ is shown in figure \ref{fig_diquark_supp}.
The antiproton multiplicity is very low, resulting in a small diquark suppression factor and a value of $\lambda_\mathrm{diquark}=0.036$ yields the best agreement with the measured antiproton rapidity spectrum. Even though the data point at mid-rapidity suggests a larger $\lambda_\mathrm{diquark}$, all other points are reproduced very well and at lower energies a higher antiproton production contradicts the measurement as shown in figure \ref{fig_pp_summary_y}.
  
\subsection{Popcorn Rate}
\label{sec_popcorn_rate}
\begin{figure}[h]
  	\centering
  	\includegraphics[width=0.5\textwidth]{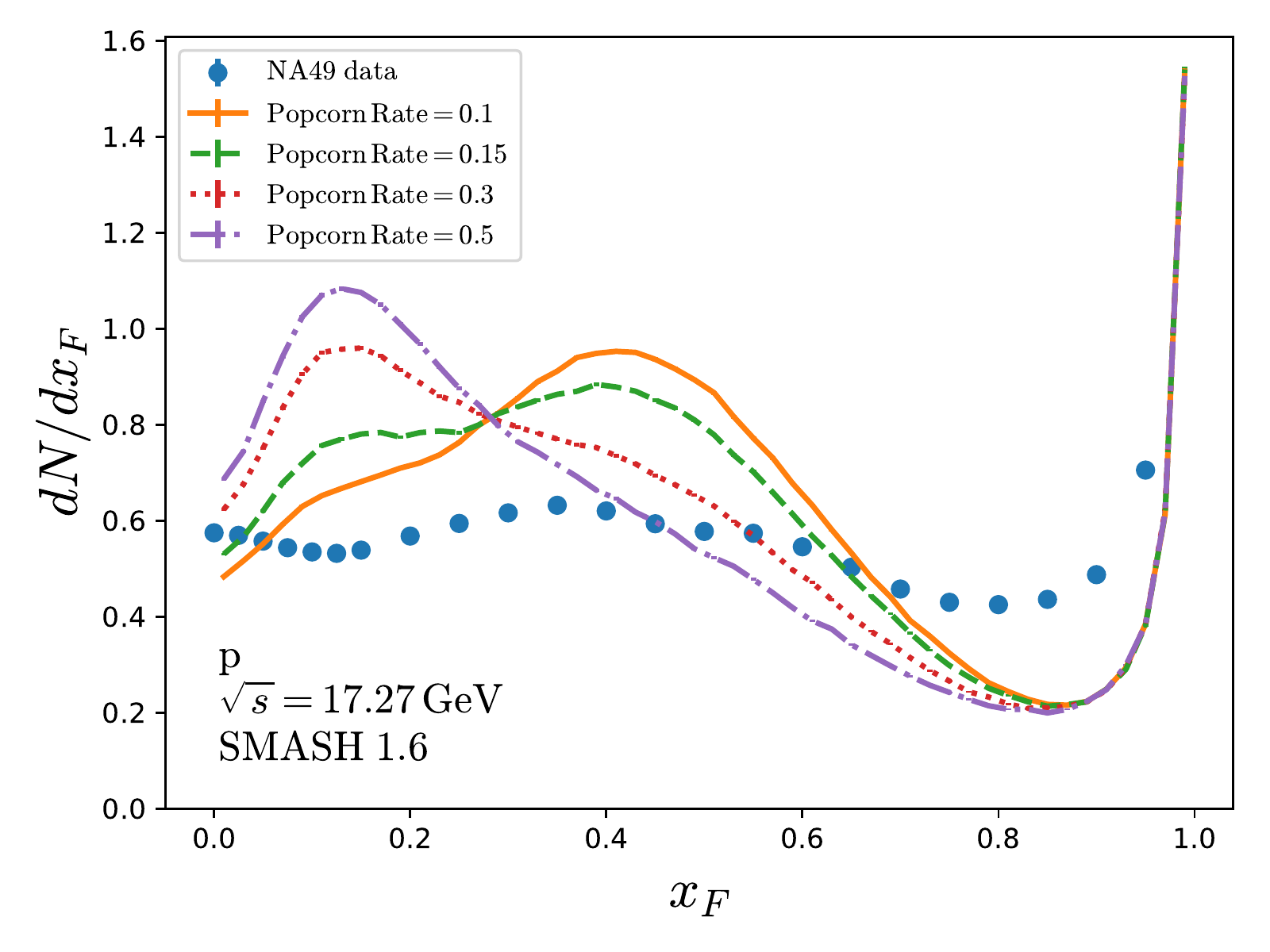}%
  	\includegraphics[width=0.5\textwidth]{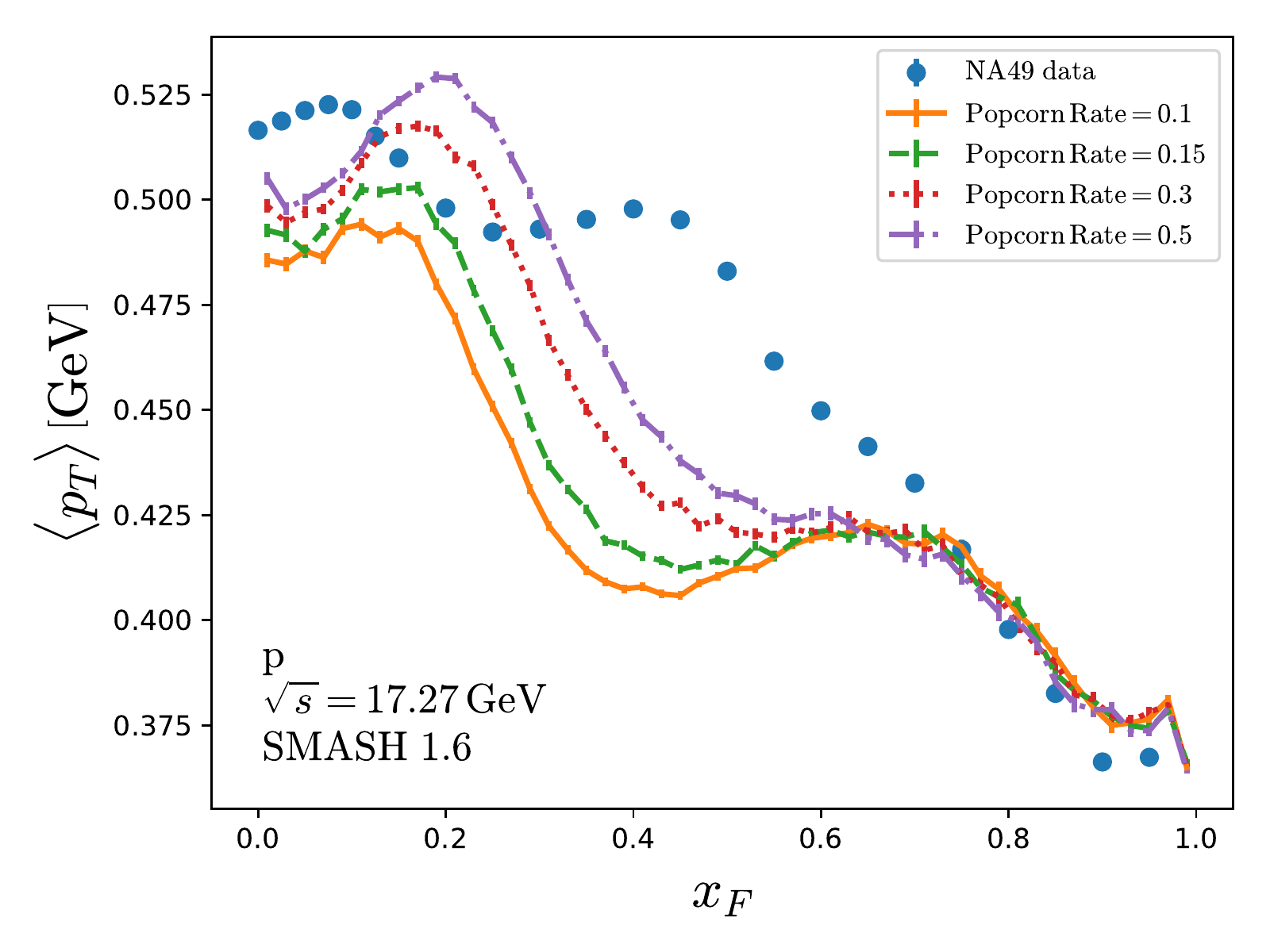}
  	\caption{$x_F$ distribution (left) and mean transverse mass (right) of protons in proton-proton collisions at $\sqrt{s}=17.27\,\mathrm{GeV}$ for different popcorn rates compared to experimental data \cite{NA49protons}.}
  	\label{fig_popcorn_rate}
  \end{figure}

When a diquark-antidiquark pair is produced, they will recombine with surrounding quarks and antiquarks, forming new baryons.
Since the diquark and the antidiquark are produced in a pair production, they are connected via their color charge.
This will in many cases lead to the two fragmented baryons to be produced next to each other in phase space.
It is however also possible to create another quark-antiquark pair in the color field spanned by the diquark and the
antidiquark. 
This leads to the production of a meson between the two baryons \cite{Andersson:1984af}.
In the case of a baryonic string, it is, within a popcorn process, possible to fragment a meson at the diquark end of the string.
Given that a diquark-antidiquark pair is created, the probability of such a process is given by the popcorn rate, which is a \textsc{Pythia} parameter that can be varied in order to reproduce the experimental data.
The effect of changing the popcorn rate on the dynamics of protons is investigated in figure \ref{fig_popcorn_rate}.
Increasing the popcorn rate leads to protons being shifted to low $x_F$.
In the transverse direction, the proton $\langle p_T\rangle$  increases with a growing popcorn rate.
While a large popcorn rate results in a better agreement for the transverse momentum, the shape of the $x_F$ distribution is not compatible with the data.
Because the $x_F$ distribution is, like the data, flat in the low $x_F$ region in the experimental data and a fair agreement in the transverse momentum can be obtained, the popcorn rate is set to 0.15.
Compared to the effect on the proton dynamics, the other particle species are only slightly affected by changing the popcorn rate.

\subsection{Tuning of Parameters}
In this section, we describe in more detail how the tuning of parameters is performed. 
As already mentioned in the beginning of section \ref{sec_proton_proton_collisions}, even though only single parameters are varied in the plots throughout section \ref{sec_proton_proton_collisions}, the interplay of the parameters was investigated by changing their values simultaneously when the final set of parameters as quoted in table \ref{tab_defaults} was determined.

\begin{figure}[h]
	\centering
	\includegraphics[width=0.5\textwidth]{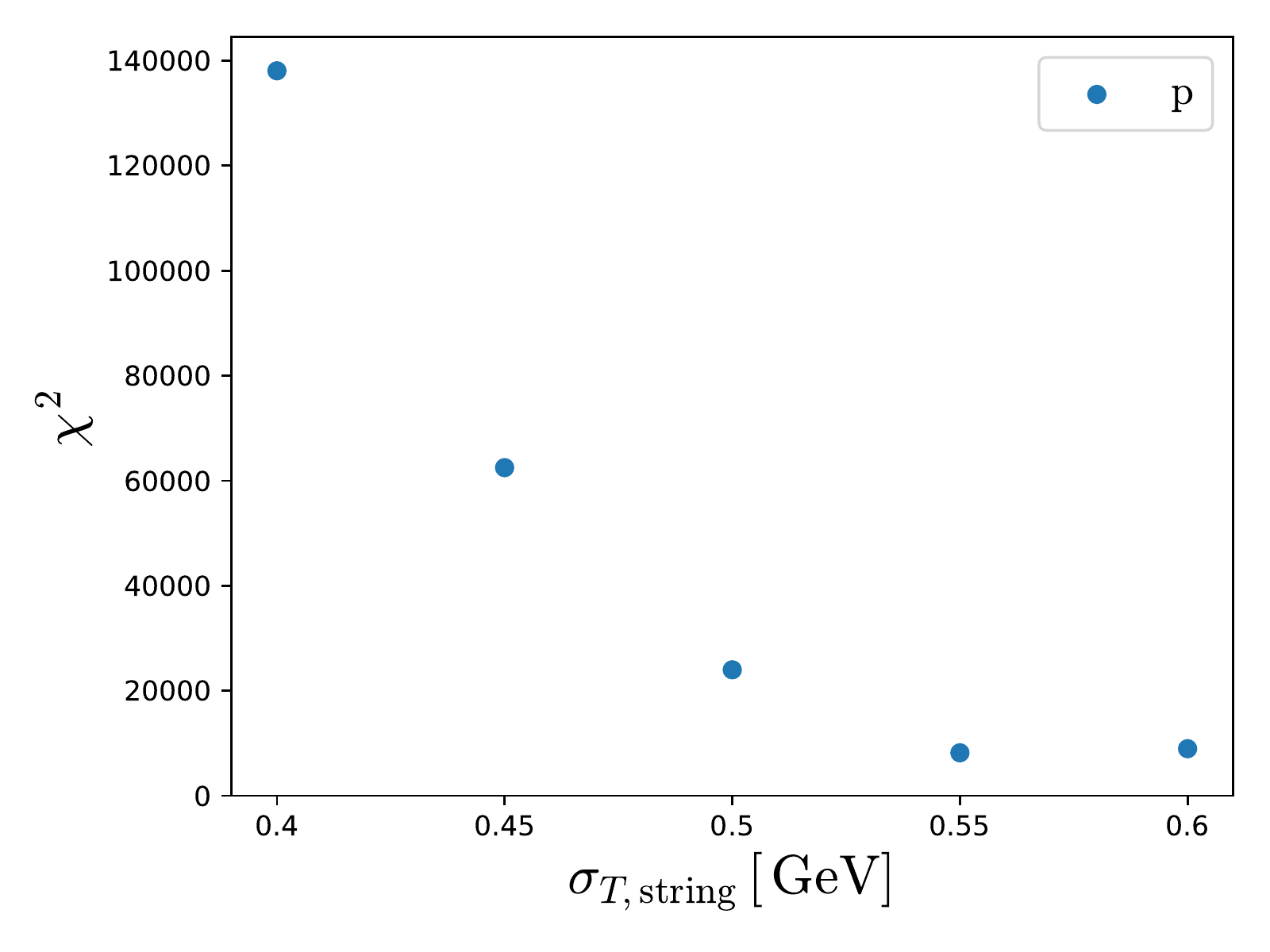}%
	\includegraphics[width=0.5\textwidth]{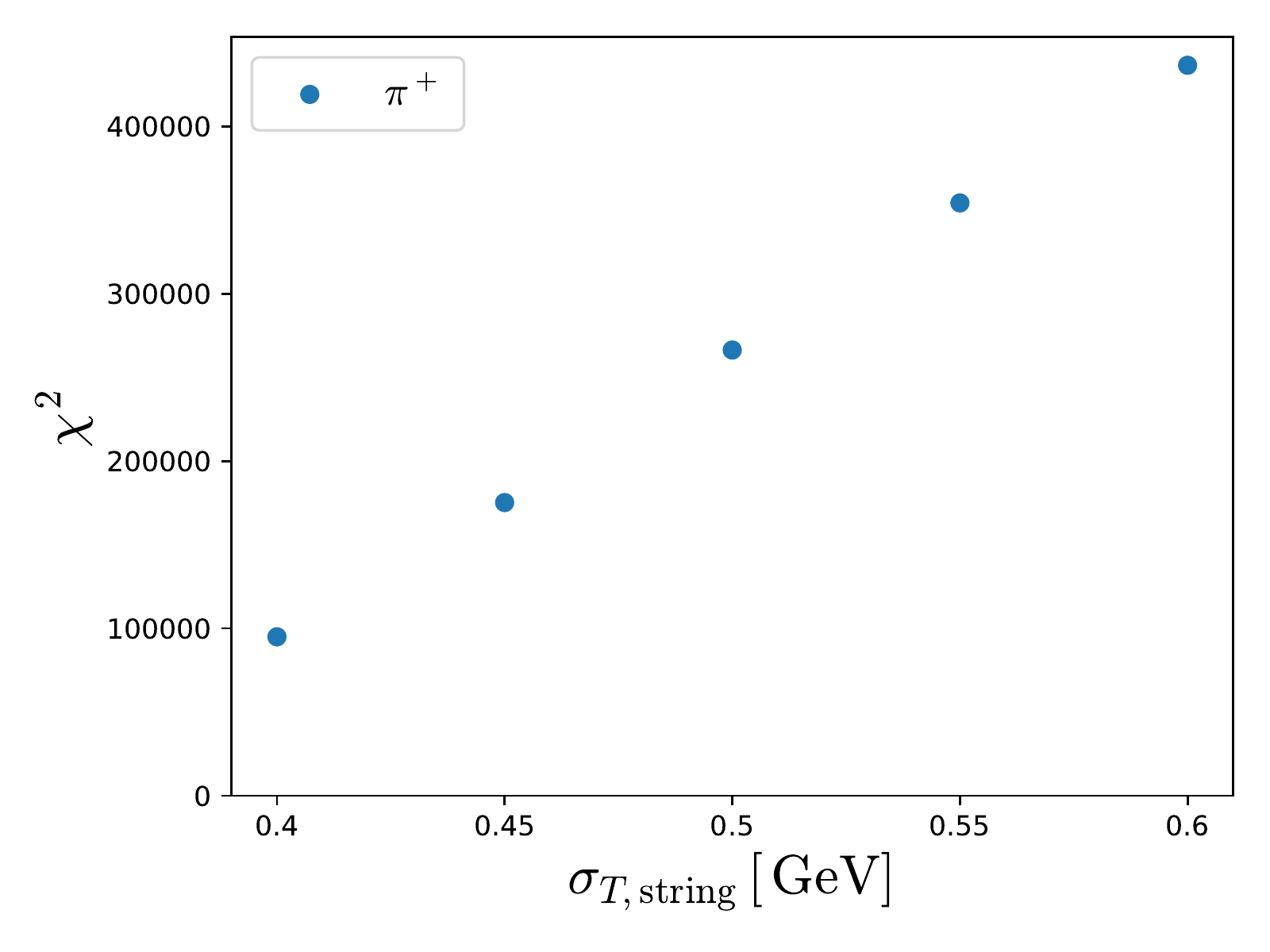}
	\caption{$\chi^2$ for the comparison between the model and experimental data shown in figure \ref{fig_string_sigma_t} as a function of the value for the parameter $\sigma_{T,\mathrm{string}}$. The plot on the left shows the $\chi^2$ for the proton $\langle p_T \rangle$ while the plot on the right shows the $\langle p_T \rangle$ of positively charged pions.}
	\label{fig_chi_sqr}
\end{figure}

The exercise of tuning parameters often is solved by minimizing the $\chi^2$ of a specific observable.
In this work we refrain from fixing the parameter values in that way because, as one can see in the previous sections, there is not one specific plot that is most important to describe but compromises need to be made between different quantities.
In the following, we show one example where the difficulties of minimizing a $\chi^2$ are evident.
Figure \ref{fig_chi_sqr} shows the difficulties when tuning $\sigma_{T,\mathrm{string}}$ in terms of the $\chi^2$ obtained from the two plots in figure \ref{fig_string_sigma_t} which show the mean transverse momentum of protons and pions respectively.

$\chi^2$ is calculated according to
\begin{equation}
	\chi^2 = \sum_i \frac{(\langle p_T\rangle(x_{Fi}) - \langle p_T\rangle_\mathrm{exp}(x_{Fi}))^2}{\sigma_i^2}\,,
\end{equation}
where $x_{Fi}$ are the bins in $x_F$ and $\sigma_i$ is the uncertainty of that bin. For the uncertainty, the experimental error and the statistical uncertainty of the calculation are added in quadrature.
The calculated values of $\chi^2$ are very large due to tiny experimental uncertainties in the proton and pion $\langle p_T \rangle$.

Figure \ref{fig_chi_sqr} reflects the tension between the two observables since for the proton $\langle p_T \rangle$ a relatively large value of $\sigma_{T,\mathrm{string}}$ is preferred by the experimental data while the $\chi^2$ can only be minimized with a small $\sigma_{T,\mathrm{string}}$ for pions.
Since there is more data for different particle species and other observables available, it is not clear how one would incorporate the knowledge about which data points are most important to describe (such as pions being more important to describe than kaons) without arbitrarily choosing weights in a $\chi^2$ analysis.
In order to get reasonable parameter values, we therefore stick to tuning the parameters by eye after studying the influence of each parameter on each plot.

As mentioned earlier, some of the parameters are correlated. This complicates the process of finding a good parameter set. We briefly describe how a parameter set is found and where one has to be especially careful in taking the correlations into account.
The available energy for the string excitation is determined by the parton distribution function. Therefore the first parameter to look at is $\beta_\mathrm{quark}$. The value of $\sigma_T$ is both important for determining the right amount of available energy for the collision and easy to fix to the $\langle p_T \rangle$ of protons at large $x_F$.
$\sigma_{T,string}$ can be determined from the mean transverse momentum of protons and pions once $\sigma_T$ is fixed.
More difficult to determine are the parameters of the fragmentation functions.
The parameters of the fragmentation function for leading baryons $a_\mathrm{leading}$ and $b_\mathrm{leading}$ are responsible for the longitudinal momentum of leading protons.
They can in principle be tuned to give a good description of protons at relatively large ($\sim 0.5$) $x_F$ but one has to take into account that the momentum taken by the leading baryons is not available for the non-leading string fragments.
If the fragmentation function for leading baryons is very hard, one would need a soft fragmentation function for the other particles to maintain the correct multiplicities.
Therefore, there is a correlation between the parameters of the fragmentation function.
The popcorn rate also affects the longitudinal dynamics of protons a lot. As a result, the fragmentation function was tuned for different popcorn rates to see what yields the best agreement with the data. $\lambda_s$ and $\lambda_\mathrm{diquark}$ are afterwards independently fixed since they only affect the kaon and antiquark multiplicities.

It is clear that tuning the parameters by eye results in a reasonable set of parameters but better parameter sets exist.
This introduces a systematic uncertainty on the following results that could be estimated by varying the parameters in a range in which a reasonable agreement is found in proton-proton collisions.
We avoid calculating this systematic uncertainty because there are more sources for systematic uncertainties that cannot be estimated in a simple way, such as the many assumptions made in
a hadronic transport approach. 


 \subsection{Proton+Proton Results Overview}
 \label{sec_pp_overview}
 
\begin{figure}[h]
  \centering
  \includegraphics[width=0.5\textwidth]{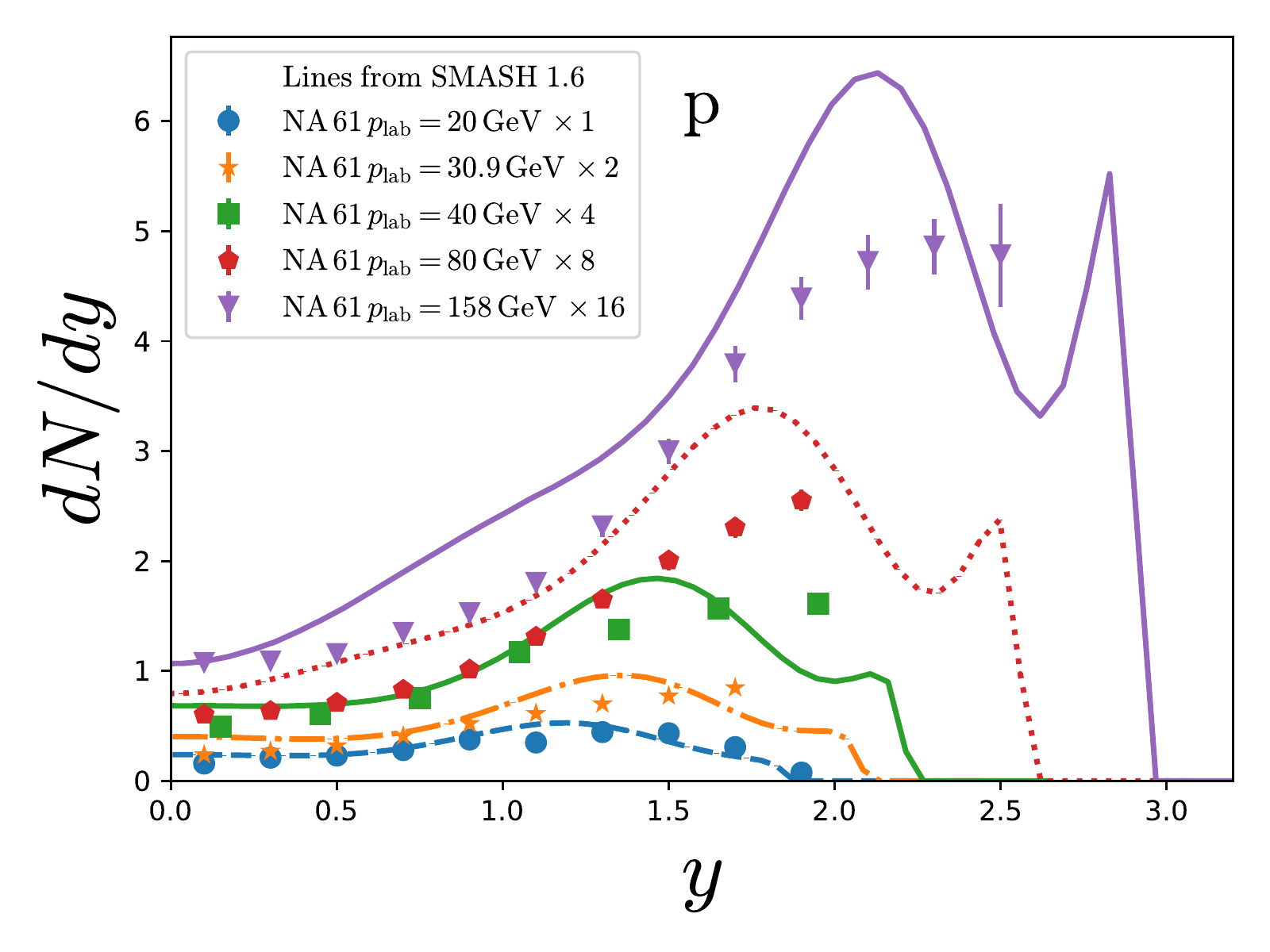}%
  \includegraphics[width=0.5\textwidth]{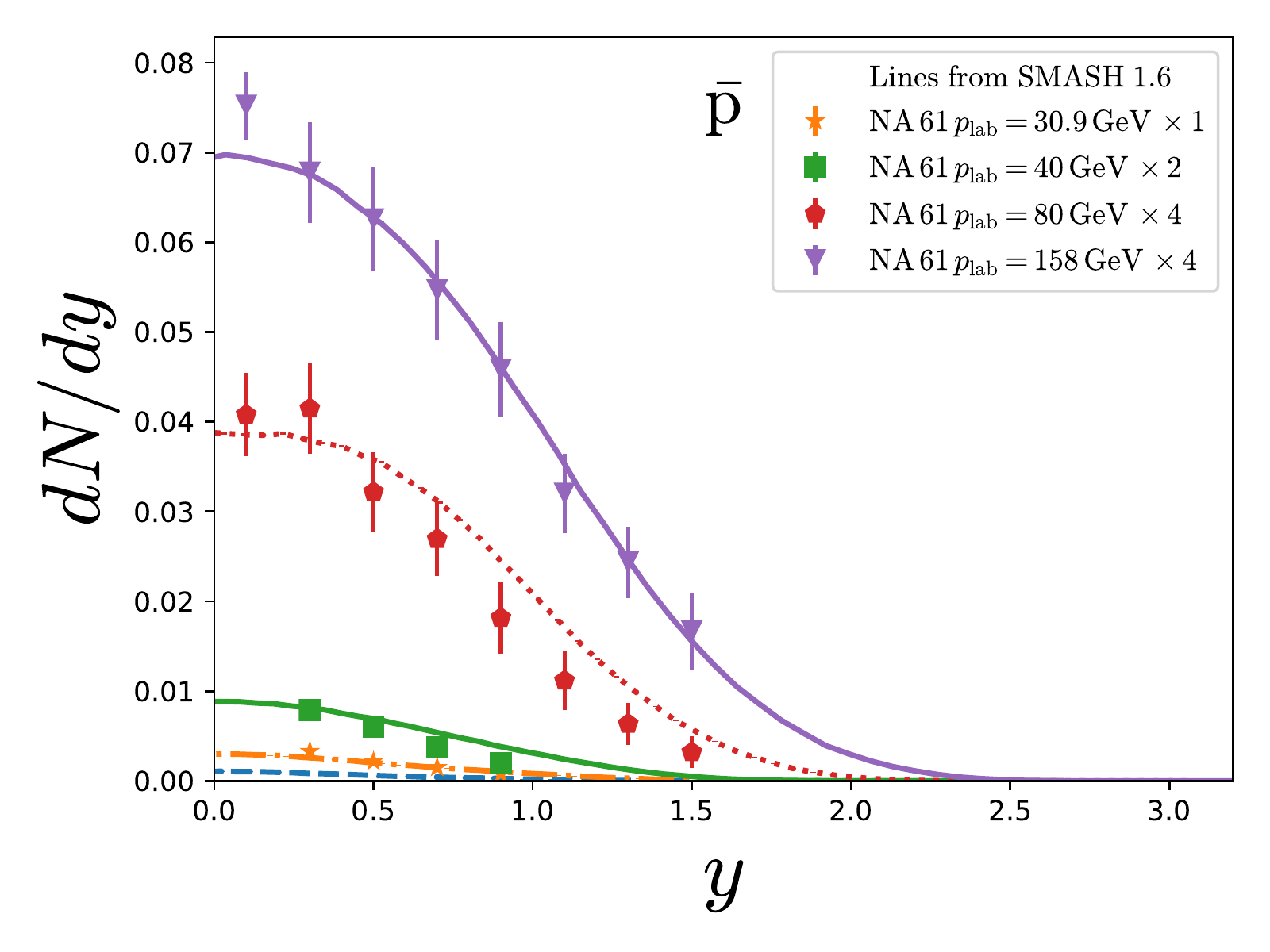}

  \includegraphics[width=0.5\textwidth]{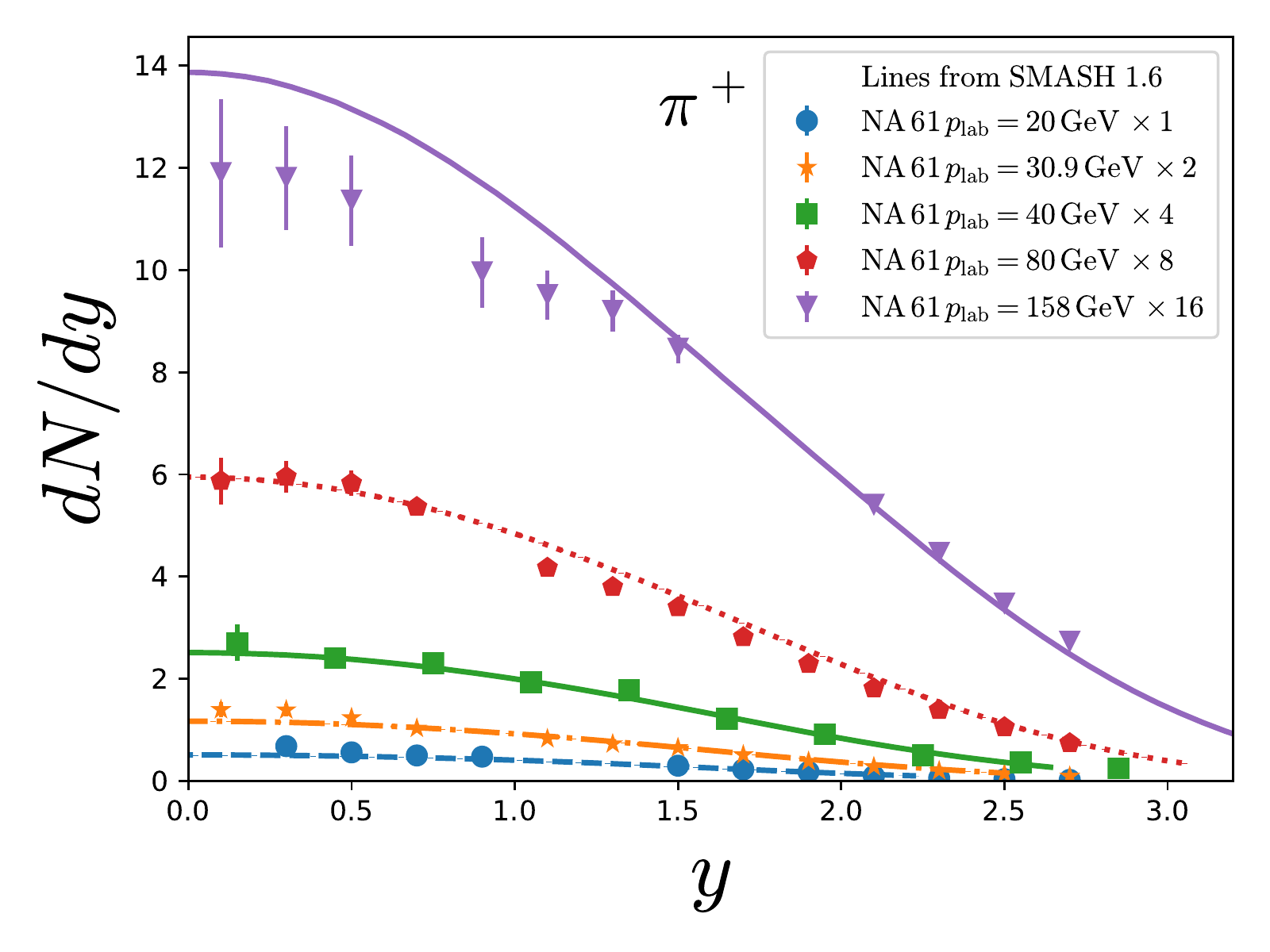}%
  \includegraphics[width=0.5\textwidth]{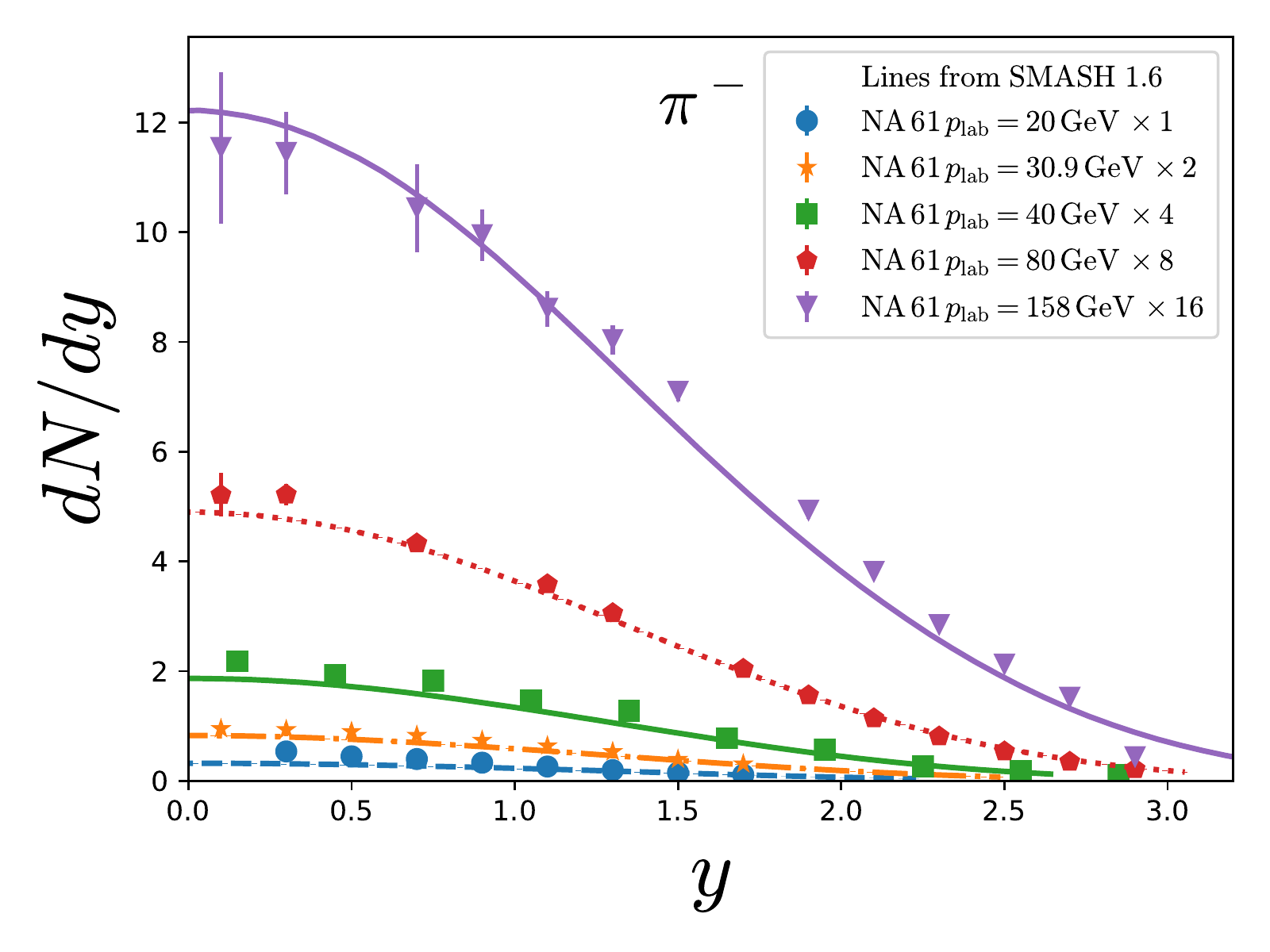}

  \includegraphics[width=0.5\textwidth]{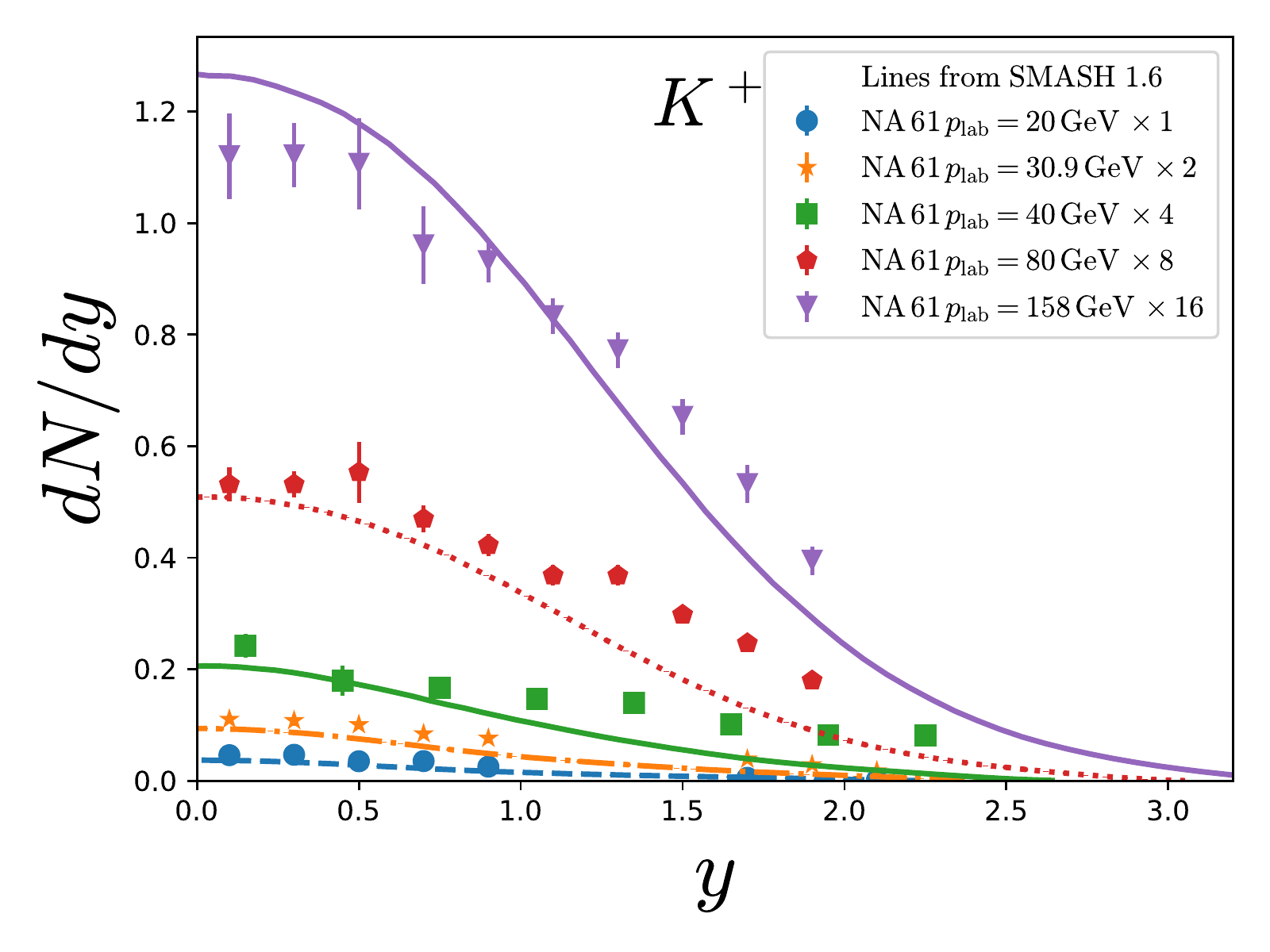}%
  \includegraphics[width=0.5\textwidth]{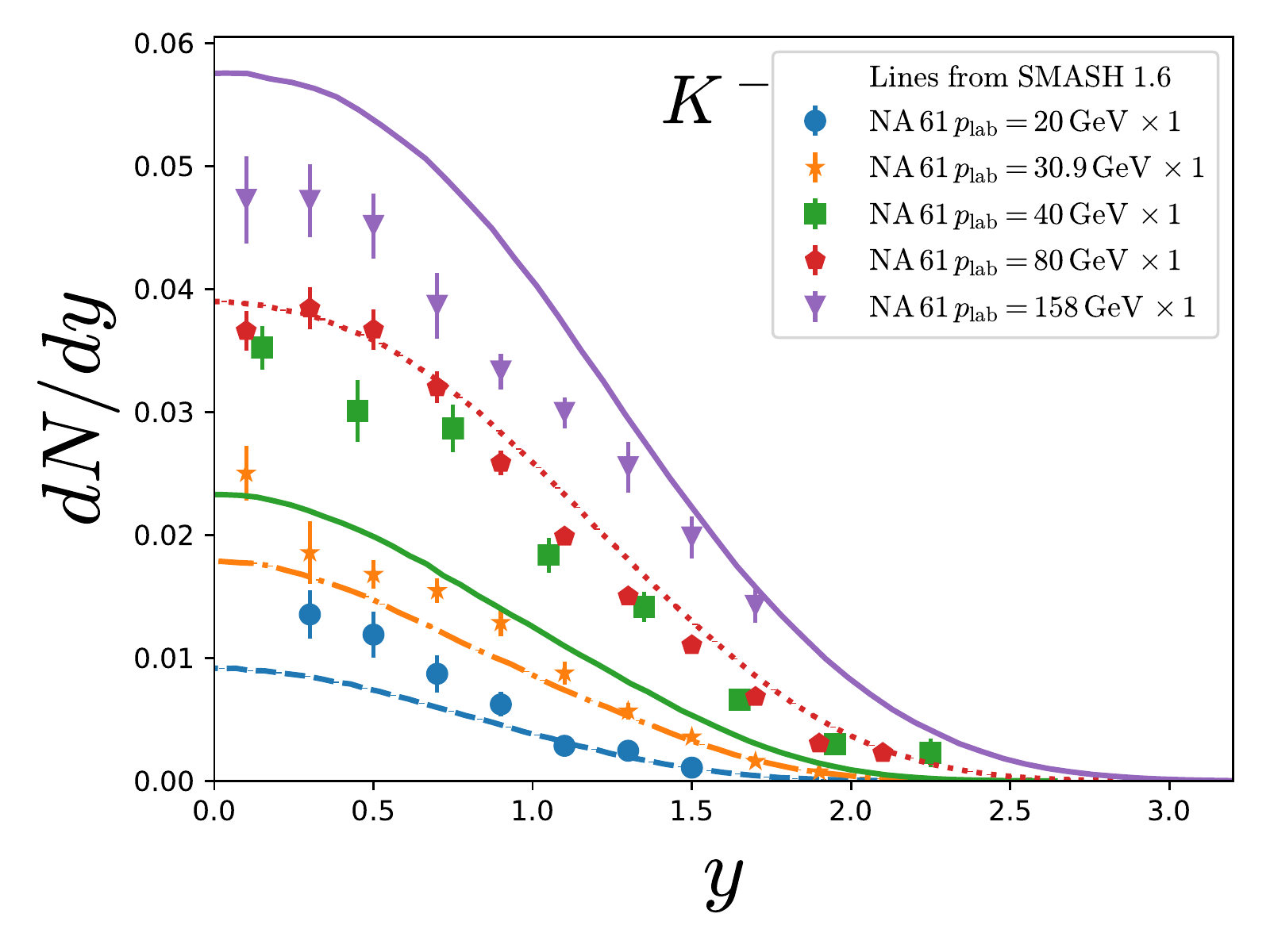}
  \caption{Rapidity spectra of (anti)protons, positively and negatively charged pions and kaons at different collision energies compared to experimental data \cite{Aduszkiewicz:2017sei}.}
  \label{fig_pp_summary_y}
\end{figure}

 The investigation of the free parameters in the previous sections resulted in the set of default parameters shown in table \ref{tab_defaults}.
 Using this set of parameters, we present the full set of final results for proton-proton collisions including mean transverse momentum as a function of $x_F$ and rapidity spectra for protons, antiprotons, positively and negatively charged pions and kaons in this section. This serves to benchmark the whole calculation in elementary collisions and provides the baseline for heavy ion calculations. 
Figure \ref{fig_pp_summary_y} shows the rapidity distribution for the mentioned particle species for different collision energies. 
A good agreement over the entire SPS energy range is achieved for pions, positively charged kaons and antiprotons.
The energy dependence of the negatively charged kaon multiplicity is too strong inside the string model compared to the measurement.
The rapidity spectrum is therefore only well reproduced at $p_\mathrm{lab}=80\,\mathrm{GeV}$, while the calculation overshoots at larger and undershoots at lower collision energies.
The proton rapidity spectrum follows the shape of the data roughly, but the longitudinal momentum of protons in proton-proton collisions in terms of $x_F$ proves to be quite challenging to describe within the string model as we expected regarding the difficulties to obtain a reasonable $x_F$ distribution. 

\begin{figure}[h]
	\centering
	\includegraphics[width=0.5\textwidth]{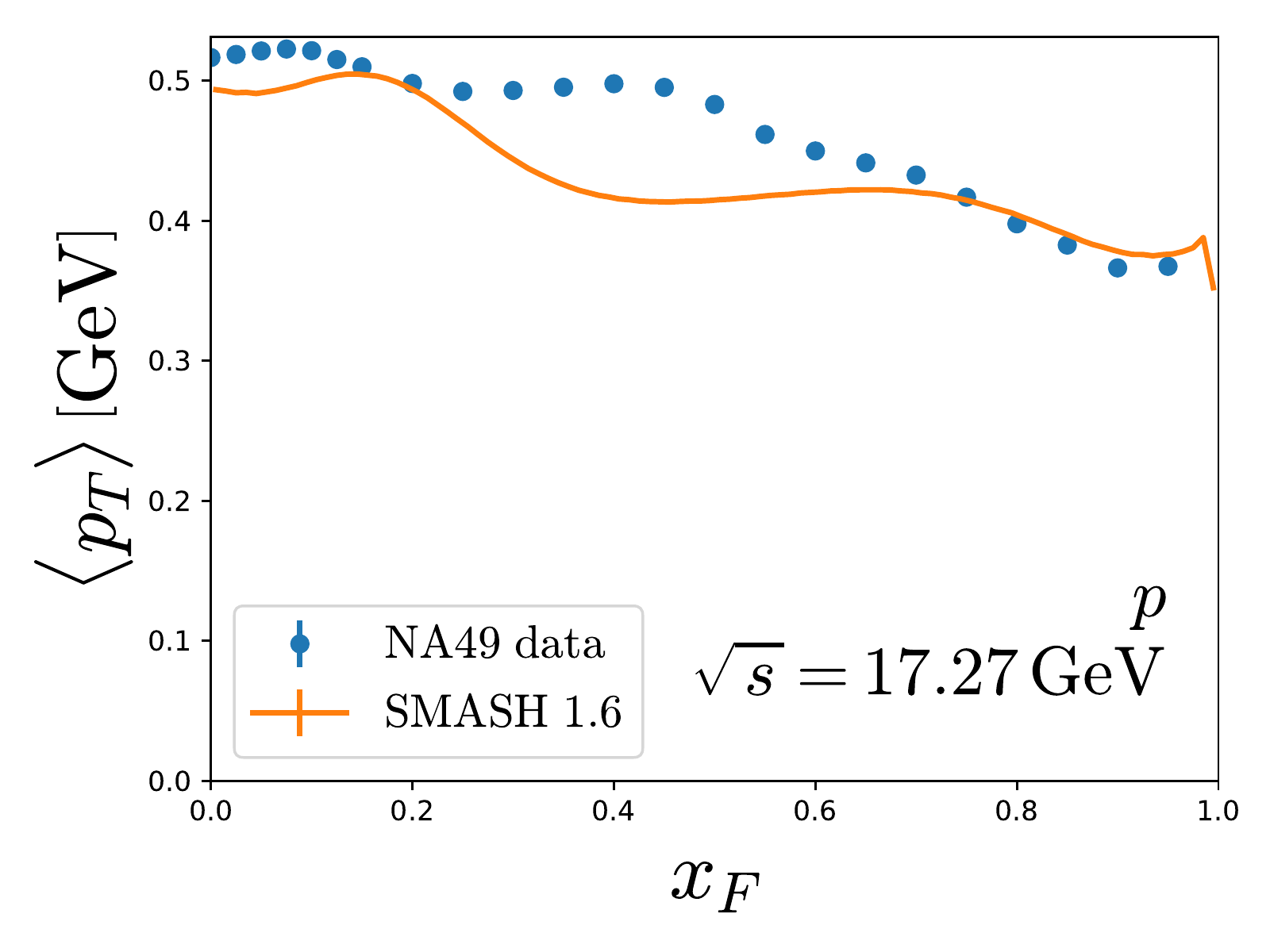}%
	\includegraphics[width=0.5\textwidth]{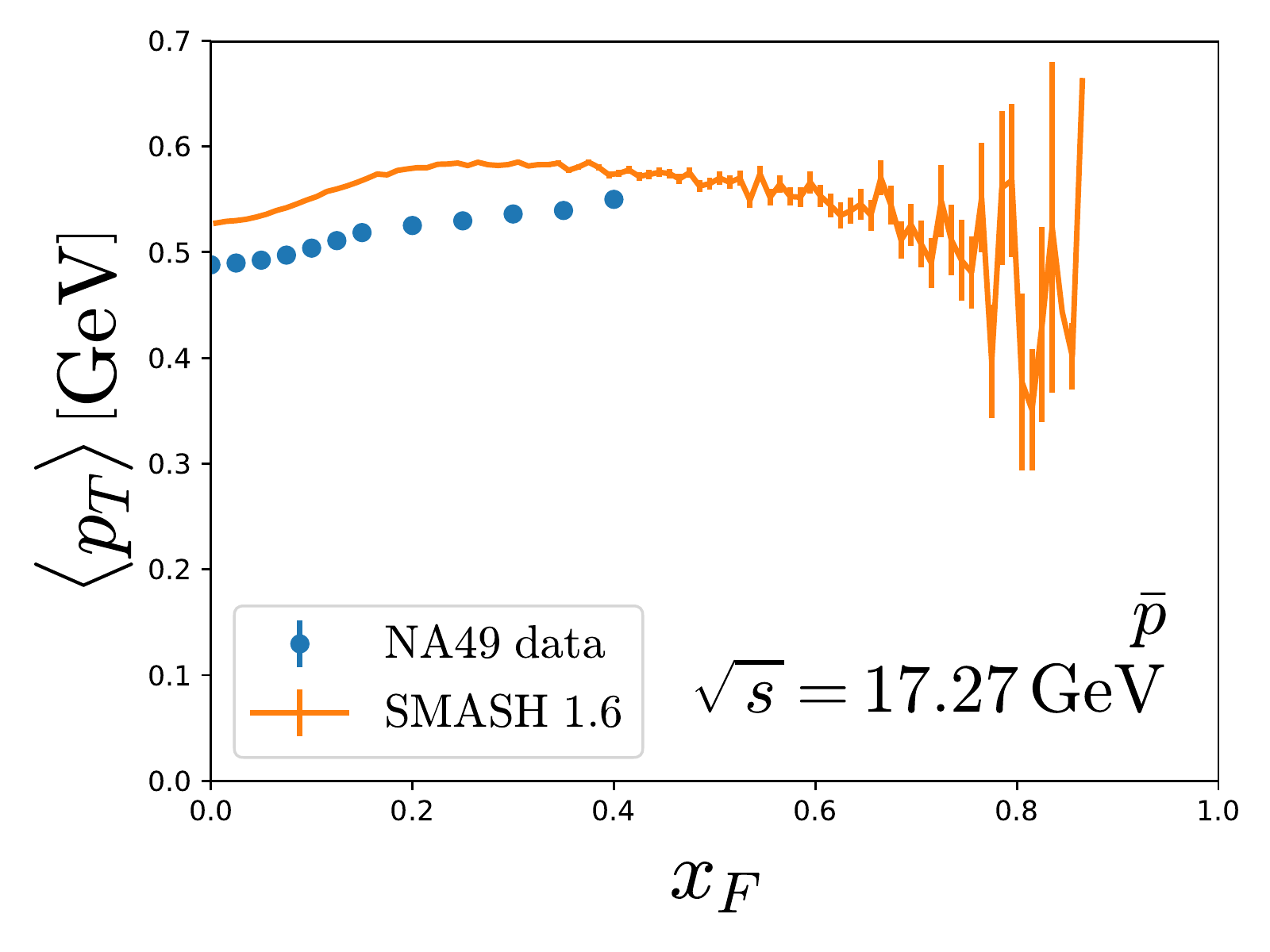}
	
	\includegraphics[width=0.5\textwidth]{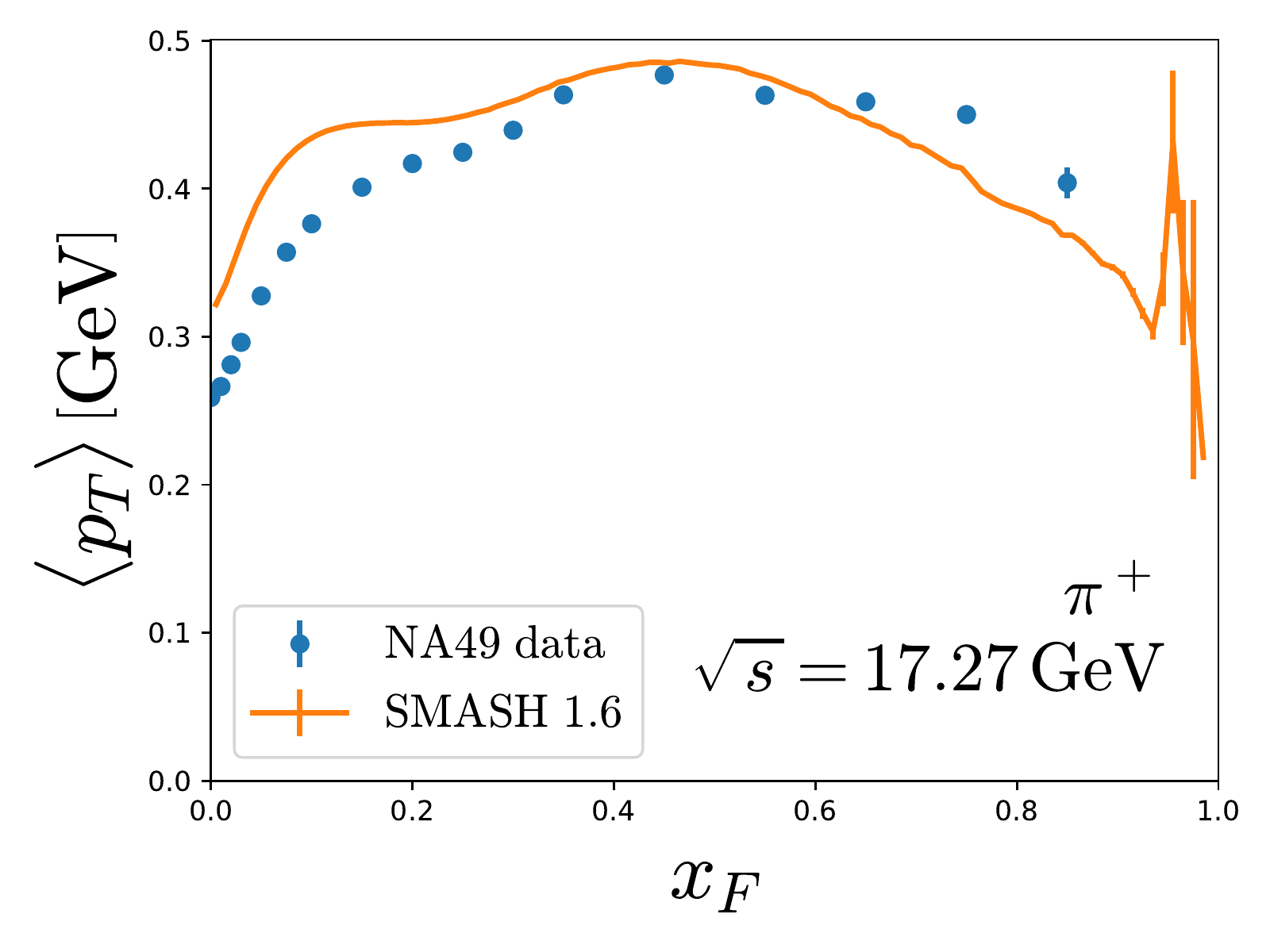}%
	\includegraphics[width=0.5\textwidth]{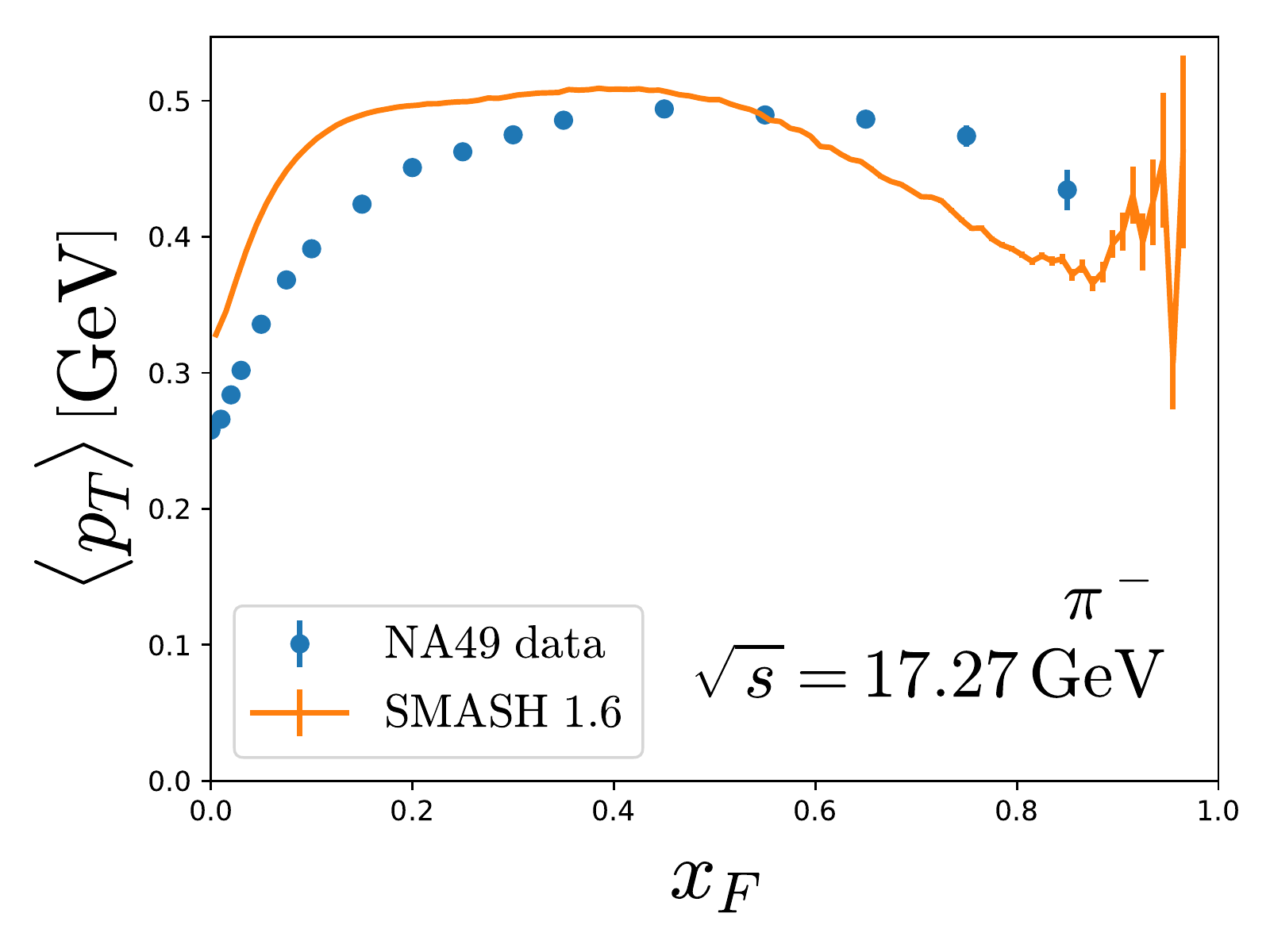}

	\includegraphics[width=0.5\textwidth]{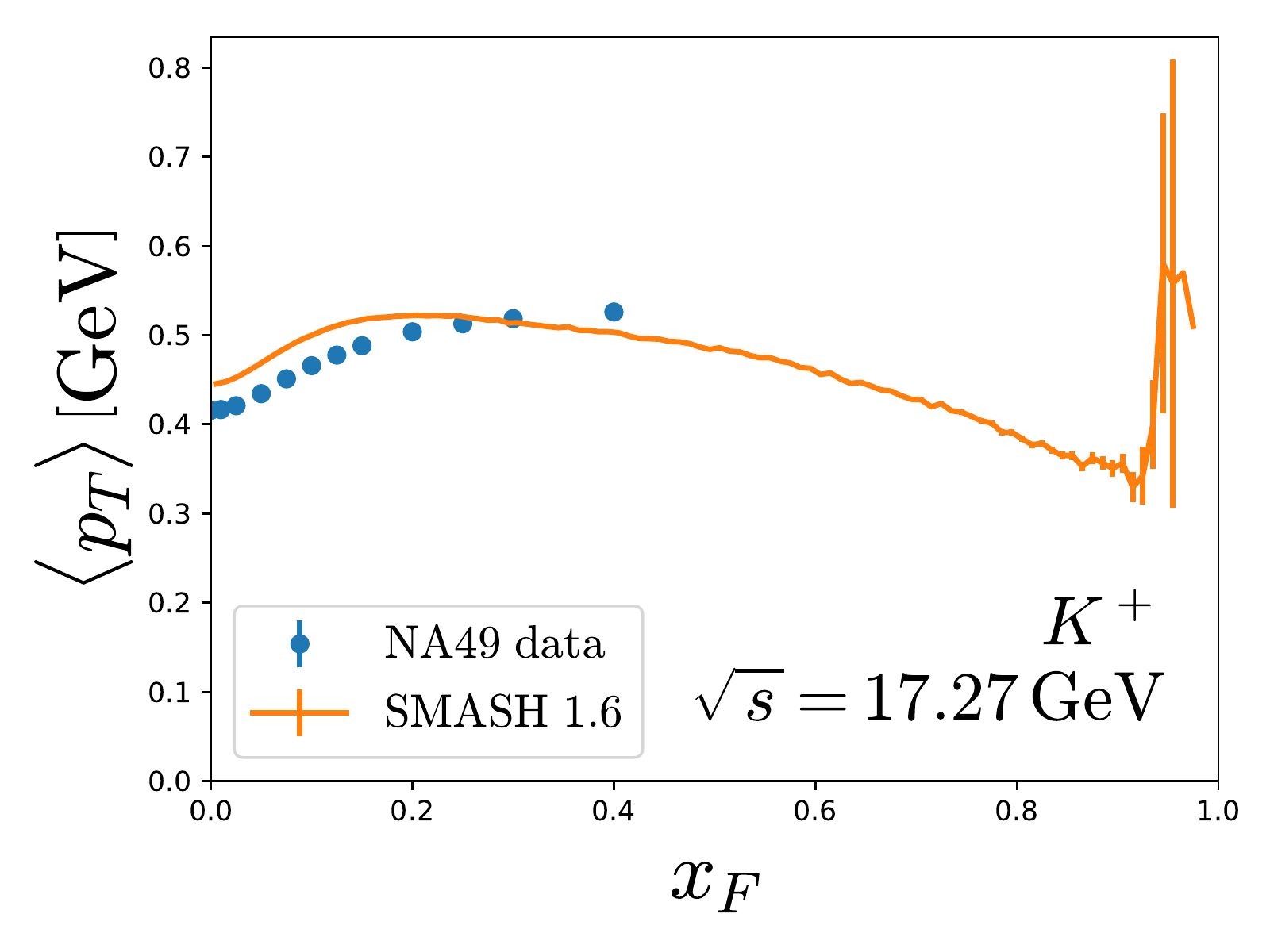}%
	\includegraphics[width=0.5\textwidth]{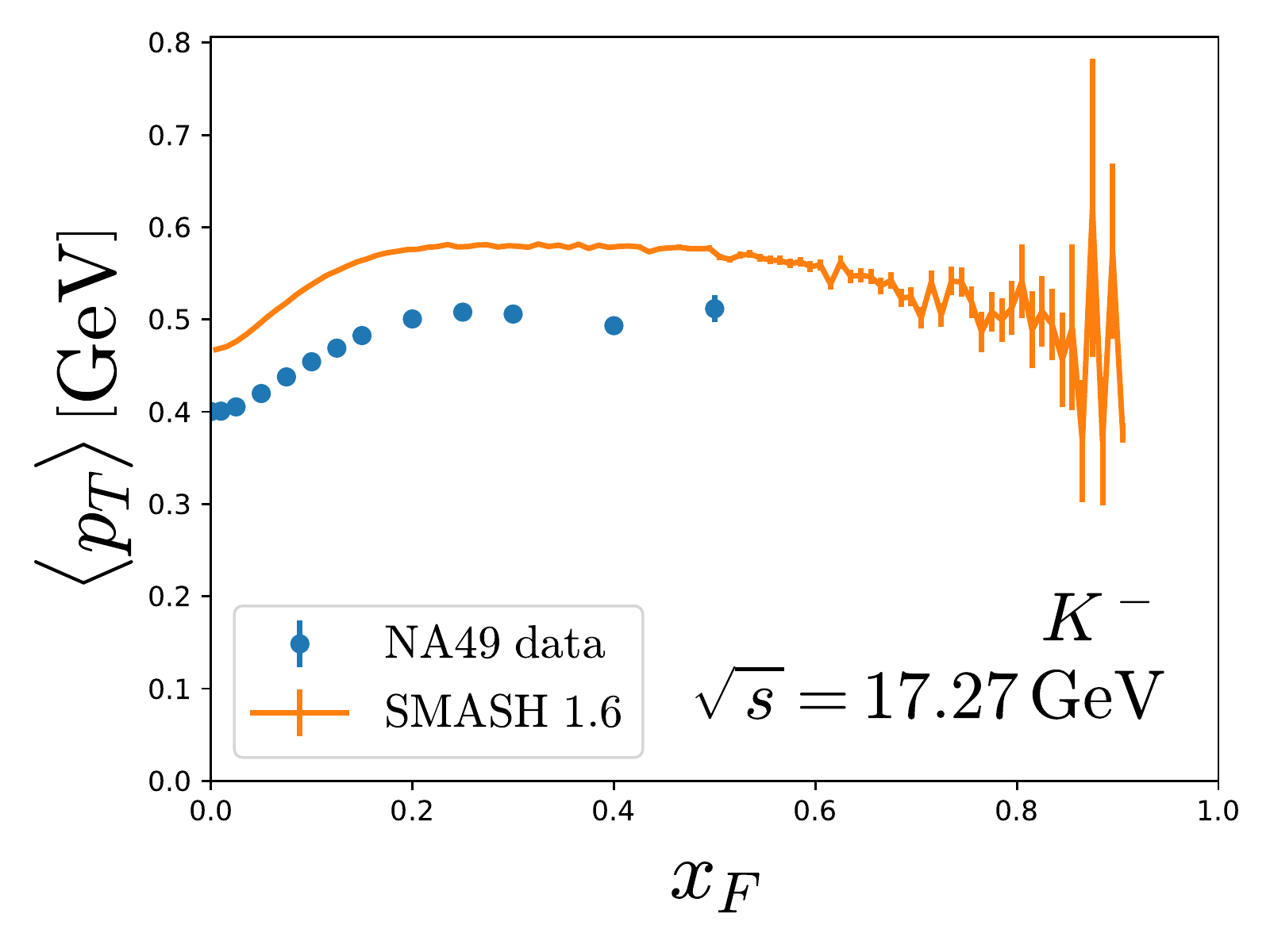}
	\caption{Mean transverse momentum of (anti)protons, positively and negatively charged pions and kaons in proton-proton collisions at $\sqrt{s}=17.27\,\mathrm{GeV}$ compared to experimental data \cite{NA49protons,NA49pions,NA49kaons}. }
	\label{fig_pp_summary_mpt}
\end{figure}

Figure \ref{fig_pp_summary_mpt} shows the mean transverse momentum of the different particle species as a function of $x_F$.
The mean transverse momentum of protons at low and intermediate $x_F$ is underestimated in the calculation while the data is reproduced at large $x_F$.
The opposite behavior is observed for the other particle species, where the mean transverse momentum at low $x_F$ is slightly overestimated while the pion mean transverse momentum undershoots the data at large $x_F$.
The mean transverse momentum for all particle species does not deviate much from the data so that overall a sufficient agreement with the measurement is found in proton-proton collisions for advancing to heavy ion collisions.

\section{Heavy Ion Collisions}
\label{sec_heavy_ion_collisions}
After adjusting the parameters for particle production in string processes to experimental observations in proton-proton collisions, we present calculations for heavy ion collisions.
Going from proton-proton collisions to heavy ion collisions in SMASH is only a change in the initialization.
Instead of colliding two protons, full lead nuclei are sampled from a Woods-Saxon distribution.
All particles are explicitly propagated throughout the simulation and hadronic interactions are determined based on the geometric interpretation of the cross section. All subsequent interactions of original and newly produced particles are treated in the microscopic non-equilibrium evolution \cite{Weil:2016zrk}.

The evolution of the shape of the proton rapidity spectrum from a single peak structure at low collision energies to a double peak structure at high energies is observed in central heavy ion collisions. Since a double peak structure corresponds to less stopping than a single peak structure, there is a trend from more stopping at low beam energies to more transparency at high beam energies.
In addition to understanding the net baryon content in the fireball, it is possible to gain insight on the formation process of string fragments.

Since the first collisions mainly take place via string excitation and fragmentation in the considered energy regime, let us first discuss the impact of the formation times and cross-section scaling factors on the results. During the fragmentation process, the particles are not immediately fully interacting hadrons but rather some pre-formed states that interact with lower cross-sections (see description in section \ref{sec_particle_formation}).

\begin{figure}[h]
	\centering
	\includegraphics[width=0.5\textwidth]{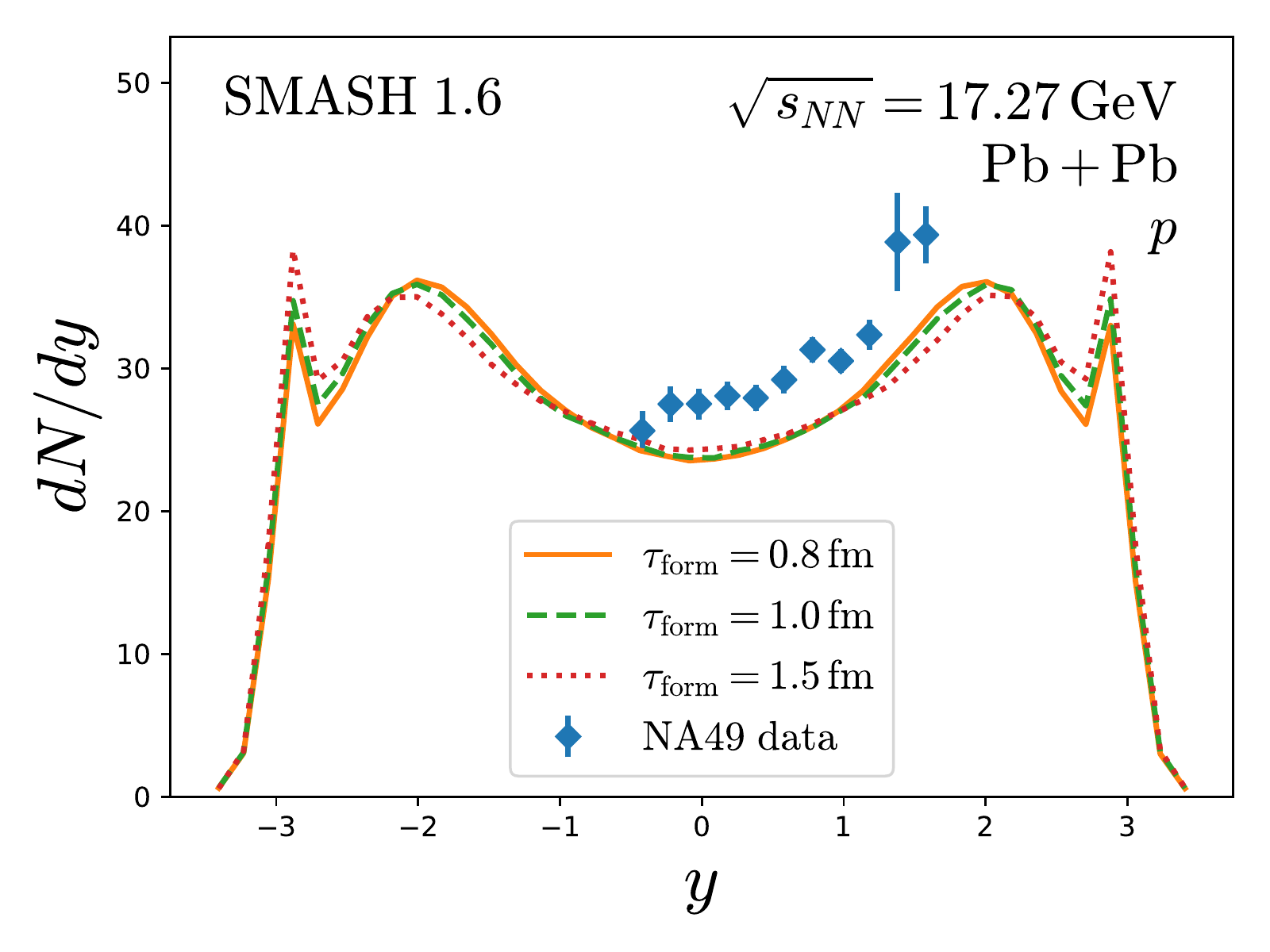}%
	\includegraphics[width=0.5\textwidth]{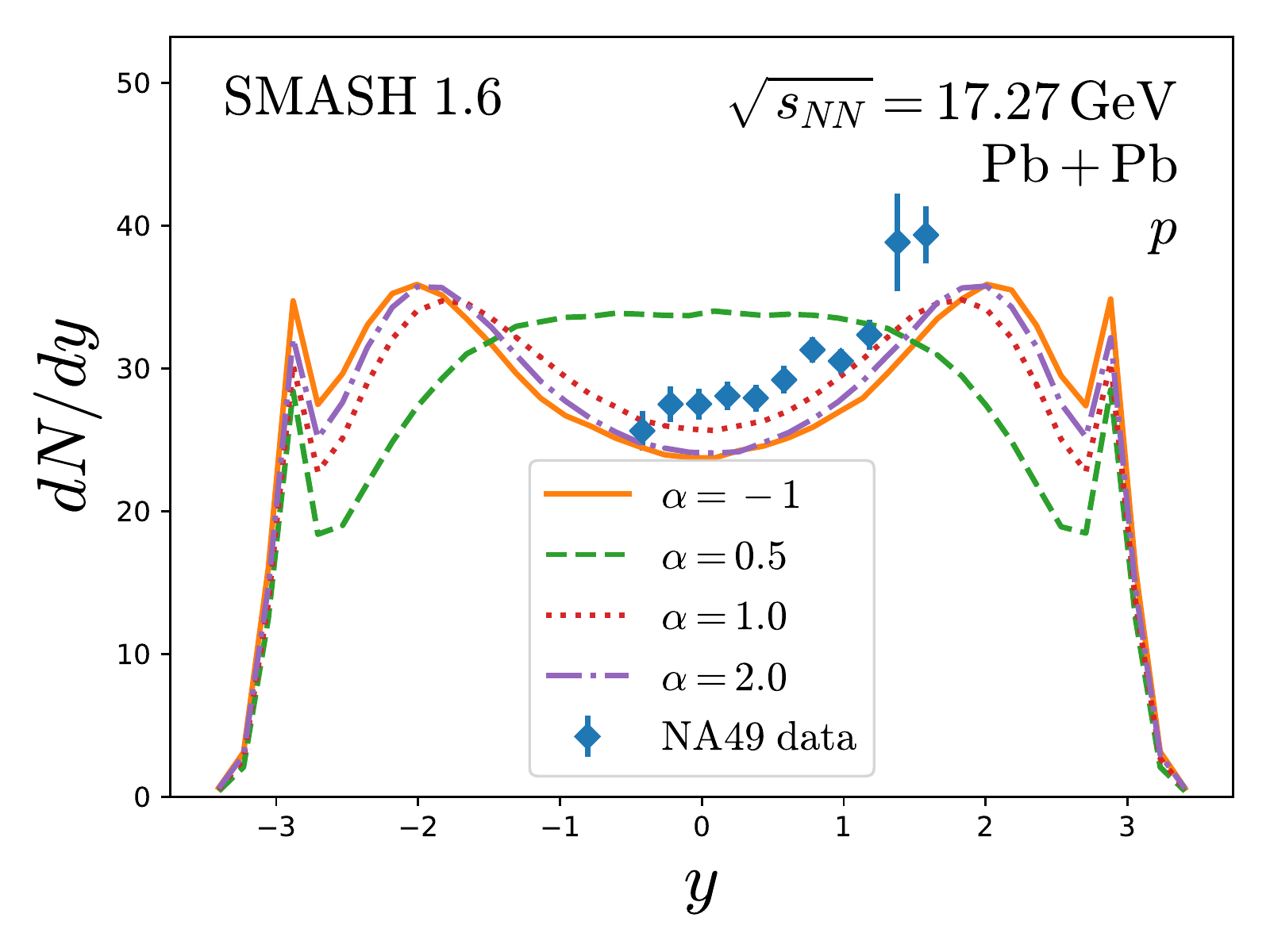}
	\caption{Rapidity spectrum of protons in central lead-lead collisions at $\sqrt{s_{NN}}=17.27\,\mathrm{GeV}$ for different formation proper times $\tau_\mathrm{form}$ (left) and for different powers $\alpha$ with which the cross section scaling factor $f_\sigma$ of string fragments grow (right) compared to experimental data \cite{NA49hicprotons}.
	Left: The cross section scaling factor $f_\sigma$ is set to be constant until it jumps to 1 at the formation time. Right: The formation proper time is set to $\tau_\mathrm{form}=1\,\mathrm{fm}$ in all calculations. The value of $\alpha=-1$ encodes a step function rather than a continuous formation in time.}
	\label{fig_hic_17_p_time}
\end{figure}

The influence of the formation time on the particle spectra is first studied in the simplest case, where the cross section scaling factor $f_\sigma$, as introduced in section \ref{sec_particle_formation}, is a step function in time and does not increase continuously.
For most string fragments, this implies that they instantly form, when the formation time has expired.
Figure \ref{fig_hic_17_p_time} (left) shows the rapidity spectrum of protons in central lead-lead collisions for different values of the formation time $\tau_\mathrm{form}$.

While all three calculations reproduce the shape of the measured rapidity spectrum, all curves fail to describe the number of stopped protons at mid-rapidity.
This reflects the fact that the formation times are too large for the string fragments to form while the nuclei still overlap.
If the cross sections continuously grow with time, there is a small probability for string fragments to immediately interact.
As shown in figure \ref{fig_hic_17_p_time} (right), this enhances the amount of stopped protons significantly. Figure \ref{fig_hic_17_p_time} (right) shows the calculation for a fixed formation time of $\tau_\mathrm{form}=1\,\mathrm{fm}$ for different powers in which the cross section scaling factor grows in time.
Using a step function ($\alpha=-1$) gives similar results as the quadratic increase.
When the cross section scaling factor grows with the square root in time, the string fragments interact too much at early times and the protons are stopped far too much.
Only for the linearly growing cross section scaling factor, the amount of stopping can be reproduced.

\begin{figure}
	\centering
	\includegraphics[width=0.6\textwidth]{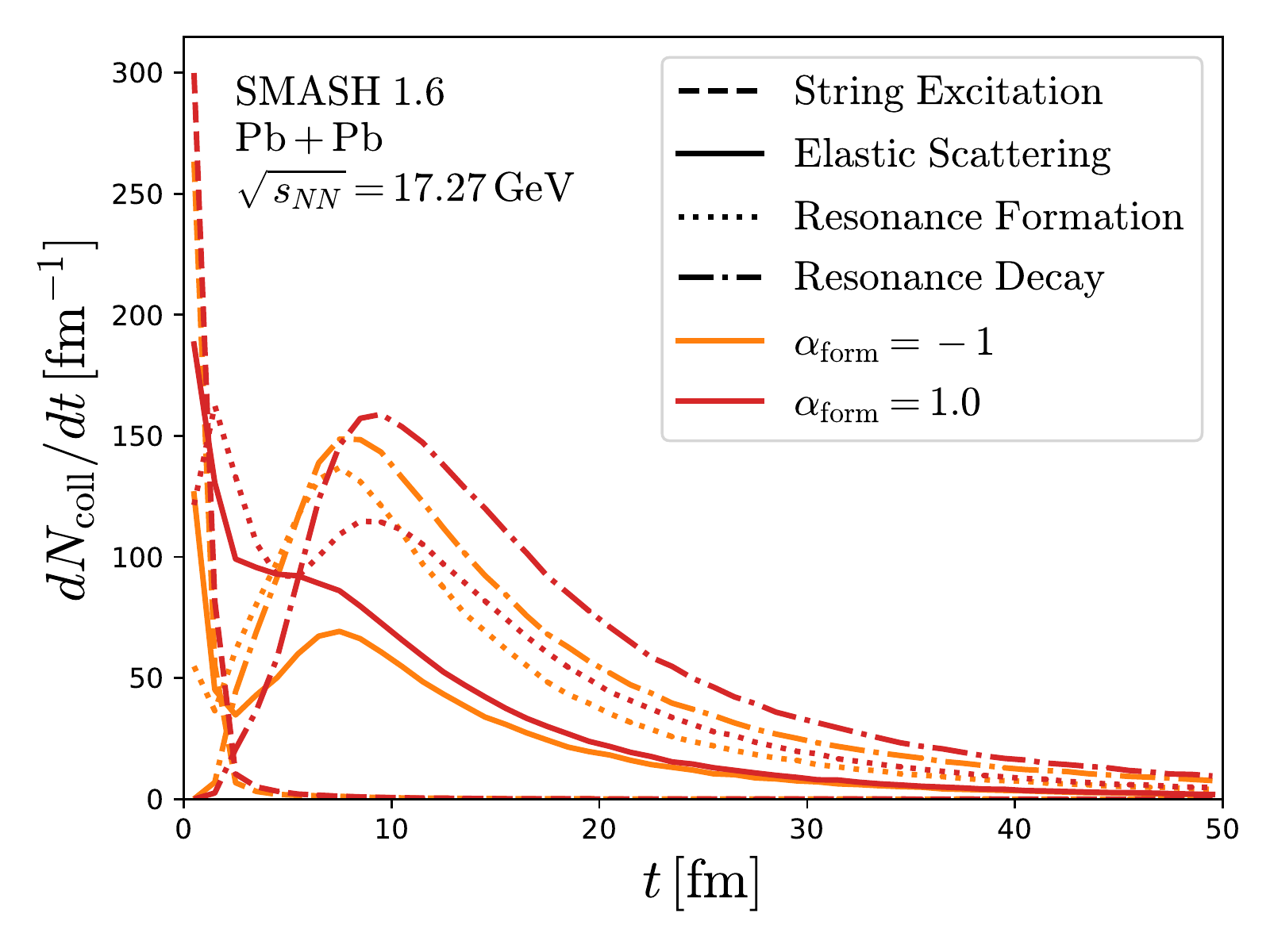}
	\caption{Rate of interactions in central lead-lead collisions at $\sqrt{s_{NN}}=17.27\,\mathrm{GeV}$ for different powers $\alpha$, in which the cross section scaling factor grows, as a function of time.
	Shown are the different contributions from string processes, elastic scatterings, resonance formations and resonance decays. The formation time in both calculations is set to $\tau_\mathrm{form} = 1\,\mathrm{fm}$.}
	\label{fig_collision_rate}
\end{figure}

A deeper understanding about how the power $\alpha$ translates to more stopping can be gained by studying the interaction rate as a function of time for the different scenarios.
Figure \ref{fig_collision_rate} shows the rate of different interactions as a function of time.

The rates are compared between a calculation with linearly growing cross section scaling factor ($\alpha=1$) and a cross section scaling factor, that does not continuously grow in time ($\alpha=-1$).
The interaction rate at early times is dominated by string processes. At later times, the energy is not sufficiently large to excite strings and the strongest contribution to the interaction rate stems from resonance formations and decays.
For the stopping, the most interesting period is right after the initial collisions.
There one can see that if the cross section immediately starts growing, the rate of elastic collisions and resonance formations is significantly increased.
These additional interactions are responsible for the higher proton yield at mid-rapidity.

\begin{figure}[h]
	\centering
	\includegraphics[width=0.5\textwidth]{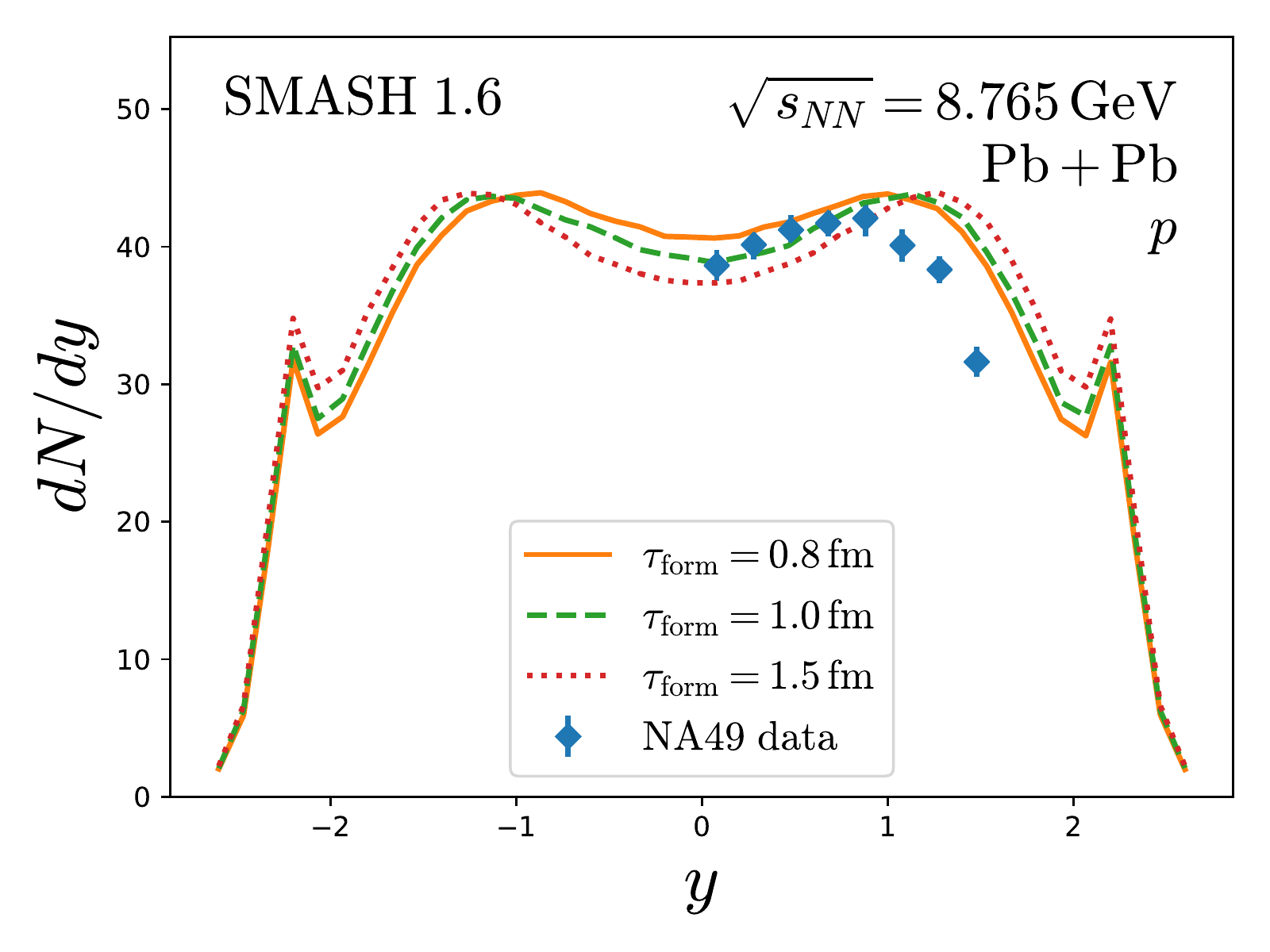}%
	\includegraphics[width=0.5\textwidth]{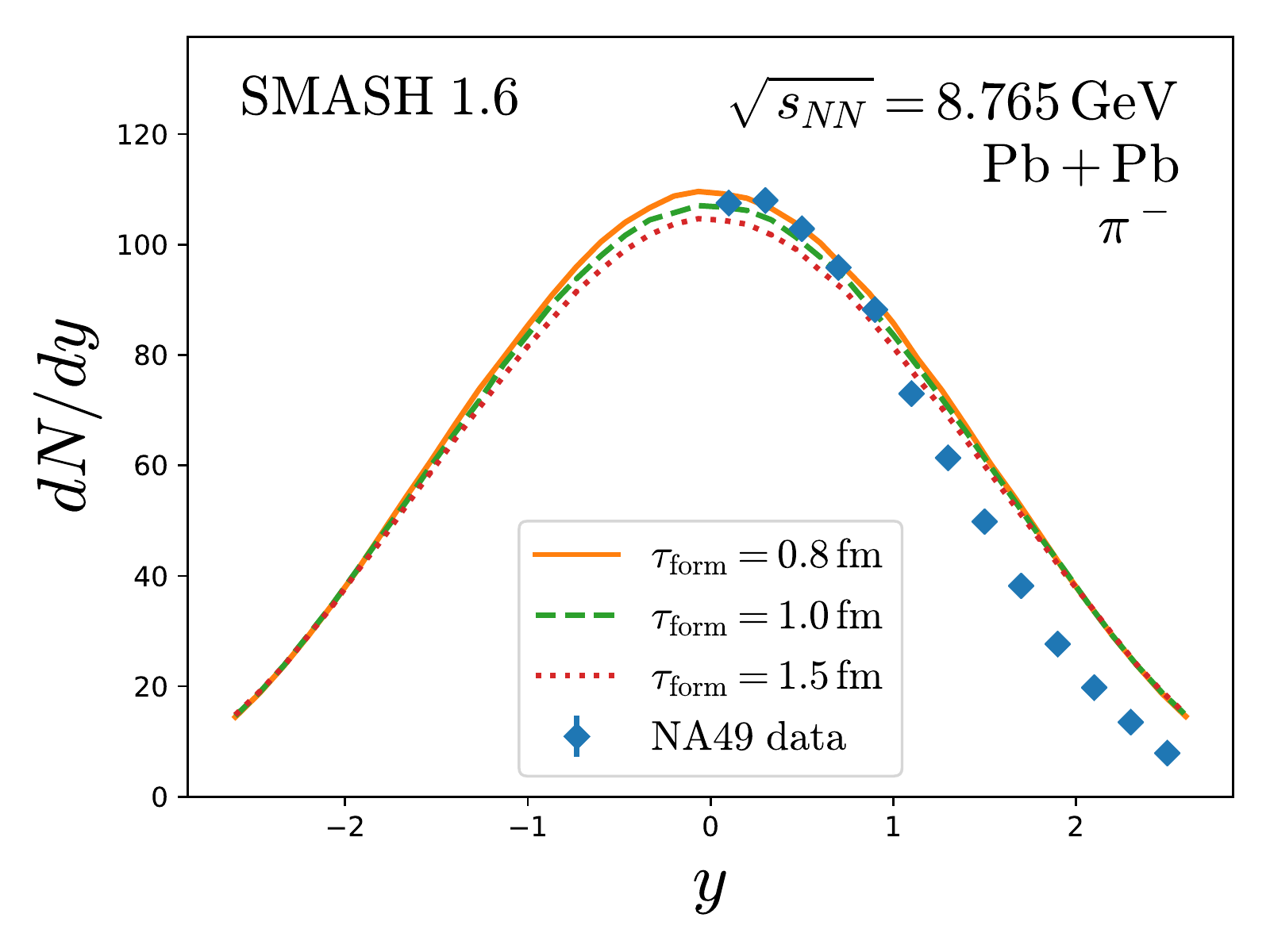}
	\caption{Rapidity spectrum of protons (left) and negatively charged pions (right) in central lead-lead collisions at $\sqrt{s_{NN}}=8.765\,\mathrm{GeV}$ for different formation times compared to experimental data \cite{NA49hicprotons, Afanasiev:2002mx}. 
	The cross section scaling factor grows linearly in time within the calculations.}
	\label{fig_hic_8_p_time}
\end{figure}

At $\sqrt{s_{NN}} = 17.27\,\mathrm{GeV}$, the formation time affects the results only slightly even when the cross sections grow linearly in time.
Going to lower collision energies, changing the formation time directly reflects in the rapidity spectra as shown in figure \ref{fig_hic_8_p_time}.
In the case of slower, and less Lorentz contracted, nuclei the passing time is on the order or larger than the formation time. 
With shorter formation times, the cross section scaling factors grow faster in time which leads to more stopping of protons and larger pion multiplicities at mid-rapidity.
A formation time of $\tau=1\,\mathrm{fm}$ is best suited for reproducing the proton and pion rapidity spectrum.

\begin{figure}[h]
	\centering
	\includegraphics[width=0.65\textwidth]{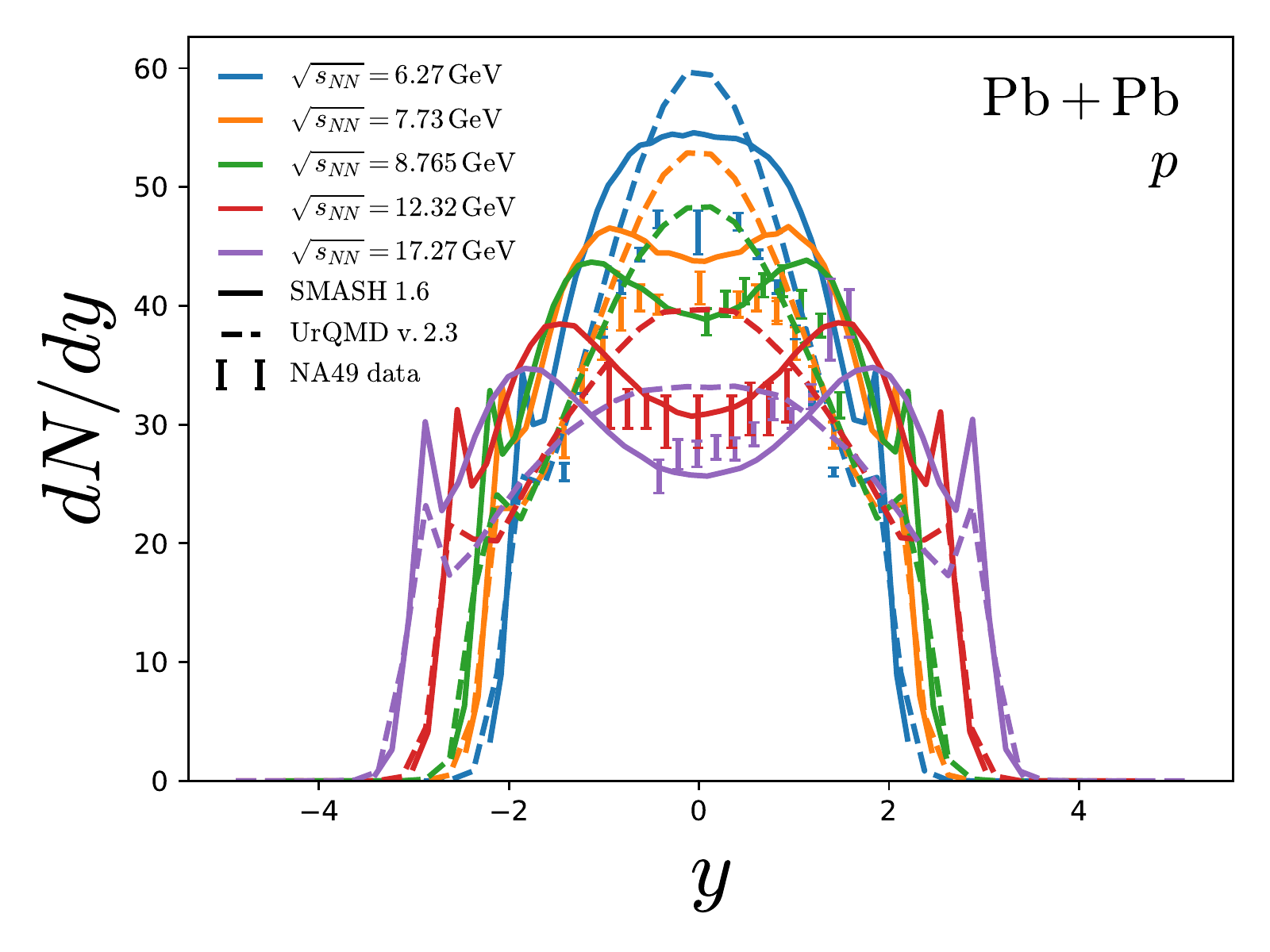}
	\caption{Rapidity spectra of protons in central lead-lead collisions at different beam energies compared to experimental data \cite{NA49hicprotons, Blume:2007kw} and UrQMD calculations \cite{Petersen:2008kb}.}
	\label{fig_hic_p_final}
\end{figure}

Figure \ref{fig_hic_p_final} shows the rapidity spectrum of protons in central heavy ion collisions compared to experimental data and UrQMD calculations for different collision energies for the final set of parameters for particle formation ($\tau=1\,\mathrm{fm}$ and $\alpha=1$).
Over the entire SPS energy range, a good agreement between the SMASH calculation and the experimental data is found.
Even though the proton multiplicity at low SPS energies is overestimated in the SMASH results, the evolution of the shape of the proton rapidity spectrum from a single peak at low energies to a double peak structure at large collision energies is well reproduced.
At low beam energies, a small fraction of protons is bound in light nuclei and should not be counted in the proton spectra. 
The clustering is not taken into account in the shown SMASH results, which might be part of the reason for the overshoot of protons at low beam energies.
Comparing figure \ref{fig_hic_p_final} to the $x_F$ distributions shown in section \ref{sec_proton_proton_collisions} might give the impression that the agreement with data in heavy ion collisions is a lot better than in proton-proton collisions, but the rapidity spectrum in proton-proton collsions in figure \ref{fig_pp_summary_y} shows a similar agreement as in heavy ion collisions.

To put our results into context, we compare to UrQMD calculations, where a very similar treatment of string processes is applied. In general, the protons within the UrQMD calculation are stopped more at mid-rapidity than the protons in the SMASH calculation. In UrQMD, the cross section of an unformed particle is kept constant until the formation time of that particle is reached. This corresponds to the SMASH calculation with $\alpha=-1$ shown in figure \ref{fig_hic_17_p_time} (right), where the least stopping is observed in the case of $\alpha=-1$, so the details of the formation process of string fragments is not the main source of the difference between SMASH and UrQMD.

\begin{figure}[h]
	\centering
	\includegraphics[width=0.65\textwidth]{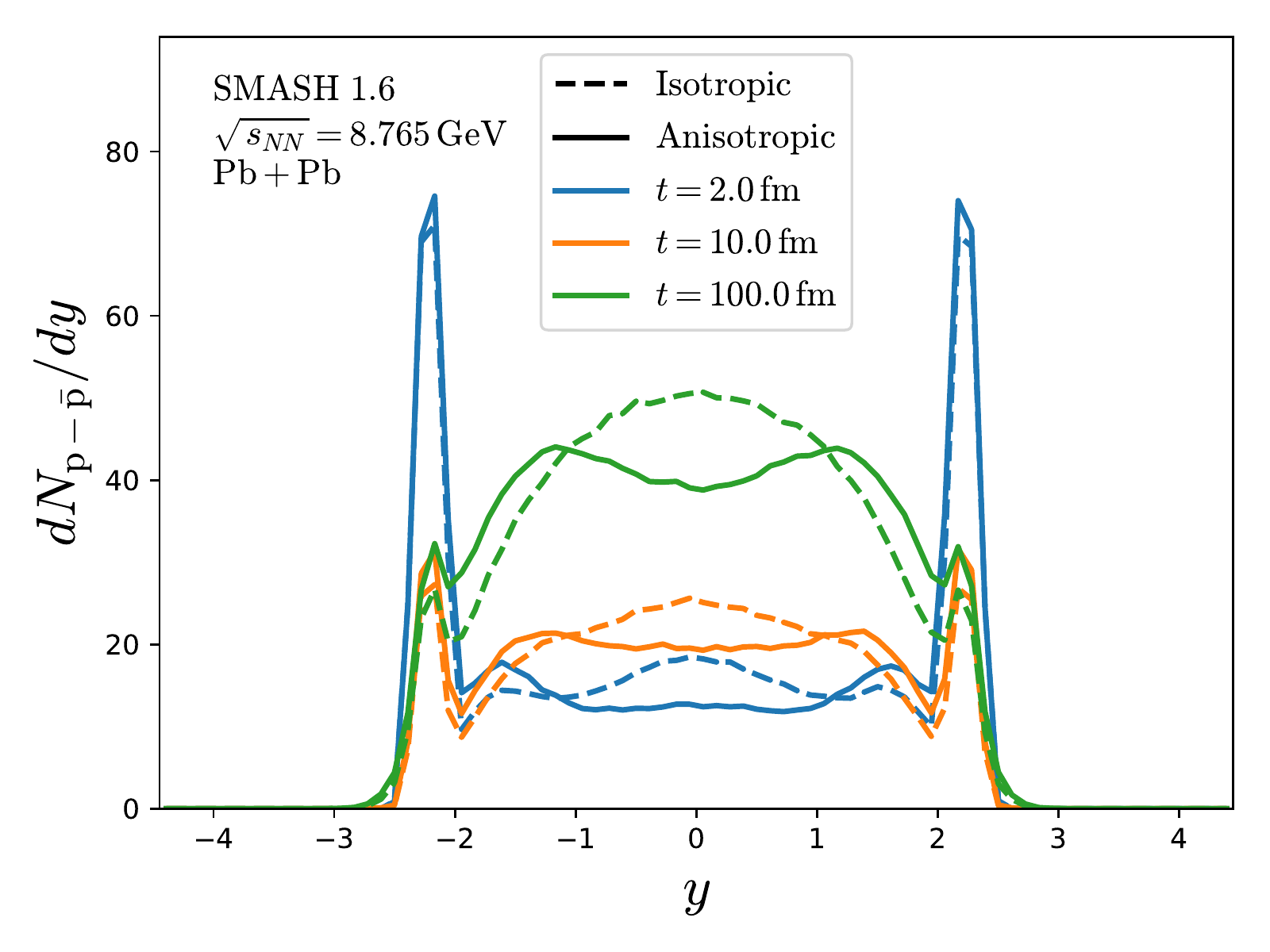}
	\caption{Rapidity spectrum of net-protons in central lead-lead collisions at $\sqrt{s_{NN}}=8.765\,\mathrm{GeV}$ at different times.
	The full lines correspond to calculations with anisotropic angular distributions for the elastic scattering, while elastic collisions in the calculations for the dashed lines are isotropic.}
	\label{fig_hic_isotropic}
\end{figure}

An important ingredient for understanding the shape of the proton rapidity spectrum are the anisotropic angular distributions for elastic collisions between all hadrons as described in section \ref{sec_elastic_collisions}.
Figure \ref{fig_hic_isotropic} shows the rapidity spectrum of net-protons in a calculation of lead-lead collisions at $\sqrt{s_{NN}} = 8.765\, \mathrm{GeV}$ at different times comparing isotropic and anisotropic angular distributions in elastic collisions.

At $t=2\,\mathrm{fm}$, the nuclei are still in the process of passing through each other, explaining why a large portion of protons are still located at beam rapidity.
Up to this point, the dynamics of the system are dominated by primary interactions between nucleons.
A significant difference is already at that time observed between the calculations with isotropic and anisotropic elastic scatterings.
Advancing in time, the net-proton number increases, mostly due to resonance decays.
The difference between the calculations with isotropic and anisotropic elastic collisions is not washed out during the evolution of the system but can be observed even after all resonances have decayed.
A double peak structure only builds up when the anisotropy of elastic scatterings is properly taken into account.

\begin{figure}[h]
	\centering
	\includegraphics[width=0.65\textwidth]{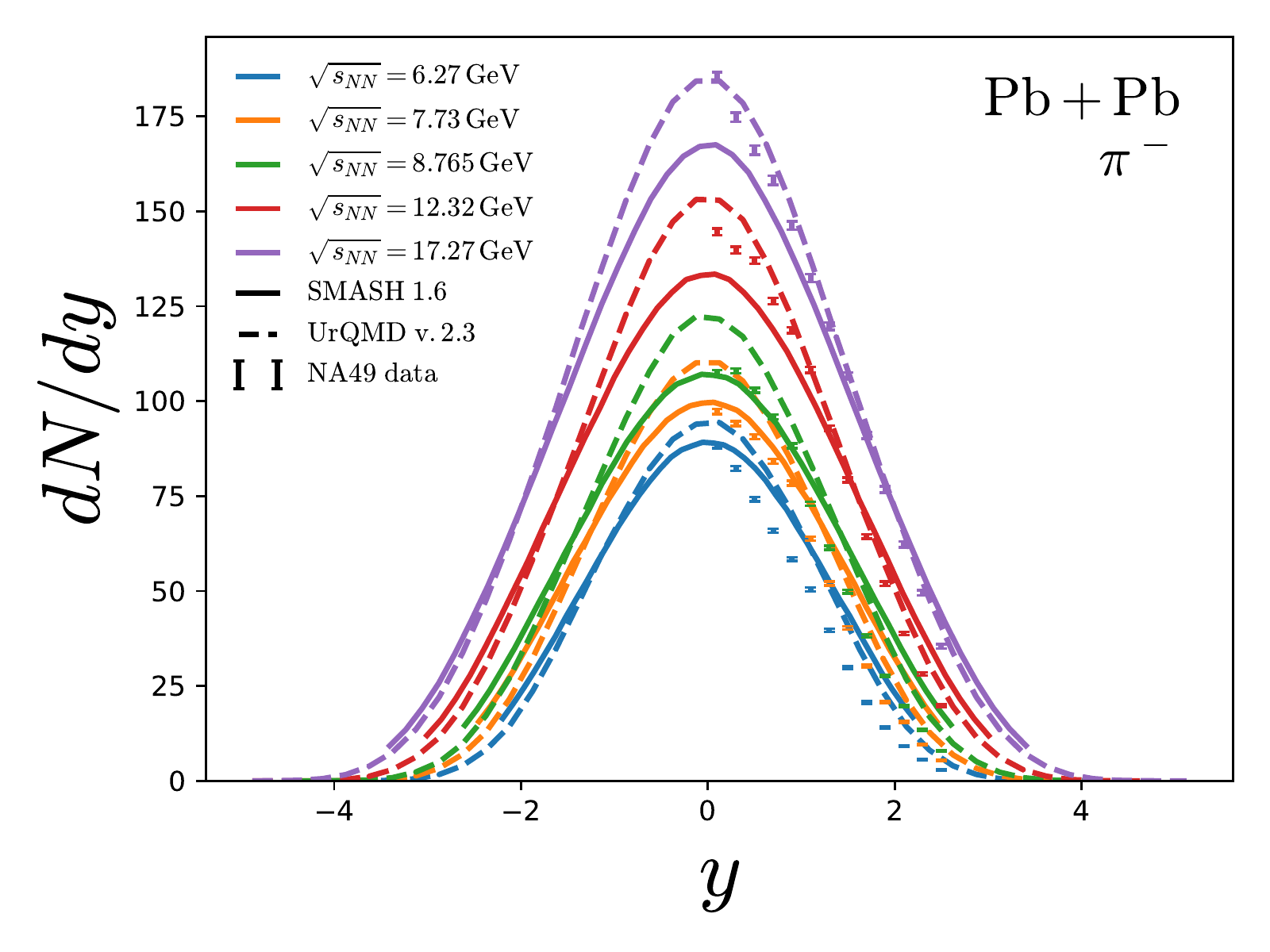}
	\caption{Rapidity spectrum of negatively charged pions in central lead-lead collisions at different beam energies compared to experimental data \cite{Afanasiev:2002mx,Alt:2007aa} and UrQMD calculations \cite{Petersen:2008kb}.}
	\label{fig_hic_pi_minus_final}
\end{figure}

To conclude the study of particle production in heavy ion collisions, figure \ref{fig_hic_pi_minus_final} shows the rapidity spectrum of negatively charged pions for different collision energies.
As shown in figure \ref{fig_hic_8_p_time} (right), the pion production is relatively well understood at intermediate SPS energies.
Similar to the intermediate energies, a good agreement with the data can be observed at the lowest collision energies in the SPS range.
At top SPS energies, the multiplicity of negatively charged pions is underestimated but still a reasonable agreement is found.
In comparison to the SMASH results, pions are more abundantly produced in the UrQMD calculations.
Compared to the data, UrQMD describes the pion production very well at high energies while SMASH gives a better description for low collision energies.

\section{Initial State Calculations}
\label{sec_initial_state}

In this last Section, we would like to show how the results from our approach can in the future be employed for initial conditions for hydroydnamic calculations. This has been very successfully done in a hybrid approach based on UrQMD initial conditions\cite{Petersen:2008dd,Karpenko:2013ama}.
More recently, a toy model for initial conditions including a 3D Glauber model with energy loss according to a string picture is developed \cite{Shen:2017bsr}.
Further, dynamical initial states are constructed based on UrQMD \cite{Du:2018mpf}.
In a similar fashion, the hadronic transport approach JAM is combined with relativistic viscous hydrodynamics \cite{Murase:2019gro}.

\begin{figure}[h]
	\centering
	\includegraphics[width=0.5\textwidth]{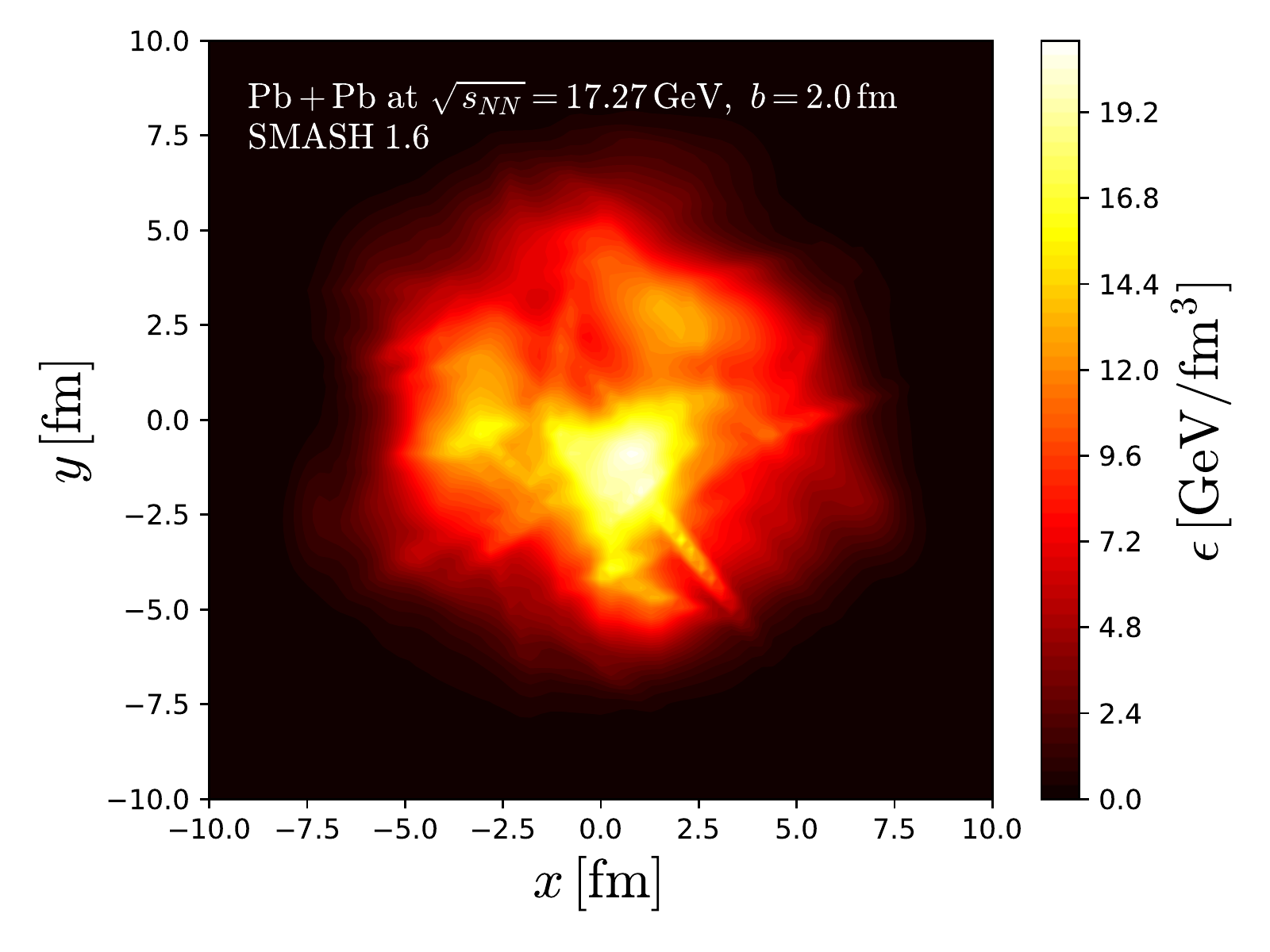}%
	\includegraphics[width=0.5\textwidth]{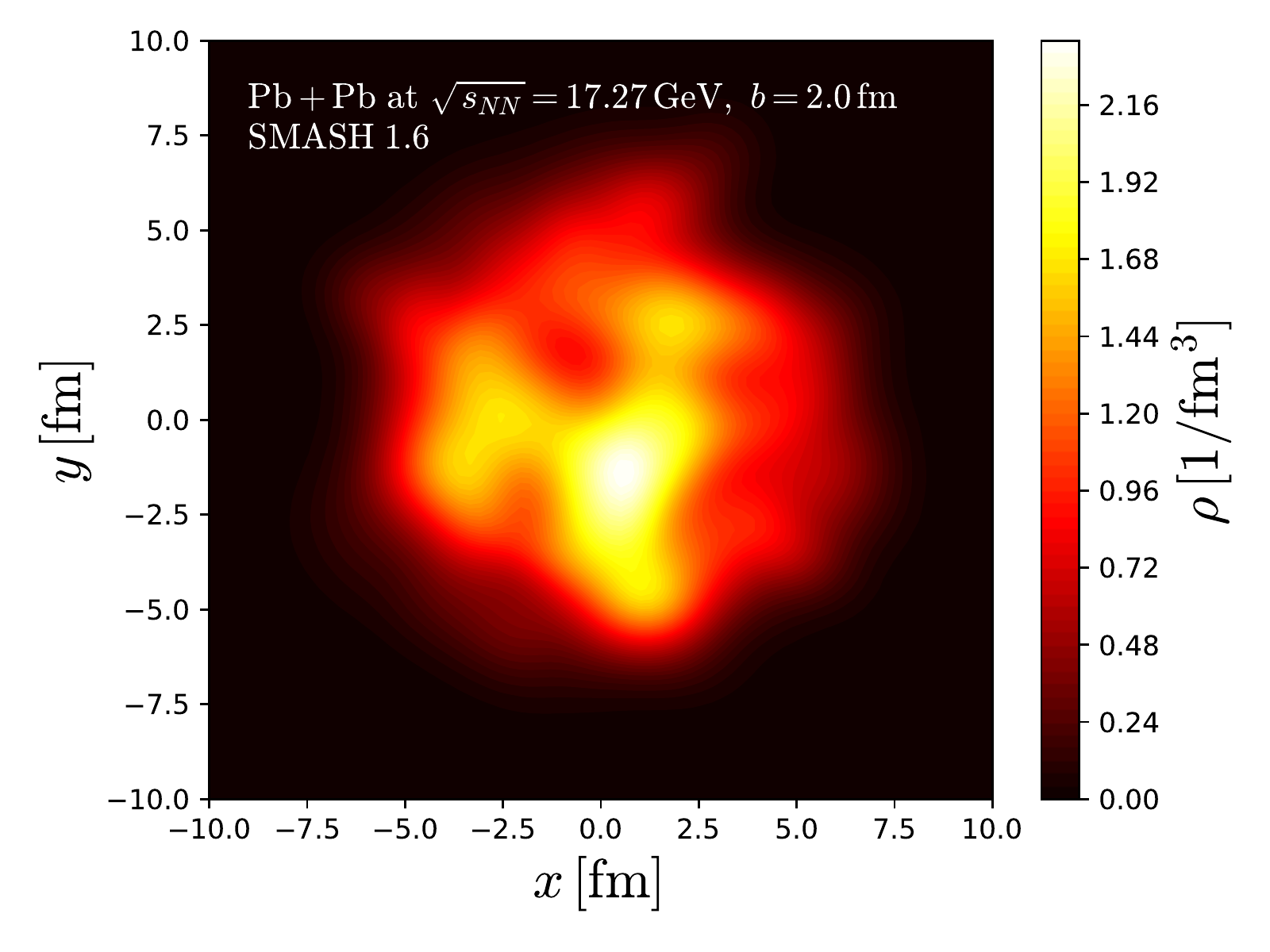}
	\caption{Energy density in the Landau rest frame including all hadrons in one heavy ion collision at $\sqrt{s_{NN}}=17.27\,\mathrm{GeV}$ with an impact parameter of $b=2.0\,\mathrm{fm}$ on the left and the net baryon density in the Eckart rest frame of the same event on the right.
	The energy density and the net baryon density are given at a specific time, exactly when the two colliding nuclei have passed through each other, and for a slice at $z=0\,\mathrm{fm}$.}
	\label{fig_2d_densities}
\end{figure}

The advantage of a dynamical approach is to include event-by-event fluctuations of all relevant quantities and having full 3D distributions for all quantum numbers available.
Figure \ref{fig_2d_densities} shows the energy density and the net baryon density in a single event for a slice at $z=0$ at the time when the two colliding nuclei have just passed through each other.

Due to secondary interactions and the produced transverse momentum, some strings are not aligned with the beam axis, which reflects in small lines of large energy density.
Since the baryon density of the string is located at the end, this structure cannot be observed on the right panel of figure \ref{fig_2d_densities}.
The scale on the left hand side of figure \ref{fig_2d_densities} ranges up to large energy densities, well in the regime where a quark-gluon plasma should be formed. Therefore, a description of the dynamical evolution in the hot and dense system in terms of hydrodynamics seems more appropriate than a pure hadronic transport approach.

\begin{figure}[h]
	\centering
	\includegraphics[width=0.5\textwidth]{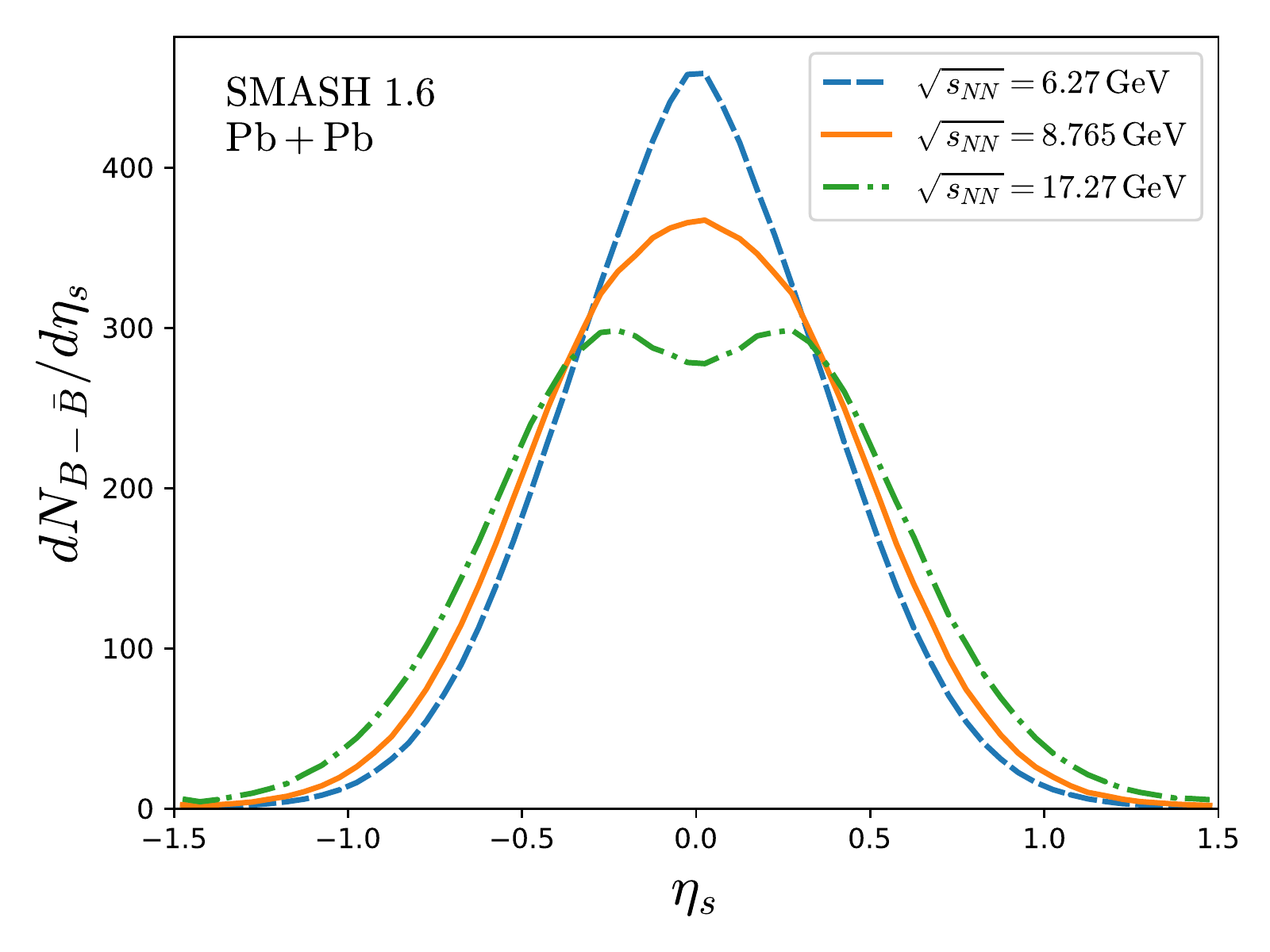}%
	\includegraphics[width=0.5\textwidth]{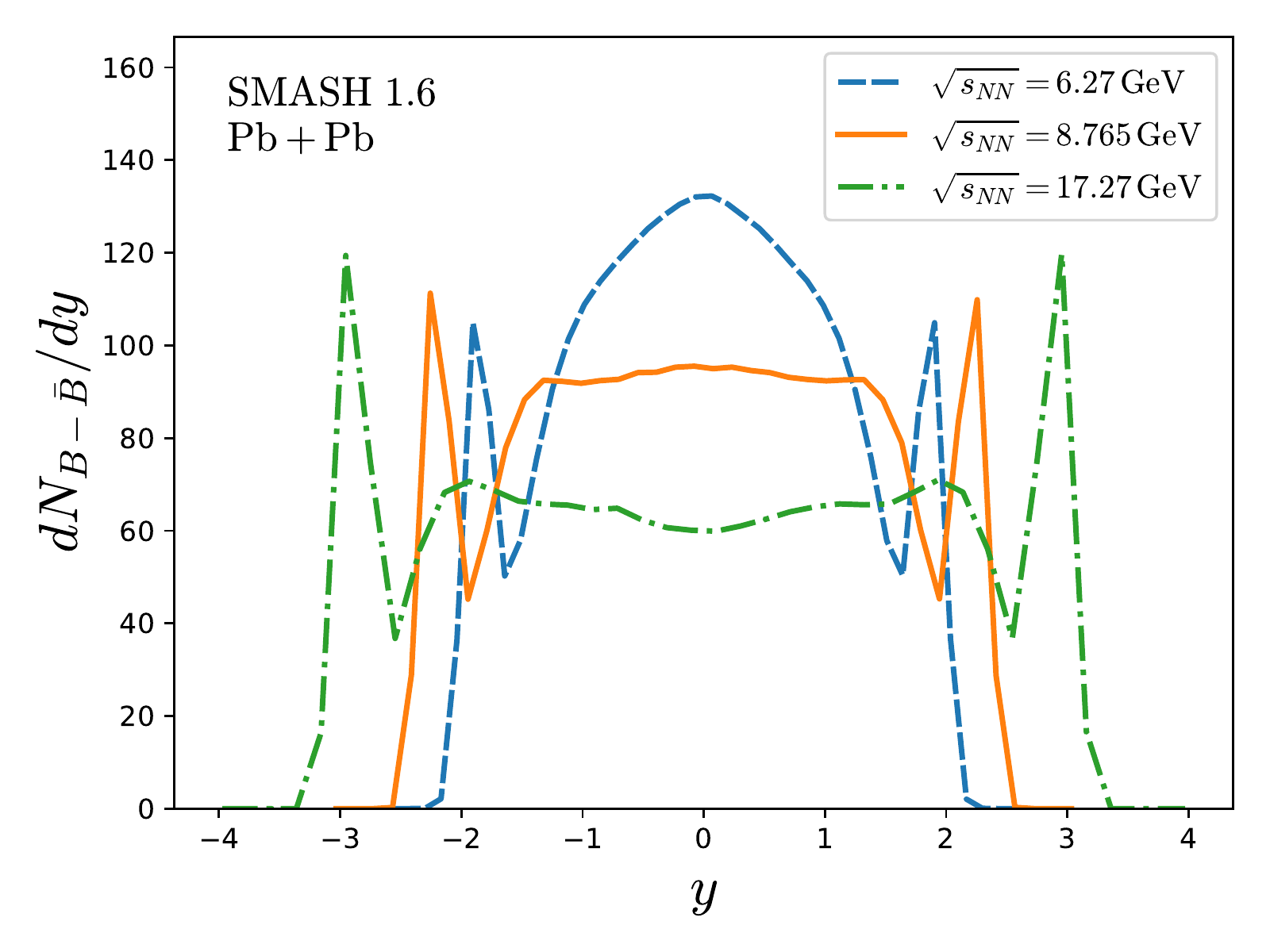}
	\caption{Spacetime rapidity (left) and momentum rapidity (right) spectra of net-baryons in central lead-lead collisions for different collision energies at the time when the nuclei have just passed through each other.}
	\label{fig_initial_y_eta}
\end{figure}

Advancing to the longitudinal dynamics of the system, figure \ref{fig_initial_y_eta} shows on the left panel the spacetime rapidity distribution of net baryons compared to the momentum space distribution on the right hand side in heavy ion collisions at different collision energies.
Again, the distributions are plotted at the time, when the nuclei have just passed through each other.
The momentum space distribution at $\sqrt{s_{NN}}=6.27\,\mathrm{GeV}$ shows a peak at mid-rapidity, while, with increasing energy, a flat plateau is developed.
Even though the spacetime rapidity spectrum shows a similar behavior with increasing energy, momentum space and coordinate space rapidity spectra differ drastically. This supports the finding, that the Bjorken assumption breaks down at lower beam energies and a full 3-dimensional initial state is more realistic. 

\section{Summary}
Baryon stopping in the SPS energy range is studied within the hadronic transport model SMASH.
Going to high collision energies, string excitations and fragmentations are the most important processes, since the contribution from resonance excitations fades out.
The string model inroduced in this work is split into soft and hard processes, where the soft processes dominate the cross section at intermediate energies while the hard processes are most important at very high energetic interactions. The soft string processes follow the UrQMD approach while hard processes are handled via \textsc{Pythia}. To take the dynamics of particle production in a string model into account, a formation time is introduced during which the cross sections of string fragments are reduced.
In order to mimic a continuous particle formation process, a mechanism is introduced to smoothly increase the cross section of forming particles over time.

The model has been benchmarked against experimental data from NA49 and NA61 in elementary proton-proton collisions and all parameters and their default values are explained. 
This comparison evidently shows that a distinct fragmentation function for leading baryons from soft non-diffractive string processes needs to be employed to get a reasonable agreement with the measured distribution of longitudinal momentum of protons.
Since many of the parameters are correlated and different observables require different values, we have presented the best possible compromise within the current approach. In the future, this multi-parameter problem might be assessed again employing Bayesian methods to allow for quantification of systematic uncertainties.

Fixing the parameters of the string routine to proton-proton collisions, a reasonable agreement over the entire SPS energy range is achieved considering that no existing theoretical model can describe all available data simultaneously in that energy range. Based on these parameters, the baryon stopping in heavy-ion collisions is investigated.
Since secondary interactions only play an important role for the dynamics of heavy ion collisions, the formation process of string fragments is studied in lead-lead collisions.
Comparing to experimental data, the best agreement was found for a formation time of $\tau_\mathrm{form}=1\,\mathrm{fm}$ in the rest frame of the respective string fragment and the cross section grows linearly in time during that period.
The proton and pion rapidity spectra closely follow the data but the proton multiplicity is overestimated at lower collision energies.
Further, the importance of non-isotropic elastic collisions is shown.
More forward-backward peaked angular distributions in elastic collisions are essential for reproducing the experimentally observed double peak structure in heavy ion collisions at top SPS energies.

Finally, SMASH is used to obtain event-by-event initial conditions for starting the evolution of the system in terms of hydrodynamics.
An energy density and net baryon density profile at the time right after the colliding nuclei have passed through each other is provided. These profiles indicate that realistic fluctuating initial conditions for all conserved charges can be obtained in the future. Another avenue for further research includes to explore dynamical initialization of hydrodynamics via source terms fed by the hadronic transport approach. 

\section{Acknowledgments}
Funded by the Deutsche Forschungsgemeinschaft (DFG, German Research Foundation) – Project number 315477589 – TRR 211.
This work was supported by the Helmholtz International
Center for the Facility for Antiproton and Ion Research (HIC for FAIR) within the
framework of the Landes-Offensive zur Entwicklung Wissenschaftlich-Oekonomischer Exzellenz (LOEWE) program launched by the State of Hesse.
Computational resources have been provided by the Center
for Scientific Computing (CSC) at the Goethe-University of Frankfurt.\\

\bibliography{bibfile}
\bibliographystyle{iopart-num}
\end{document}